\documentclass[%
reprint,
nofootinbib,
 amsmath,amssymb,
 aps,
]{revtex4-2}

\usepackage{graphicx}
\usepackage{dcolumn}
\usepackage{bm}
\usepackage{amsmath,amssymb}
\usepackage{amsthm}
\usepackage{mathtools}
\usepackage{appendix}
\usepackage{comment}
\usepackage{braket}
\usepackage{grffile}
\usepackage{mathrsfs}
\usepackage{url}
\usepackage{ulem}
\usepackage{xcolor}
\usepackage{microtype}   
\usepackage{xurl}        
\usepackage{hyperref}
\usepackage{booktabs}
\hypersetup{
    colorlinks=true,        
    linkcolor=magenta,      
    citecolor=magenta,      
    urlcolor=magenta        
}
\usepackage{orcidlink}
\usepackage{journal-abbrev}

\newcommand{\I}{\mathbf{I}}

\newcommand{\Gam}{\boldsymbol{\Gamma}}
\newcommand{\dd}{\mathop{}\!\mathrm{d}}
\newcommand{\btheta}{\boldsymbol{\theta}}
\newcommand{\bptheta}{\boldsymbol{p}}
\newcommand{\bxi}{\boldsymbol{\xi}}
\newcommand{\bk}{\boldsymbol{k}}
\newcommand{\bn}{\boldsymbol{n}}
\newcommand{\bx}{\boldsymbol{x}}
\newcommand{\by}{\boldsymbol{y}}
\newcommand{\bphi}{\boldsymbol{\phi}}
\newcommand{\bQ}{\boldsymbol{Q}}
\newcommand{\Tr}{\mathrm{Tr}}

\newcommand{\uu}{|u|^2}
\newcommand{\vv}{|v|^2}

\newcommand{\Zeven}{\mathbb{Z}^2_{\mathrm{even}}}
\newcommand{\bae}[1]{%
  \begin{align}
    #1
  \end{align}
}
\begin{document}

\preprint{APS/123-QED}


\title{Wigner function shapelets: Symplectic representation of astronomical images}

\author{Shun Arai\,\orcidlink{0000-0002-2527-3705}}
\email{shunarai@kmi.nagoya-u.ac.jp}
\affiliation{Kobayashi-Maskawa Institute for the Origin of Particles and the Universe (KMI), Nagoya University, Nagoya, 464-8602, Japan}
 
\date{\today}

\begin{abstract}
\noindent
We extend shapelets for the analysis of astronomical images to be available in a phase space, introducing \textit{Wigner function shapelets} (WFSs). Whereas conventional shapelets expand images separately in configuration or Fourier space using Hermite--Gaussian or Laguerre--Gaussian modes, WFSs represent images directly in the four-dimensional phase space with symplectic group $\mathrm{Sp}(4,\mathbb{R})$, which is quantized by a phase-space cell $2\pi\lambdabar$ that determines a resolution limit of a telescope. WFSs consist of a bilinear form of the cross-Wigner function of the Laguerre-Gaussian modes as an orthogonal and complete basis for the Wigner function of an image,  carrying out $\mathrm{SU}(2)$ irreducible representations of the phase space with the Hopf tori. We introduce a scalar function $\mathcal{W}_{k\ell} (Q_0,Q_2)$ from the $\mathrm{U}(1)\times \mathrm{U}(1)$ - covariant tori to a two-dimensional space of constants of motion $(Q_0,Q_2)$---the harmonic energy and axial angular momentum---thereby yielding a natural phase-space ``band structure'', given a pair of winding numbers $(k,\ell) \in \mathbb{Z}^2$.
The WFSs leverage key properties of the Wigner function for image analysis: (i) it encodes full information of an image in a symmetry-preserving way; (ii) its trasport equation naturally evolves with a Liouville equation at $\lambdabar \rightarrow 0$, (iii) it admits positive/negative oscillatory patterns on the $(Q_0,Q_2)$ plane that can be a sensitive spatial coherent structure of galaxy morphology and cosmological imprints; and (iv) systematics and noise can be manipulated as a quantum channel operation.  This paper aims to bring together all the formulae related to the Wigner function in the context of astrophysics and cosmology, formally organizing them in terminologies of both astronomy and quantum information theory.  
\end{abstract}

\maketitle

\tableofcontents
\clearpage

\section{Introduction}\label{sec:intro}
The shapelets formalism \cite{Refregier2003a, RefregierBacon2003, MasseyRefregier2005,MasseyEtAl2007} has been widely employed to analyze the morphology of astronomical bodies in two-dimensional configuration space. One of the major applications of the shapelets is to extract the shape distortions of galaxies induced by weak gravitational lensing, i.e., cosmic shear (see the latest review in \cite{Mandelbaum2018}) and flexions \cite{Bacon:2005qr}. In 2018,  \cite{LiEtAl2018} employed the shapelets to project the Fourier power function of a galaxy image in two-dimensional Fourier space,  introducing the Fourier power function shapelets (FPFS) method and accomplishing robust estimation of cosmic shear against measurement biases \cite{LiEtAl2022}. With the advent of high-resolution imaging surveys such as the James Webb Space Telescope, the Euclid mission, the Vera C. Rubin Observatory Legacy Survey of Space and Time, the Nancy Grace Roman Space Telescope (Roman), and the Square Kilometre Array Observatory \cite{Gardner2006,Laureijs2011,LSST2009,Braun2019,Spergel2015},  the analyses with shapelets continue to serve fiducial modeling so that any advanced methods can be validated and benchmarked \cite{TagoreKeeton2016,Ephremidze2025,Yatawatta2024}.

The beauty of using shapelets, regardless of whether in configuration space or Fourier space, lies in their symmetry-preserving representation. In fact, a linear response to gravitational lensing is directly analogous to quantum transitions between angular-momentum eigenstates in the $\mathrm{SU}(2)$ representation\footnote{We describe the mode functions and the selection rules with $\mathrm{SU}$(2) representation in the whole paper. One can choose $\mathrm{SO}$(3) representation alternatively.}. In this framework, Hermite-Gaussian (HG) and Laguerre-Gaussian (LG) modes are simply two bases of the two-dimensional isotropic oscillator, labeled by total angular momentum $j = N/2$ (with $N$ the total mode order) and spin quantum number $s$ with $-j \leq s \leq j$ \footnote{Throughout this paper, we use $s$ instead of $m$ that has been commonly used in other papers in order to avoid the confusion when the spherical harmonics is labeled in Sec.\ref{sssec:stats_corr_cosmo_ens}.}\cite{Nienhuis1993,Edmonds1957,Varshalovich1988}. 
Gravitational lensing couples these states according to familiar selection rules: cosmic shear, generated by quadratic operators, induces transitions with $\Delta s = \pm 2$ at fixed $j$  , while flexion, generated by cubic operators, produces $\Delta s = \pm 1$ and $\Delta s = \pm 3$. \cite{RefregierBacon2003,MasseyRefregier2005,MasseyEtAl2007}. 
Because both HG and LG modes form infinite, orthogonal, and complete bases, related by $\mathrm{SU}(2)$ rotations with amplitudes given by Wigner $d$-matrix elements \cite{Wigner1959,Edmonds1957,Varshalovich1988,Allen1992,Nienhuis1993}, it is mathematically natural to employ shapelets to represent the information encoded in galaxy images with either HG or LG modes. This construction reflects the same completeness and orthogonality principles that underlie the Hilbert space of quantum mechanics \cite{Dirac1930,Cohen1997}.

In this paper, we construct shapelets in the phase space described by the transverse coordinate on the image plane $\boldsymbol{\theta} = (\theta_x,\theta_y)$ and its conjugate momentum by $\boldsymbol{p} = (p_x,p_y)$, given the symplectic structure as 
$(\btheta,\bptheta) \in \mathrm{Sp}(4,\mathbb{R})$.  In the phase space, an astronomical image is represented by its Wigner function \cite{Wigner1932}\footnote{The Wigner function is often referred to as the Wigner-Ville distribution, following the work of J.~Ville in \cite{Ville1948}.}, and then we will claim that the shapelets that expand the Wigner function of the image, \textit{Wigner function shapelets} (WFSs), provides a natural basis with preserving symplectic geometry \cite{Wigner1932,Schleich2001,Leonhardt1997,LvovskyRaymer2009,Weedbrook2012}. The Wigner function $W(\btheta,\bptheta)$ plays a central role in quantum physics as a quasiprobability distribution, namely the phase-space representation of the density operator which governs the evolution of a system in quantum mechanics. Hence, it obeys the quantum Liouville (Moyal) flow that exhibits the conservation of information \cite{Groenewold1946,Moyal1949}, whose classical limit with $\hbar \rightarrow 0$\footnote{In this paper, we introduce $\lambdabar$ instead of $\hbar$ that denotes the scale where "quantization" of the image is given, following the notation \cite{SimonAgarwal2000}. We will describe this in more detail in Sec.~\ref{ssec:choice_lambdabar}.} is nothing but the classical Liouville equation. Crucially, because the Wigner function is bilinear in the underlying field, the Wigner-Weyl transform provides a unitary correspondence, i.e., the Stratonovich-Weyl correspondence\cite{Weyl1927,Stratonovich1956,VarillyGraciaBondia1989,BrifMann1999}, between Hilbert-Schmidt operators on the underlying Hilbert space and square-integrable functions on phase space \cite{HilleryEtAl1984,Folland1989, Schleich2001}. Within this representation, one can construct a Hilbert-Schmidt space spanned by the bilinear products of HG or LG modes. These modes, being orthogonal and complete by construction, offer a natural basis in which both image structure and its lensing-induced distortions can be decomposed and quantified, namely the WFS.

To extract the full information in the Wigner function, it is helpful to employ various mathematical tools in group representation in SU(2). For instance, through the associated Hopf fibration of phase space \cite{Hopf1931}, the Hilbert-Schmidt space decomposes naturally into diagonal sectors, depending only on the invariants $(Q_0,Q_2)$ ---the harmonic energy and axial angular momentum---and off-diagonal sectors, which arrange into torus harmonics. The diagonal sector encodes number-state content, while the off-diagonal coherences form band structures that isolate spin components relevant for image deformations, all while retaining locality in $(\btheta,\bptheta)$. The nonlocal or topological information isis embedded to Hopf-torus phases distinguished by a pair of torus-winding numbers $(k,\ell) \in \mathbb{Z}^2$. Thus, the Wigner representation of the image not only embeds galaxy images in phase space but also exposes the underlying group-theoretic architecture of their shape decomposition, including nonlocal and topological information. In other words, the Wigner function encapsulates correlations between configuration and Fourier space that are inaccessible to analyses carried out in either domain alone. 

It is important to note that the decomposition of the Wigner function in phase space is not a new idea: it has long been developed in quantum optics to characterize laser beam morphology \cite{KogelnikLi1966,Allen1992,Siegman1986, Nienhuis1993,Schleich2001}, within the paraxial approximation to the Helmholtz equation derived from Maxwell’s equations. The novelty of our work lies in transposing that formalism into the morphological analysis of galaxy images, which can likewise be modeled by paraxial propagation equations in observations of distant galaxies by telescopes\cite{BornWolf1999,Schroeder2000,Hecht2002}, and in introducing the concept of the WFS into the astrophysical context. We anticipate that this framework may open fertile avenues of interplay between quantum optics and astrophysics ahead of the future astronomical observations.

The rest of the paper is organized as follows. In Sec.~\ref{sec:wigner_for_galaxies} we will describe what the Wigner function of a galaxy image is, given the basic properties of the Wigner function. We will mention statistics and modulation of the Wigner function under systematics and noises, clarifying the relation between the Wigner function supported by the quantum information theory and the conventional astrophysical concepts in image analysis. In Sec.~\ref{sec:wfs}, we will introduce the formalism of WFSs and introduce the topologically sensitive observable $\mathcal{W}_{k\ell}$. In Sec.~\ref{sec:cosmo_info_wigner}, we will make use of the Wigner function and the WFSs to characterize the linear response to weak gravitational lensing and galaxy shape responses to any higher spin modes. We will give an observable that is sensitive to parity-violating signatures in galaxy-shape angular correlations. In Sec.~\ref{sec:practical_usage_in_image_analyses}, we will argue how the WFS is applied for practical image analysis, prescribing modest choices of the galaxy size parameter $\sigma$ and the one-dimensional phase-space cell $2\pi\lambdabar$.
Finally, we will conclude our paper. Technical materials are collected in the Appendices, supplying sufficient information that is necessary to understand the formulae in the main text. 


\section{Wigner function of galaxy images}\label{sec:wigner_for_galaxies}
Our theme is to analyze a galaxy image in phase space with the Wigner function, thereby extracting morphological information beyond what is accessible in configuration or Fourier space separately. In this section we define the Wigner function for an image field of a measured galaxy and summarize its essential mathematical properties given the Wigner function. We clarify how the Wigner function is related to the conventional astrophysical statistics, systematics, and noises.

\subsection{Basic properties}\label{ssec:basic_properties}
Let us consider the image plane with the coordinate $\btheta = (\theta_x,\theta_y)$. The phase space associated with the image plane is defined by introducing the conjugate momentum $\bptheta=(p_x,p_y)$ and the symplectic structure as $\mathrm{Sp}(4,\mathbb{R}):{\big\{ S \in GL(4,\mathbb{R}) \; \big|\; S^T \Omega S = \Omega \big\}}$ where 
\bae{
\label{eq:symplectic_map}
\Omega \;=\;
\begin{pmatrix}
0 & I_2 \\
- I_2 & 0
\end{pmatrix},
}
with $I_2$ being the $2\times 2$ identity. $\mathrm{Sp}(4,\mathbb{R})$ contains translations, rotations, squeezes, and more general linear canonical transforms, 
all of which relates to physical deformation of an image.

Let us consider an image field of a galaxy $\psi(\btheta)$. In practice, $\psi(\btheta)$ is often a real positive function as it loses phase information of electromagnetic fields, but we treat $\psi(\btheta)$ as an arbitrary field that can be complex in general. 
The Wigner function of the image is defined as
\bae{
\label{eq:wigner-def}
W(\btheta,\bptheta) \equiv \frac{1}{(2\pi \lambdabar)^2} \int{\dd^2\bxi} \psi(\btheta+\bxi/2)\psi^*(\btheta-\bxi/2)e^{-\frac{i\bptheta\cdot\bxi}{\lambdabar}}\,.
}
Here $\lambdabar$ is the scale of the discretization of phase space that determines the canonical commutation relations
\bae{
\label{eq:xp_commutation_relations}
[\hat{\theta}_i,\hat{p}_j] = i \lambdabar\delta_{ij}\,, [\hat{\theta}_i,\hat{\theta}_j] = [\hat{p}_i, \hat{p}_j] = 0\,,
}
with $i,j = x, y$.
The Wigner function defined by Eq.~\eqref{eq:wigner-def} obeys several well-known properties as follows.
\paragraph*{Reality}
$W(\btheta,\bptheta) \in \mathbb{R}$ for any $\psi(\btheta)$
\paragraph*{Parity}
The Wigner function is invariant under the parity transform in phase space as
\bae{
\hat{\Pi} W(\btheta,\bptheta)
= W(\btheta,\bptheta),
}
where $\hat{\Pi}$ acts as reflection in phase space,
$\hat{\Pi} A(\btheta,\bptheta) = A(-\btheta,-\bptheta)$.
\paragraph*{Chirality}
The Wigner function is invariant under the chiral transform in the phase space; 
\bae{
\hat{C} W(\btheta,\bptheta)
= W(\btheta,\bptheta),
}
where $\hat{C}$ acts as reflection in phase space,
$\hat{C} A(\btheta,\bptheta) = A(\btheta,-\bptheta)$.  Since the Wigner function of the single image is always real, the image itself preserves the chirality. 

\paragraph*{Symplectic covariance}
Let $S\in \mathrm{Sp}(4,\mathbb{R})$ act on phase-space points
$z=(\btheta,\bptheta)$ as $z\mapsto S z$.
We adopt the pullback convention
\(
\hat{\mathcal{S}}F(z)\equiv F(S^{-1}z).
\)
Then the Wigner map is covariant under any linear canonical (symplectic) transformation:
\[
W \;\mapsto\; \hat{\mathcal{S}}W, \qquad {S} \in \mathrm{Sp}(4,\mathbb{R}).
\]
Note that translations, rotations, shears, and squeezes are all included.

\newcommand{\ddmu}{\dd^2\btheta\,\dd^2\bptheta}

\paragraph*{Weyl symbols and expectations} Let us define the state $\hat{\rho}$ for the galaxy field with the kernel $\rho(\btheta_1,\btheta_2) = \psi(\btheta_1)\psi^*(\btheta_2)$. An observable $\hat A$ with kernel $A(\btheta_1,\btheta_2)$
defines the Weyl symbol of  $\hat{A}$ as 
\bae{
A_W(\btheta,\bptheta)
&=
\int \dd^2\bxi\;
A\!\left(\btheta+\tfrac{\bxi}{2},\,\btheta-\tfrac{\bxi}{2}\right)
e^{-\frac{i\bptheta\cdot\bxi}{\lambdabar}}\,.
}
Then the expectation value is the phase-space inner product
\bae{
\langle \hat A\rangle \;=\; \mathrm{Tr}(\hat\rho\,\hat A)
\;=\; \int{\dd \mu(\btheta,\bptheta)}A_W(\btheta,\bptheta)\,W(\btheta,\bptheta)\,.
}
Here $\dd \mu(\btheta,\bptheta) \equiv \dd^2\btheta \dd^2 \bptheta$ denotes the Liouville measure.  If $\mathrm{Tr}\,\hat\rho=1$ then $\int\dd\mu W=1$. Although $W$ may take negative values,
$W$ is called as a ``quasi" probability density.

\paragraph*{Wigner transport}
Suppose that the galaxy field can evolve along an optical path $s$, namely $\psi(\btheta,s)$. Let $\hat H$ be the generator of evolution along an optical path parameter $s$, and
$H_W(\btheta,\bptheta,s)$ its Weyl symbol. The Moyal (star) product on the phase space is defined between the two Weyl symbols as 
\bae{
A_W\star B_W \;&=\; A_W\,\exp\!\Big[\frac{i\lambdabar}{2}\overleftrightarrow{\Lambda}\Big]\,B_W, \cr
\qquad
\overleftrightarrow{\Lambda}
&\equiv \overleftarrow{\nabla}_{\btheta}\!\cdot\!\overrightarrow{\nabla}_{\bptheta}
 - \overleftarrow{\nabla}_{\bptheta}\!\cdot\!\overrightarrow{\nabla}_{\btheta},
}
and the associated Moyal bracket is
\bae{
\{A_W,B_W\}_{\rm MB}=\tfrac{2}{\lambdabar}\,A_W\,
\sin\!\big(\tfrac{\lambdabar}{2}\overleftrightarrow{\Lambda}\big)\,B_W.
}
The von Neumann equation $i\lambdabar\,\partial_s\hat\rho=[\hat\rho,\hat H]$
maps to the exact Wigner transport as
\bae{
\label{eq:wigner_transport}
\partial_s W(\btheta,\bptheta,s) \;=\; \{W, H_W\}_{\rm MB}.
}
Expanding the sine in the limit of $\lambdabar \rightarrow 0$ gives the classical Liouville equation as
\bae{
\partial_s W \;&=\; \{W, H_W\}_{\rm PB}. \;+\; \mathcal{O}(\lambdabar^{\,2}),\cr
\{W, H\}_{\rm PB}.&=\nabla_{\btheta}W\!\cdot\!\nabla_{\bptheta}H_W
        -\nabla_{\bptheta}W\!\cdot\!\nabla_{\btheta}H_W\,.
}
Note that this is a generalization of the radiative-transfer equation of an image, including the phase-space information. In Appendix.~\ref{app:classical_lim_wigner}, we derive that $H_W$ matches the classical Hamiltonian and the Wigner function approaches the spectral radiance $L(\btheta,\bptheta)$. 
\paragraph*{Duality to characteristic function}
The Wigner function $W(\btheta,\bptheta)$ is the phase-space representation
dual to the Weyl characteristic function, which serves as the fundamental
generating an object of the image density matrix.
Defining the characteristic function as the expectation value of the
phase--space displacement operator,
\bae{
\chi(\boldsymbol{\eta},\boldsymbol{\zeta})
:=
\Tr\!\left[
\hat\rho\,
\exp\!\Big(
\frac{i}{\lambdabar}
(\boldsymbol{\eta}\!\cdot\!\hat\btheta
+\boldsymbol{\zeta}\!\cdot\!\hat\bptheta)
\Big)
\right]\,.
}
One finds that $\chi$ generates all symmetrized moments of the canonical
variables and encodes both the probabilistic content (diagonal part) and
the coherence structure (off-diagonal part) of the density matrix.
The Wigner function is obtained by a Fourier transform of $\chi$ with respect to
to the source variables,
\bae{
W(\btheta,\bptheta)
=
\frac{1}{(2\pi\lambdabar)^2}
\int \dd^2\boldsymbol{\eta}\,\dd^2\boldsymbol{\zeta}\;\chi(\boldsymbol{\eta},\boldsymbol{\zeta})
e^{\!
-\frac{i}{\lambdabar}
(\boldsymbol{\eta}\!\cdot\btheta
+\boldsymbol{\zeta}\!\cdot\bptheta)
}\,,
}
while the inverse relation reconstructs $\chi$ from $W$ by the inverse Fourier
transformation.
This establishes a one-to-one duality between the density matrix $\hat\rho$,
its characteristic function $\chi$, and the Wigner function $W$, showing that
the latter is not an independent object but a coordinate representation of
the same underlying statistical state.

\paragraph*{Marginals} 
The marginalization of the Wigner function in Eq.~\eqref{eq:wigner-def} derives the well-known quantities. The integration over the configuration space gives the Fourier power spectrum of $\psi(\btheta)$ : $F(\bk) = |\tilde{\psi}(\bk)|^2$ at $\bk = \bptheta/\lambdabar$ as,
\bae{
F(\bptheta/\lambdabar) = \int{\dd^2\btheta}{W(\btheta,\bptheta)}\,,
}
whereas the integration over the momentum space gives the intensity of $\psi(\btheta)$ : $I(\btheta) = |\psi(\btheta)|^2$ as
\bae{
I(\btheta) = \int{\dd^2\bptheta}{W(\btheta,\bptheta)}\,.
}
Note that the normalization for the intensity is fixed at the limit $\lambdabar \rightarrow 0$, so that corresponding to the standard normalization.
One can see that the information in the phase space described by the Wigner function is marginalized into either $F(\bptheta/\lambdabar)$ or $I(\btheta)$. In other words, the direct analysis of the Wigner function in the phase space can possess the information that is integrated out in these marginals.

The Wigner function in Eq.~\eqref{eq:wigner-def} is computed with the single function.  In other words, we assume a galaxy image at a certain filter in colors or fixed polarization and, therefore, the image is a single function. In general, one can obtain multiple images of the same source galaxy with different color bands and polarizations. Moreover, even for a single image, the expansion of the Wigner function with a certain basis e.g. HG modes or LG modes, creates cross-terms in different quantum numbers and indeed these are nonzero at local points in the phase space. To make the image analysis complete in the phase space, such cross-terms must be taken into account, and we must be careful, as some leakage occurs in image reconstruction. After all, it is significant to introduce the cross-Wigner function as follows.
The cross-Wigner function is defined with the two different functions $\psi_a(\btheta)$ and $\psi_b(\btheta)$ as
\bae{
\label{eq:xwigner-def}
W_{ab}(\btheta,\bptheta) \equiv \frac{1}{(2\pi \lambdabar)^2} \int{\dd^2\xi} \psi_a(\btheta+\bxi/2){\psi_b}^*(\btheta-\bxi/2)e^{-\frac{i\bptheta\cdot\bxi}{\lambdabar}}\,.
}
Note that the abstract indices $a,b$ can be the pair of the quantum numbers as $a=(j,s)$, as well as denoting multiple states of a source galaxy. It is sesquilinear (linear in $ \psi_a $ and conjugate-linear in $ \psi_b $), satisfies $ W_{ba}=W_{ab}^* $, and reduces to the ordinary Wigner function when $ a=b $. Note that the cross-Wigner function is no longer parity invariant by definition unless $a=b$.

The covariance of the Wigner function is not that simple compared to Eq.~\eqref{eq:wigner-def}, but it is still defined in a covariant manner.  For any unitary $U$ acting on fields [so the indices transform as $(a,b)\mapsto(Ua,Ub)$], the cross-Wigner funciton covariantly transforms as
\bae{
W_{Ua,\,Ub}(z)=
\begin{cases}
W_{ab}(z-\zeta), & U=D(\zeta)\\
W_{ab}(S^{-1}z), & U=\mu(S),
\end{cases}
}
where $D(\zeta)$ and $\mu(S)$ denote the Heisenberg-Weyl displacement and the metaplectic transformations, respectively. Note that the linear transformation between HG and LG modes via the Wigner $d$-matrix is one of the metaplectic transformations.  There is another deformation of polarization is described by a unitary transform $U$ in $\mathrm{SU}$(2). For instance, the Faraday rotation is described by this group of unitary operations, realising the covariance of the cross-Wigner function.  The transform between images in colors of the single galaxy, in contrast, is not in general unitary because it may include diffusion, absorption, emission, and convolution with instruments and atmospheric effects. Nevertheless these effects may be built in the cross-Wigner transport equation along the optical path $s$ as a straightforward extension of the Wigner transport, Eq.~\eqref{eq:wigner_transport}, with open system, e.g., the Lindbladian ( see a short review in \cite{2020AIPA...10b5106M}). In this paper, we focus in employing the cross-Wigner function to complete the basis representation in the phase space.

The marginals reproduce cross-kernels:
\bae{
\int\!\dd^2\bptheta\, W_{ab}(\btheta,\bptheta)&= \psi_a(\btheta)\,\psi_b^*(\btheta),\cr
\int\!\dd^2 \btheta\, W_{ab}(\btheta,\bptheta)&=\tilde\psi_a(\bptheta/\lambdabar)\,\tilde\psi_b^*(\bptheta/\lambdabar),
}
and, with the normalization in \eqref{eq:xwigner-def}, the total integral yields the inner product
\bae{
\label{eq:xwigner_marginals_to_innerproduct}
\int\!\dd \mu(\btheta,\bptheta)\, W_{ab}(\btheta,\bptheta)=\langle \psi_b\,|\,\psi_a\rangle,
}
where the inner product is defined as
\bae{
\label{eq:inner_product_in_config_sp}
\langle \psi_b\,|\,\psi_a\rangle \equiv \int\!\dd^2\btheta\, \psi_a(\btheta)\,\psi^*_b(\btheta)\,.
}
The Moyal/Plancherel identity reads
\bae{
\label{eq:moyal_plancherel_id}
{\langle\,W_{cd} | W_{ab} \rangle}_{\rm HS}
=\frac{1}{(2\pi\lambdabar)^2}\,\langle \psi_a\,|\,\psi_c\rangle\,\langle \psi_b\,|\,\psi_d\rangle\,,
}
where we introduce the inner product in a Hilbert-Schmidt space as
\bae{
\langle B | A \rangle_{\rm HS} \equiv \mathrm{Tr}(\hat{B}^\dagger \hat{A}) = \int{\dd \mu A_W(z)B^*_W(z)}\,.
}
The identity Eq.~\eqref{eq:moyal_plancherel_id} is used to project a Wigner function onto the basis function in the next section. 

Let us comment on the uniqueness of the Wigner function. Given the Stratonovich-Wigner axioms: translation covariance, Hermiticity/self-duality, standardization, and metaplectic covariance, the kernel is uniquely fixed (up to the sign set by exact marginals) to be the displaced parity. Hence, the Stratonovich-Weyl correspondence \cite{Stratonovich1956} built from parity-reflected displacements is identical to the Wigner transform used in Eq.~\eqref{eq:wigner-def}:
\bae{
\hat w(z)=\hat D(z)\,\hat\Pi\,\hat D(z)^\dagger\,,
}
and this defines the Wigner function as $W =  {\rm Tr}[\hat{\rho}\hat{w}]$ given the state $\hat{\rho}$.
%
%
\subsection{Statistics for the Wigner function}\label{sec:stats_wigner}
We describe several statistical quantities derived from the Wigner function, recalling that the Wigner function is a phase-space quasiprobability distribution of an underlying state described by a density matrix of a galaxy image. We aim to organize this section so that the concept of the Wigner function fits into the conventional astrophysical quantities.

\subsubsection{Van Cittert--Zernike theorem, Wigner function, and speckle statistics}
\label{sssec:VCZ_Wigner_speckle}

We summarize how the Van~Cittert--Zernike (VCZ) theorem \cite{vanCittert1934,Zernike1938},
statistical optics \cite{Goodman1985}, and speckle statistics \cite{Goodman1976, Goodman2005} are unified in the Wigner function
representation adopted in this work, following a concise derivation in \cite{Cerbino2007}. In the context of astronomical imaging, the VCZ theorem states how electromagnetic waves emitted from an astronomical body turn into an observable at a detector or a telescope. Statistics optics treats the persistence of the same astronomical body on the image plane by averaging a chunk of images that can fluctuate on the image plane over observational time. Speckle statistics is one of the popular representations of statistical optics, which formalizes the statistical displacement of each image of the same astronomical body in a resolved time period. In what follows, we describe how naturally these quantities are harnessed by the Wigner function.

Let us assume that astronomical sources emit spatially incoherent (thermal)
electromagnetic wave radiation\footnote{One may wonder whether the spatial structure of an observed astronomical object is spatially coherent-- spirals, bars, jets etc. What we note here is the incoherence of the radiation, and the spatial structures remain spatially coherent with dependence of resolution of images.}.  The image field $\psi(\btheta)$ is treated as a product through random complex
processes, and its second-order statistics are encoded in the mutual intensity
(cross-spectral density)
\bae{
\Gamma(\btheta_1,\btheta_2)
\;\equiv\;
\big\langle
\psi(\btheta_1)\,\psi^\ast(\btheta_2)
\big\rangle_{\rm obs.}\,,
}
where $\langle\cdots\rangle_{\rm obs.}$ denotes an ensemble (or ergodic time) average over
statistically equivalent field realisations. Inserting the image field $\psi$ into the definition of the Wigner function generates the Wigner function. Ensemble averaging over the time-varying electromagnetic field yields the Weyl transform of the mutual intensity,
\bae{
\label{eq:Wigner_ensemble}
\big\langle W(\btheta,\bptheta)\big\rangle_{\rm obs.}
=
\frac{1}{(2\pi\lambdabar)^2}
\int \dd^2\bxi\;
\Gamma\!\left(
\btheta+\frac{\bxi}{2},
\btheta-\frac{\bxi}{2}
\right)
e^{-\frac{i\bptheta\cdot\bxi}{\lambdabar}}.
}
Equation.~\eqref{eq:Wigner_ensemble} is inverted by the Fourier transform, and we obtain
\bae{
\label{eq:mutual_intensity_from_Wigner_ensemble}
\Gamma\!\left(
\btheta+\frac{\bxi}{2},
\btheta-\frac{\bxi}{2}
\right) = 
\int \dd^2\bptheta\;
\big\langle W(\btheta,\bptheta)\big\rangle_{\rm obs.}\,.
}
Equation.~\eqref{eq:mutual_intensity_from_Wigner_ensemble} indicates that the mutual information is a marginal of the ensemble-averaged Wigner function.
Let us further assume that the observed image is no longer spatially incoherent due to strong image blurring in observational time, namely,  the detailed structures are smoothed away. In this extreme case, the mutual intensity is reduced to the brightness distribution
$I_S(\boldsymbol{\beta})$, and the VCZ theorem implies that the mutual intensity at
the telescope is the Fourier transform of the source intensity,
\bae{
\label{eq:VCZ_angular}
\Gamma\!\left(
\btheta+\frac{\bxi}{2},
\btheta-\frac{\bxi}{2}
\right)
=
\int \dd^2\beta\;
I_S(\boldsymbol{\beta})\,
e^{
-\frac{i}{\lambdabar}\,
\boldsymbol{\beta}\cdot\bxi
},
}
where $\bxi$ is the angular separation variable.  The propagation distance does
not appear explicitly, as the geometry is fully encoded in angular coordinates.
Substituting Eq.~\eqref{eq:VCZ_angular} gives
\bae{
\label{eq:VCZ_Wigner_result}
\big\langle W(\btheta,\bptheta)\big\rangle_{\rm obs.}
\propto
I_S(\bptheta),
}
showing that, for an incoherent source, the ensemble-averaged Wigner function is
independent of $\btheta$ and directly encodes the source brightness in the conjugate-
momentum space.

A single observed image corresponds to one realization of the random field
$\psi(\btheta)$ and exhibits speckle due to random phase interference.  Let us define the intensity of the image field as
\bae{
I(\btheta) \equiv |\psi(\btheta)|^2 .
}
Its ensemble mean is given by the diagonal of the mutual intensity,
\bae{
\label{eq:mean_intensity}
\langle I(\btheta)\rangle_{\rm obs.}
=
\Gamma(\btheta,\btheta)
=
\int \dd^2\bptheta\;
\big\langle W(\btheta,\bptheta)\big\rangle_{\rm obs.} .
}
Speckle fluctuations are quantified by the intensity-intensity correlation
function
\bae{
G^{(2)}(\btheta_1,\btheta_2)
\equiv
\langle I(\btheta_1) I(\btheta_2)\rangle_{\rm obs.} .
}
If the field $\psi$ is a circular complex Gaussian random process (fully developed speckle), the fourth-order moments reduce via the Siegert relation \cite{Siegert1943, Lax1960, HBT1956,HBT1957} \footnote{In optics, the Siegert relation is often written with a normalized version in which the first term in the right-hand side of Eq.~\eqref{eq:Siegert} is unity.},
\bae{
\label{eq:Siegert}
\langle I(\btheta_1) I(\btheta_2)\rangle_{\rm obs.}
=
\langle I(\btheta_1)\rangle_{\rm obs.}\,
\langle I(\btheta_2)\rangle_{\rm obs.}
+
\big|
\Gamma(\btheta_1,\btheta_2)
\big|^2 .
}
The Siegert relation can be written entirely in terms of the ensemble-averaged
Wigner function,
\bae{
\label{eq:Siegert_Wigner}
\langle I(\btheta_1) I(\btheta_2)\rangle
&=
\langle I(\btheta_1)\rangle\,
\langle I(\btheta_2)\rangle\, \cr
&+
\left|
\int \dd^2p\;
\big\langle
W\!\left(\tfrac{\btheta_1+\btheta_2}{2},\bptheta\right)
\big\rangle_{\rm obs.}
e^{+i\frac{\bptheta\cdot(\btheta_1-\btheta_2)}{\lambdabar}}
\right|^2\,. \cr
}
Eqs.
~\eqref{eq:VCZ_Wigner_result}--~\eqref{eq:Siegert_Wigner}
make explicit that the VCZ theorem fixes the ensemble-averaged background
Wigner distribution, while speckle arises from higher-order statistics of
single realisations and is governed by the modulus-squared coherence encoded in
$\langle W\rangle_{\rm obs.}$.
Throughout the rest of the main text, we implicitly consider the statistical ensemble over an observational time window, and we abbreviate the notation of the ensemble until it is necessary to show.

%
\subsubsection{Statistical correlations over cosmological image ensemble}
\label{sssec:stats_corr_cosmo_ens}

We consider a cosmological ensemble of galaxy images whose centroids are located at directions
$\boldsymbol{n}_c \in S^2$.  Around each centroid, we introduce local tangent-plane coordinates and define
two nearby directions
\begin{align}
\boldsymbol{n}_\pm \equiv \exp_{\boldsymbol{ n}_c}\!\left(\boldsymbol{\theta} \pm \frac{\boldsymbol{\xi}}{2}\right),
\qquad
|\boldsymbol{\theta}|,|\boldsymbol{\xi}|\ll 1,
\end{align}
where $\boldsymbol{\theta}$ and $\boldsymbol{\xi}$ denote small angular separations orthogonal to
$\boldsymbol{n}_c$.
The local Wigner function associated with the image field $\psi(\boldsymbol{n}) = \psi(\btheta;\boldsymbol{n}_c)$ is rewritten as
\begin{align}
W(\boldsymbol{\theta},\boldsymbol{p};{\boldsymbol{n}_c})
=
\frac{1}{(2\pi\lambdabar)^2}
\int \dd^2\bxi\;
\psi(\boldsymbol{n}_+)\psi^\ast(\boldsymbol{n}_-)e^{-\frac{i\bptheta\cdot\bxi}{\lambdabar}}\,,
\label{eq:wigner_local}
\end{align}
which is the Fourier transform, with respect to the relative angular separation $\boldsymbol{\xi} = \boldsymbol{n}_+ - \boldsymbol{n}_-$
of the local configuration-space bilinear.
Taking an ensemble average over realizations of the cosmological image field yields
\begin{align}
\langle W(\boldsymbol{\theta},\boldsymbol{p};{\boldsymbol{n}_c}) \rangle_{\rm cosmo.}
=
\frac{1}{(2\pi\lambdabar)^2}
\int \dd^2\xi\;
C(\boldsymbol{n}_+,\boldsymbol{n}_-)\,
e^{-\frac{i\bptheta\cdot\bxi}{\lambdabar}}\,,
\label{eq:wigner_ensemble_avg}
\end{align}
where
\begin{align}
C(\boldsymbol{n},\boldsymbol{n}')
\equiv
\big\langle \psi(\boldsymbol{ n})\psi^\ast(\boldsymbol{n'}) \big\rangle_{\rm cosmo.}\,,
\end{align}
is the two-point angular correlation function of the ensemble. The correlation function on the sphere admits a complete expansion as
\begin{align}
C(\boldsymbol{n},\boldsymbol{n'})
=
\sum_{\ell_1\ell_2}\sum_{LM}
A^{LM}_{\ell_1\ell_2}\,
\Big\{Y_{\ell_1}\otimes Y_{\ell_2}\Big\}_{LM}(\boldsymbol{n},\boldsymbol{n'}),
\label{eq:biposh_expansion}
\end{align}
where $\ell_{1,2}$ are the multipole moments and $m_{1,2}$ are magnetic quantum numbers, involving a total azimuthal quantum number $L$ and a total magnetic quantum number $M$ as $|\ell_1 - \ell_2| \leq L \leq \ell_1 + \ell_2$ and $ M = m_1 + m_2$. $\{Y_{\ell_1}\otimes Y_{\ell_2}\}_{LM}$ denotes the bipolar spherical harmonics (BiPoSHs) \cite{1988qtam.book.....V, 2010PhRvD..81h3012J,2012PhRvD..85b3010B} that is defined as
\bae{
&\{Y_{\ell_1}\otimes Y_{\ell_2}\}_{LM} (\boldsymbol{n},\boldsymbol{n}') \cr
& \qquad = \sum_{m_1,m_2}\langle\ell_1 m_1 \ell_2 m_2|LM \rangle Y_{\ell_1 m_1}(\boldsymbol{n})Y_{\ell_2 m_2}(\boldsymbol{n'})\,. \cr
}
Note that $\langle\ell_1 m_1 \ell_2 m_2|LM \rangle$ are Clebsch-Gordan coefficients. The inversion of the bipolar spherical harmonics is derived as
\bae{
Y_{\ell_1,m_1}(\boldsymbol{n})Y_{\ell_2 ,m_2}(\boldsymbol{n'}) = \sum_{LM} \langle\ell_1 m_1 \ell_2 m_2|LM \rangle \{Y_{\ell_1}\otimes Y_{\ell_2}\}_{LM}\,.
}
After projecting out the basis, $A^{LM}_{\ell_1 \ell_2}$ is represented as
\bae{
\label{eq:ALM_l1l2}
A^{LM}_{\ell_1 \ell_2} = \sum_{m_1, m_2}\langle a_{\ell_1 m_1} a^*_{\ell_2 m_2}\rangle_{\rm cosmo.}\langle\ell_1 m_1 \ell_2 -m_2|LM \rangle
}
where $a_{\ell m}$ denotes the multipole coefficient for the expansion of $\psi$.
$A^{LM}_{\ell_1\ell_2}$ encode the statistical anisotropy of the ensemble. 
In the statistically isotropic limit, only the $L=0$ component survives, reproducing the usual
angular power spectrum. Substituting Eq.~\eqref{eq:biposh_expansion} into
Eq.~\eqref{eq:wigner_ensemble_avg}, we obtain
\begin{align}
\langle W(\boldsymbol{\theta},\boldsymbol{p};\boldsymbol{n}_c) \rangle_{\rm cosmo.}
=
\sum_{\ell_1\ell_2}\sum_{LM}
A^{LM}_{\ell_1\ell_2}\,
\mathcal{K}^{LM}_{\ell_1\ell_2}
(\boldsymbol{\theta},\boldsymbol{p};\boldsymbol{n}_c),
\label{eq:wigner_biposh}
\end{align}
where the kernel
\begin{align}
\mathcal{K}^{LM}_{\ell_1\ell_2}
\equiv
\frac{1}{(2\pi\lambdabar)^2}
\int \dd^2\bxi\;
\Big\{Y_{\ell_1}\otimes Y_{\ell_2}\Big\}_{LM}(\boldsymbol{n}_+,\boldsymbol{n}_-)\,
e^{-\frac{i\bptheta\cdot\bxi}{\lambdabar}}\,,
\end{align}
acts as a phase-space response function relating bipolar angular correlations to the local Wigner
representation. Equation.~\eqref{eq:wigner_biposh} demonstrates that the ensemble-averaged Wigner function naturally
inherits decomposition with BiPoSHs. 
This provides a direct bridge between cosmological two-point statistics on the sphere and local
phase-space descriptions of galaxy images. One can apply the same logic for the cross-Wigner function.

At last, let us briefly mention the two-point angular correlation of the Wigner function composed of the two galaxy images, $\langle W_{11}(\bn_{c1}) W_{22}(\bn_{c2}) \rangle_{\rm cosmo.}$, setting the same lab frame of the two images. Note that are the galaxies are located at $\bn_{c1}$ and $\bn_{c2}$, respectively. We obtain that $\langle W_{11}(\bn_{c1}) W_{22}(\bn_{c2}) \rangle_{\rm cosmo.}$ is composed of the four-point correlation functions of the galaxy images, including the trispectrum of the galaxy image.  We shall not show any detailed expression of the formula to keep the presentation concise, whereas we adhere to the essential meaning of such an ensemble. In practice, one may be interested in the configuration in which the two galaxy samples are distant compared to their own size, namely $|\bn_{c1}-\bn_{c2}| \gg |\btheta|$. In this case, the four-point correlation function takes its collapsed limit cf. \cite{2025arXiv250908787K} in Fourier space. 
In Sec.~\ref{sec:cosmo_info_wigner}, we will utilize $\langle W_{ab}\rangle_{\rm cosmo.}$ to measure the cosmic shear as the standard analysis for weak gravitational lensing does. In addition, we will mention that one can build a parity-violating estimator within $\langle W_{11}(\bn_{c1}) W_{22}(\bn_{c2}) \rangle_{\rm cosmo.}$, after projecting into the spin-sensitive observables that are defined in Sec.~\ref{ssec:gal_shape_response_spin_modes}.
\subsection{Systematics and noise in the Wigner function}\label{ssec:systematics_noise_wigner}
We introduce systematics and noises on a galaxy image in terms of the Wigner function. Respecting that the Wigner function is the phase-space representation of an image state
$\hat{\rho}$, we describe systematics and noises with corresponding algebraic operations in quantum information theory. To begin with this subsection, we provide a generic operation that characterizes systematics and noises. Then, we examplify the two typical cases divided by whether the point spread function (PSF) is deterministic or stochastic.

A physical process acting on a state $\hat{\rho}$ is represented by a
completely positive trace-preserving (CPTP) map,
\bae{
\mathcal{E}:\ \hat{\rho}\ \longmapsto\ \mathcal{E}(\hat{\rho}) .
}
Any CPTP map admits a Kraus representation,
\bae{
\mathcal{E}(\hat{\rho})
=
\sum_{a} \hat{K}_{a}\,\hat{\rho}\,\hat{K}_{a}^{\dagger},
\qquad
\sum_{a}\hat{K}_{a}^{\dagger}\hat{K}_{a}=\hat{\mathbb{I}},
\label{eq:kraus_representation}
}
where $\{\hat{K}_a\}$ are Kraus operators.
The representation may involve a discrete or continuous index. In particular, the continuous case is defined as
\bae{
\mathcal{E}(\hat{\rho}) = \int{\dd \mu}\hat{K}(z)\hat{\rho}(z)\hat{K}^\dagger(z)\,, \ \int{\dd \mu}\hat{K}(z)\hat{K}^\dagger(z) = \hat{\mathbb{I}}\,.
}
\sout{Hereafter, we do not mention whether the operation is discrite or continuous except it should be in necessity.}

Given two CPTP maps $\mathcal{E}_1$ and $\mathcal{E}_2$, their product is
defined as composition,
\bae{
(\mathcal{E}_2 \circ \mathcal{E}_1)(\hat{\rho})
\equiv
\mathcal{E}_2\!\left(\mathcal{E}_1(\hat{\rho})\right).
\label{eq:channel_product}
}
This product is associative,
\bae{
\mathcal{E}_3\circ(\mathcal{E}_2\circ\mathcal{E}_1)
=
(\mathcal{E}_3\circ\mathcal{E}_2)\circ\mathcal{E}_1,
}
and closed on the space of CPTP maps.
In general, CPTP maps are not invertible, so the resulting structure is a
semigroup rather than a group.
The product of the two CPTP maps is represented with the Kraus operators as
\bae{
\mathcal{E}_1(\hat{\rho})
=
\sum_{a}\hat{K}_a\,\hat{\rho}\,\hat{K}_a^{\dagger},
\qquad
\mathcal{E}_2(\hat{\rho})
=
\sum_{b}\hat{L}_b\,\hat{\rho}\,\hat{L}_b^{\dagger},
}
then their product is
\bae{
(\mathcal{E}_2\circ\mathcal{E}_1)(\hat{\rho})
=
\sum_{a,b}
(\hat{L}_b\hat{K}_a)\,
\hat{\rho}\,
(\hat{L}_b\hat{K}_a)^{\dagger},
\label{eq:composed_kraus}
}
so that the composed channel has Kraus operators
\bae{
\hat{M}_{ba}=\hat{L}_b\hat{K}_a .
}
For any observable $\hat{O}$, the action of a channel on expectation values
is given by
\bae{
\mathrm{Tr}\!\left[\mathcal{E}(\hat{\rho})\,\hat{O}\right]
=
\mathrm{Tr}\!\left[\hat{\rho}\,\mathcal{E}^\dagger(\hat{O})\right],
}
where $\mathcal{E}^\dagger$ denotes the adjoint (Heisenberg picture) map,
\bae{
\mathcal{E}^\dagger(\hat{O})
=
\sum_a \hat{K}_a^\dagger\,\hat{O}\,\hat{K}_a .
}
This duality is employed to represent a recovery map for image reconstruction in Sec.~\ref{ssec:entanglement_fidelity_recovery}.

\subsubsection{Deterministic PSF}\label{sssec:deterministic_psf}
First we take care of the effect of the image blurring with the quasistationary PSF as
\bae{
\psi_O(\btheta) = \int{\dd^2\btheta^I} K_{\rm PSF}(\btheta-\btheta^I)\psi_I(\btheta^I)\,.
}
By changing the variables for the integration (see the direct derivation in Appendix.~\ref{app:psf_kernel_derivation}), we represent the observed Wigner function as
\bae{
W_O(\btheta,\bptheta)
&\equiv \frac{1}{(2\pi\lambdabar)^2}\int{\dd^2\bxi}\psi_O(\btheta+\bxi/2)\psi^*_O(\btheta-\bxi/2)e^{-\frac{i\bptheta\cdot\bxi}{\lambdabar}}\cr
&=
\int
\dd^2\btheta^I
\Pi_{\rm PSF}
\!\left(
\btheta-\btheta^I,\,
\bptheta^I
\right)
W_I(\btheta^I,\bptheta^I)\,, 
\label{eq:WO_with_PSF}
}
where
\bae{
\Pi_{\rm PSF}(\btheta,\bptheta) &= \int{\dd^2\bxi} K_{\rm PSF}(\btheta+\bxi/2) K^*_{\rm PSF}(\btheta-\bxi/2) e^{-\frac{i\bptheta\cdot\bxi}{\lambdabar}}\,,\cr
W_I(\btheta,\bptheta)
&\equiv \frac{1}{(2\pi\lambdabar)^2}\int{\dd^2\bxi}\psi_I(\btheta+\bxi/2)\psi^*_I(\btheta-\bxi/2)e^{-\frac{i\bptheta\cdot\bxi}{\lambdabar}}.
}
Note that in Eq.~\eqref{eq:WO_with_PSF} the translational invariance of the point spread function is preserved.
One can express Eq.~\eqref{eq:WO_with_PSF} algebraically in terms of a channel interaction
\bae{
\hat{\rho}_O &= \mathcal{E}_{\rm PSF} (\hat{\rho}_I)\,,\cr
\mathcal{E}_{\rm PSF} (\hat{\rho}_I) &= \int{\dd\mu} \hat{K}(z)\hat{\rho}_I\hat{K}(z)^{\dagger}\,.
}
Note that a channel is described by a continuous Kraus operator $\hat{K}(z) = \sqrt{\Pi_{\rm PSF}}\hat{D}(\btheta,\boldsymbol{0})$. Then, we present
\bae{
W_O = {\mathrm{Tr}}[\hat{\rho}_O\hat{w}]
= \int{\dd\mu} {\mathrm{Tr}}[\hat{K}(z)\hat{\rho}_I\hat{K}(z)^\dagger\hat{w}]
}
Provided a Gaussian PSF as
\bae{
K_{\rm PSF}(\btheta) = \frac{e^{-\frac{|\btheta|^2}{2\sigma^2_\theta}}}{2\pi\sigma^2_\theta}\,,
}
we obtain 
\bae{
\label{eq:Gaussian_PSF_in_phase_space}
\Pi_{\rm PSF}(\btheta,\bptheta)
=
\frac{1}{\pi\sigma^2_\theta}\,
\exp\!\left(
-\frac{|\btheta|^2}{\sigma^2_\theta}
-\frac{\sigma^2_\theta|\bptheta|^2}{\lambdabar^2}
\right).
}
As we will argue in Sec.~\ref{ssec:choice_lambdabar}, Eq.~\eqref{eq:Gaussian_PSF_in_phase_space} explicitly yields that an uncertainty relation $\lambdabar = \sigma_\theta \sigma_p$ where $\sigma_p = \lambdabar/\sigma_\theta$\footnote{In this paper, we have yet to be strict in following the uncertainty relations in quantum mechanics, cf. \cite{1927ZPhy...43..172H,1927ZPhy...44..326K,1929PhRv...34..163R}. We would rather simply argue that the Wigner function follows an uncertainty relation, as shown here and Sec.~\ref{ssec:choice_lambdabar}.}.
\subsubsection{Stochastic PSF}\label{sssec:stochastic_psf}
We describe the other extreme case where a galaxy image is blurred by a nonstationary PSF in position space, resulting in stochastic displacements of the image structure by a certain probability $\Pi_{PSF}(\Delta)$. In this case, we obtain the PSF operation as
\bae{
\mathcal{E}_{\rm PSF}(\hat\rho)
&=
\int \dd^2\Delta\,
\Pi_{\rm PSF}(\Delta)\,
\hat T(\Delta)\hat\rho\hat T^\dagger(\Delta)\,,
}
where $\hat{T}$ denotes the translation operator. Then the Wigner function is obtained as
\bae{
W_O(\btheta,\bptheta) = \int{\dd^2 \Delta} \Pi_{\rm PSF}(\Delta)W_I(\btheta-\Delta,\bptheta)\,.
}
We can choose a Gaussian distribution function for $\Pi_{\rm PSF}(\Delta)$ as one of the simplest choices. We find that the deterministic PSF creates a momentum coherence on the Wigner function of the source image, which reflects the fact that the PSF is stationary over the galaxy image. Such coherence is absent in the stochastic PSF as the source image is locally blurred at random, imprinting incoherent patterns on the source Wigner function. In Sec.~\ref{ssec:systematics_noise_wigner}, we argue how the PSF mixes the morphology and the rotational pattern of a galaxy image,  featuring how the two PSFs work in such destruction in different ways.

Let us remark what is relevant to our study. We make use of the feature that the Wigner function is real but not positive everywhere. Unlike a probability density, an intensity map, or a Fourier power spectrum, it shows sign oscillations and structured zero sets that encode interference at the phase-space resolution scale set by $\lambdabar$. For a galaxy image field $\psi(\btheta)$, the associated Wigner function $W[\psi]$ inherits these features and thus carries information beyond the intensity or Fourier power spectrum, however its four-dimensional nature makes direct analysis cumbersome. We, therefore, introduce \textit{Wigner function shapelets}: an $\mathrm{Sp}(4,\mathbb{R})$-covariant decomposition organized by the $\mathrm{SU}(2)$ irreducible representation of the two-mode oscillator. The procedure to make the analytic form of the (cross-)Wigner function with the mode functions is well controlled by the linear metaplectic transformation given by the Wigner $d$-matrix. In the rest of the paper, we assume that the image is static in observational time in the single color band and consider only the intensity without making use of any polarization for simplicity. 
Despite such simplification it is already sufficient to show how the Wigner function captures the information of a galaxy image, which is novel in the astrophysical literature.

\section{Wigner Function Shapelets}\label{sec:wfs}
In this section we construct the Wigner Function Shapelets (WFSs), which is one of the main results of this paper.  In the first half of this section, we define the Wigner function shapelets following  \cite{SimonAgarwal2000}. This paper presents the analytic form of the Wigner function of the LG modes. We extend their computation to the cross-Wigner function via the Hopf spinor (or the two-dimensional Schwinger harmonic oscillators). Note that we follow the normalization of the Wigner function in \cite{SimonAgarwal2000} as shown in Eq.~\eqref{eq:wigner-def} and Eq.~\eqref{eq:xwigner-def}. We introduce the fundamental properties of the WFS in the following three subsections. In the first, we briefly describe the WFS as a novel basis of the phase space, introducing its orthogonal and complete nature as a Hilbert-Schmidt space. In the second, we derive the analytic forms of the WFS via $\mathrm{SU}$(2) algebra. We introduce a key geometric feature of the WFS: oscillation and zeros in the two-dimensional $\mathrm{U}$(1)-invariant space of $\mathrm{SU}$(2) --- the harmonic energy $Q_0$ and the angular momentum $Q_2$.  Note that $Q_0$ and $Q_2$ are used in \cite{SimonAgarwal2000} as dimensionless quantities. Lastly, we introduce a mathematical procedure that embeds the phase space into $\mathrm{SU}$(2) quadrants, making use of the Hopf fibration\cite{Hopf1931}. This procedure makes the WFS a unique quantity to capture the phase-space structure in a symmetry-preserving manner. The physical dimensions are then defined by a typical size of the image in the angular scale $\sigma$ and the discretization scale $\lambdabar$. We will discuss how to determine $\sigma$ and $\lambdabar$ in practical image analysis in Sec.~\ref{sec:practical_usage_in_image_analyses}.
Throughout this section, we omit the arguments of the functions for simple presentation unless we need to show them explicitly. Also we do not show the derivation of the Wigner function, suggesting readers to see \cite{SimonAgarwal2000}.

\subsection{Definition of Wigner function shapelets}
\label{ssec:def_wigner}
Shortly speaking, the WFS is the cross-Wigner function Eq.~\eqref{eq:xwigner-def} between the two LG modes $\Psi^{\rm LG}_{j,s}$ and $\Psi^{\rm LG}_{j',s'}$, namely $W^{\rm LG}_{(j,s),(j's')}$. At the polar coordinate $\btheta = (|\btheta|{\rm cos}(\varphi),|\btheta|{\rm sin}(\varphi))$, the LG modes are defined as
\bae{
\Psi^\mathrm{LG}_{j,s}(|\btheta|,\varphi;\sigma)
&= \mathcal{N}_{j,s}
\left(\tfrac{2|\btheta|^2}{\sigma^2}\right)^{|s|}\;\cr
& \times L_{j-|s|}^{2|s|}\!\Big(\tfrac{2\,|\btheta|^2}{\sigma^2}\Big)\;
e^{-\tfrac{|\btheta|^2}{\sigma^2}}\;e^{i2s\varphi}\,,\cr
\mathcal{N}_{j,s} &= \sqrt{\frac{2}{\pi\,\sigma^2\,}}\left[\frac{(j-|s|)!}{(j+|s|)!}\right]^{1/2}\;\,.
}
Note that the LG modes are normalized as $\langle \Psi^{\rm LG}_{j,s} | \Psi^{\rm LG}_{j',s'} \rangle = \delta_{jj'}\delta_{ss'} $, respectively. 
The Fourier transforms of the LG modes have self-similar expression as
\(
\sigma \mapsto 2/\sigma
\),
up to a phase depending only on the total order \(2j\):
\bae{
\widetilde{\Psi}^{\mathrm{LG}}_{j,s}(|\bk|,\varphi_{\bk};\sigma)
=
\pi\,\sigma^2\,(-i)^{2j}\;
\Psi^{\mathrm{LG}}_{j,s}\!\Big(|\bk|,\varphi_{\bk};\tfrac{2}{\sigma}\Big).
}
As it immediately follows, introducing the orthogonality of $\Psi^{\mathrm{LG}}_{j,s}$ into the Moyal/Plancherel identity Eq.~\eqref{eq:moyal_plancherel_id} gives that the set of $W^{\rm LG}_{(j,s),(j',s')}$ makes an orthogonal and complete set of functions. This is our starting point, where the WFS is well defined to decompose the Wigner function of a galaxy image in the phase space. Note that this orthogonality and completeness of the set of $W^{\rm LG}_{(j,s),(j',s')}$ is a realization of a Hilbert-Schmidt space as
\bae{
\big\langle W^{\rm LG}_{(j_1,s_1),(j_2,s_2)}\,|
W^{\rm LG}_{(j_3,s_3),(j_4,s_4)}\big\rangle_{\mathrm{HS}}
&=\frac{\delta_{j_1j_3}\,\delta_{s_1s_3}\,\delta_{j_2j_4}\,\delta_{s_2s_4}}{(2\pi\lambdabar)^2}\,.
}
Thanks to this orthogonality, we can represent any function in the phase space that is quadratically integrable as a complete form of series expansion of WFSs. In particular, the Wigner function $W$ is expanded as
\bae{
\label{eq:wigner_expansion_wrt_wfs}
W
&=\sum_{(j,s),(j',s')}\!
c_{(j,s),(j',s')}\;
W_{(j,s),(j',s')}.
}
where
\bae{
c_{(j,s),(j',s')} \equiv (2\pi\lambdabar)^2\big\langle
W_{(j,s),(j',s')}\,| W\big\rangle_{\mathrm{HS}}
}
The series expansion Eq.~\eqref{eq:wigner_expansion_wrt_wfs} is mathematically related to the conventional polar shapelets \cite{Refregier2003a} as follows. Let us expand the field $\psi$ in terms of the LG modes
\bae{
\label{eq:polar_shapelets}
\psi = \sum_{j,s}\psi_{js}\Psi^{\rm LG}_{j,s}\,.
}
Note that the reality condition of the image requires $\psi_{j,-s} = \psi^*_{j,s}$ thanks to the relation $\Psi^{\rm LG}_{j,-s} = \Psi^{\rm LG *}_{j,s}$. 
Subtracting Eq.~\eqref{eq:polar_shapelets} in the definition of the Wigner function Eq.~\eqref{eq:wigner-def}, we find that the relation
\bae{
c_{(j,s),(j',s')} = \psi_{js}\psi^*_{j's'}\,,
}
which is nothing but the consequence of the identity in Eq.~\eqref{eq:moyal_plancherel_id}. The coefficients are positive when $j=j'$ and $s=s'$, which is consistent with the fact that the diagonal elements of the state exhibit the probability distribution in modes.
In practice, the Wigner function of a single galaxy image is real, and thus the coefficient is Hermite.
One can extend these formulae in multiple types of fields as discussed in Sec.~\ref{ssec:basic_properties}, which we leave as future work.

\subsection{Closed analytic form of the WFS}\label{ssec:closed_analytic_form_wfs}
The WFS has the closed analytic form as a function of the phase space, which enables us to investigate the structure of the phase space rigorously.
In particular, given $j=j'$ and $s = s'$, which is nothing but the Wigner function of a single LG mode, it carries $U(1) \times U(1)$ invariance as it only depends on the pair of $\mathrm{U}$(1)-invariant variables $(Q_0,Q_2)$. In what follows, we review \cite{SimonAgarwal2000} that derives the Wigner function of the single LG modes, adding the extra derivation of the cross-Wigner function with the Hopf spinor that consists of the two-dimensional bosonic harmonic oscillators\cite{Jordan1935SymmetricLinearGroups, Schwinger1952AngularMomentum}. In this paper, being aware that we are working in angular coordinates, we employ the notation $(\btheta,\bptheta)$ that is different from what \cite{SimonAgarwal2000} describe while the contents are equivalent.

Let us start with $\mathrm{SU}$(2)
algebra. Given the definition of commutation relations Eq.~\eqref{eq:xp_commutation_relations} there is $\mathrm{SU}$(2)
algebra in quadratic forms of $\hat{\theta}_i$ and $\hat{p}_i$ as
\bae{
[\hat{T}_a,\hat{T}_b]=i\,\epsilon_{abc}\,\hat{T}_c,\qquad
[\hat{T}_a,\hat{T}_0]=0,
}
with $a,b,c\in\{1,2,3\}$ and the Levi-Civita tensor $\epsilon_{abc}$ with $\epsilon_{123}=1$.
\bae{
\hat{T}_0&=\frac{1}{4\lambdabar}\Big[
\alpha^{-1}(\hat{\theta}^2_x+\hat{\theta}^2_y)+\alpha(\hat{p}_x^2+\hat{p}_y^2)\Big]-\frac{1}{2},\\
\hat{T}_3&=\frac{1}{4\lambdabar}\Big[
\alpha^{-1}(\hat{\theta}^2_x-\hat{\theta}^2_y)+\alpha(\hat{p}_x^2-\hat{p}_y^2)\Big],\\
\hat{T}_1&=\frac{1}{2\lambdabar}\Big[
\alpha^{-1}\hat{\theta}_x\hat{\theta}_y+\alpha\,\hat{p}_x\hat{p}_y\Big],\qquad \\
\hat{T}_2&=\frac{1}{2\lambdabar}\Big(\hat{\theta}_x\hat{p}_y-\hat{\theta}_y\hat{p}_x\Big),
}
where $\alpha = \sigma^2/2\lambdabar$. There are scalar variables corresponding to the operators as
\bae{
Q_0&=\frac12\!\left(\frac{\theta^2_x+\theta^2_y}{\sigma^2}
+\frac{\sigma^2(p_x^2+p_y^2)}{4\lambdabar^2}\right),\\
Q_1&=\frac{\theta_x\theta_y}{\sigma^2}+\frac{\sigma^2 p_x p_y}{4\lambdabar^2},\qquad \\
Q_2&=\frac{\theta_x p_y-\theta_y p_x}{2\lambdabar},\\
Q_3&=\frac12\!\left(\frac{\theta_x^2-\theta_y^2}{\sigma^2}
+\frac{\sigma^2(p_x^2-p_y^2)}{4\lambdabar^2}\right).
}
$Q_a$ satisfies the quadratic relations $Q^2_0 = Q^2_1 + Q^2_2 + Q^2_3$ where $Q_0$ represents a Casimir constant that is invariant in any $\mathrm{SU}$(2) transformation. In addition to this feature, $Q_0$ and $Q_2$ carry the invariance with the linear symplectic transform. 

Let us introduce dimensionless phase-space variables
$\tilde{\btheta}\equiv\btheta/\sigma$ and $\tilde{\bptheta}\equiv \sigma\,\bptheta/(2\lambdabar)$. Then we introduce the two complex variables $(u,v)$ induced from $(\tilde{\btheta},\tilde{\bptheta})$ as
\begin{align}
    u = \frac{\tilde{\theta}^* + i \tilde{p}^*}{\sqrt{2}},\,\ v = \frac{\tilde{\theta} + i \tilde{p}}{\sqrt{2}}\,,
\end{align}
where $\tilde{\theta} =\tilde{\theta}_x + i\tilde{\theta}_y$ and $\tilde{p} = \tilde{p}_x + i\tilde{p}_y$.
Then we define the Hopf spinor as
\bae{
\label{eq:Hopf_spinor}
z \equiv
\begin{pmatrix}
u\\
v
\end{pmatrix}
\in \mathbb C^2 .
}
such that the Casimir variables $Q_{0,1,2,3}$ are expressed by Pauli matrices $\sigma_i$ as
\bae{
Q_0 &= \frac{z^\dagger z}{2} = \frac{|u|^2+|v|^2}{2},\\
Q_2 &= \frac{z^\dagger\sigma_3 z}{2} = \frac{|u|^2-|v|^2}{2},\\
Q_3 &= \frac12\,z^\dagger\sigma_1 z = \Re(u^*v),\\
Q_1 &=\frac12\,z^\dagger\sigma_2 z = \Im(u^*v).
}
Hence the radii of the two circular modes are fixed by
\bae{
|u|^2 = Q_0+Q_2,\qquad
|v|^2 = Q_0-Q_2 .
}
For fixed $(Q_0,Q_2)$ with $Q_0>0$ and $|Q_2|<Q_0$, the remaining degrees of freedom
are two angular phases $(\phi_u,\phi_v)\in[0,2\pi)^2$ defined by
\bae{
u=\sqrt{Q_0+Q_2}\,e^{i\phi_u},\qquad
v=\sqrt{Q_0-Q_2}\,e^{i\phi_v}.
}
At the operator level, the $\mathrm{SU}(2)\oplus\mathrm{U}(1)$ generators
$\hat T_a$ ($a=1,2,3$) and $\hat T_0$ can be realized by two bosonic modes
(the Schwinger construction).  In particular, one may choose the commuting pair
\bae{
\hat T_0\pm \hat T_2 \ \propto\ \hat N_u,\ \hat N_v,
\qquad [\hat N_u,\hat N_v]=0,
}
where $\hat N_u=\hat a_u^\dagger \hat a_u$ and $\hat N_v=\hat a_v^\dagger \hat a_v$
are the number operators of the two circular modes associated with the complex
coordinates $u$ and $v$. Note that $(\hat a_u, \hat a_u^\dagger)$ and $(\hat a_v, \hat a_v^\dagger)$ are related to the original operators for the canonical coordinates $\hat{\theta}_x, \hat{\theta}_y, \hat{p}_x, \hat{p}_y$ as
\begin{align}
    &\hat{\theta}_x + \alpha\hat{p}_y = \frac{\sigma}{\sqrt{2}}\left(\hat a_u + \hat a_u^\dagger\right)\,,\\
    &\hat{p}_x - \alpha^{-1}\hat{\theta}_y = \frac{\sqrt{2}\lambdabar}{i\sigma}\left(\hat a_u - \hat a_u^\dagger\right)\,,\\
    &\hat{\theta}_x - \alpha\hat{p}_y = \frac{\sigma}{\sqrt{2}}\left(\hat a_v + \hat a_v^\dagger\right)\,,\\
    &\hat{p}_x + \alpha^{-1}\hat{\theta}_y = \frac{\sqrt{2}\lambdabar}{i\sigma}\left(\hat a_v - \hat a_v^\dagger\right)\,.
\end{align}
Consequently, the LG modes at fixed
$j$ are simultaneous eigenstates of $(\hat N_u,\hat N_v)$ and can be labeled as
\bae{
\ket{\Psi^{\rm LG}_{j,s}} \equiv \ket{n_u}\otimes\ket{n_v},
\qquad
n_u\equiv j+s,\quad n_v\equiv j-s,
}
(and similarly $n_u'=j'+s'$ and $n_v'=j'-s'$ for $\ket{\Psi^{\rm LG}_{j',s'}}$). Recalling that the eigenfunctions for $\ket{n_u}$ and $\ket{n_v}$ are nothing but the Hermite-Gaussian function labeled by $n_u$ and $n_v$, respectively, we obtain the one-to-one correspondence of $\Psi^{\rm LG}_{j,s} \rightarrow \Psi^{\rm HG}_{n_u n_v}$.
This algebraic identification is the foundation of the Hopf-spinor factorization
of phase-space objects built from LG modes.
In the Hopf variables $(u,v)$ in Eq.~\eqref{eq:Hopf_spinor}, the phase-space point is equivalently
parametrized by $(Q_0,Q_2;\phi_u,\phi_v)$ at fixed $(Q_0,Q_2)$ on the Hopf torus.
Because LG modes are product Fock states in the circular basis, the two-mode Wigner kernel
factorizes into a tensor product of one-mode kernels.  Equivalently, the
cross-Wigner function factorizes into a product of two one-mode cross-Wigners,
one for the $u$ mode and one for the $v$ mode, 
\bae{
W\!\big[\Psi^{\rm LG}_{j,s},\Psi^{\rm LG}_{j',s'}\big](u,v)
=
W^{(1)}_{n_u n_u'}(u)\;
W^{(1)}_{n_v n_v'}(v),
\label{eq:W_factorisation_uv}
}
with $n_u=j+s$, $n_v=j-s$ and $n_u'=j'+s'$, $n_v'=j'-s'$. Note that none of $n_u, n'_u, n_v, n'_v$ are negative due to the constraints $-j\leq s \leq j$ and $-j'\leq s' \leq j'$.
A convenient closed form for the one-mode cross-Wigner between number states
$\ket{n_u}$ and $\ket{n_v}$ is
\bae{
W^{(1)}_{nn'}(w) &= e^{i(n-n')\phi}R^{(1)}_{nn'}(|w|)\, \cr
R^{(1)}_{nn'}(|w|) &= \frac{(-1)^{n_<}}{\pi\lambdabar}
\sqrt{\frac{n_<!}{n_>!}}\;
(\sqrt2\,|w|)^{|n-n'|}\, \cr
& \qquad \times e^{-2|w|^2}\,
L_{n_<}^{(|n-n'|)}\!\big(4|w|^2\big), \cr
\label{eq:W1_nnp_closed}
}
where $n_<\equiv \min(n,n')$ and $n_>\equiv \max(n,n')$. Note that $R^{(1)}_{nn'} = R^{(1)}_{n'n}$.
Applying Eq.~\eqref{eq:W1_nnp_closed} to $w=u$ and $w=v$ and using
$|u|^2=Q_0+Q_2$ and $|v|^2=Q_0-Q_2$ yields the explicit closed form
\bae{
\begin{aligned}
&W\!\big[\Psi^{\rm LG}_{js},\Psi^{\rm LG}_{j's'}\big](Q_0,Q_2;\phi_u,\phi_v)\\
&=
\frac{(-1)^{n_{u<}+n_{v<}}}{\pi^2\lambdabar^2}\,
\sqrt{\frac{n_{u<}!\,n_{v<}!}{n_{u>}!\,n_{v>}!}}\;
(\sqrt2\,|u|)^{|\Delta_u|}\,(\sqrt2\,|v|)^{|\Delta_v|}\; \cr
&\hspace{2.2cm}\times e^{-2(|u|^2+|v|^2)}\;
e^{i(\Delta_u\phi_u+\Delta_v\phi_v)}\cr
&\hspace{2.2cm}\times
L_{n_{u<}}^{(|\Delta_u|)}\!\big(4|u|^2\big)\;
L_{n_{v<}}^{(|\Delta_v|)}\!\big(4|v|^2\big),
\end{aligned}
\label{eq:W_LG_cross_closed_hopf}
}
with the ``charges''
\bae{
\Delta_u&\equiv n_u-n_u'=(j+s)-(j'+s'),\cr
\Delta_v&\equiv n_v-n_v'=(j-s)-(j'-s')\,.
\label{eq:Delta_uv_def}
}
Eq.~\eqref{eq:W_LG_cross_closed_hopf} makes it manifest that the dependence
on the Hopf-torus angles $(\phi_u,\phi_v)$ is a single torus harmonic,
while the remaining dependence is purely radial through $(Q_0\pm Q_2)$. Note that this derivation extends the original derivation in \cite{SimonAgarwal2000} of the Wigner function of the single LG modes to the cross-Wigner function of the LG modes.
The diagonal case $j=j'$ and $s=s'$ gives $\Delta_u=\Delta_v=0$ and collapses
to the well-known torus-invariant Wigner function (product of ordinary Laguerre
polynomials) \cite{SimonAgarwal2000}.

\subsection{Hopf fibration of Wigner function}\label{ssec:Hopf_wf}
We introduce a representation of the phase-space coordinate $(\btheta,\bptheta)$ via the $\mathrm{SU}(2)$-invariant subspace labeled by a pair of $U$(1) invariants $(Q_0,Q_2)$, i.e., energy and angular momentum. We construct a compact band map that preserves the essential symplectic content while encapsulating the full four-dimensional information. This band structure exposes the sign-alternating oscillations and the web of zeros (interference) of the Wigner function $W(\btheta,\bptheta)$, and we use it in the next section to construct morphology diagnostics directly from the $(Q_0,Q_2)$ map.

The pair $(\phi_u,\phi_v)$, therefore,  parametrizes an invariant
two-torus $T_{Q_0,Q_2}\simeq U(1)\times U(1)$, i.e., a Hopf torus, embedded in the normalized phase space. Given the coordinate with the Hopf spinor $z$, we obtain several properties of the Hopf torus as follows:
\paragraph*{Disjointness (foliation by tori).}
For $(Q_0,Q_2)\neq(Q_0',Q_2')$,
\bae{
T_{Q_0,Q_2}\cap T_{Q_0',Q_2'}=\varnothing,
}
since a point $z\in\mathbb C^2$ cannot satisfy two distinct sets of level-set
conditions for $(|u|^2,|v|^2)$.

\paragraph*{Torus harmonics as $U(1)$ gauge modes.}
On $T_{Q_0,Q_2}$ the residual symmetry is $U(1)\times U(1)$ acting as
$(\phi_u,\phi_v)\mapsto(\phi_u+\alpha,\phi_v+\beta)$.
The corresponding characters (torus harmonics) are
\bae{
\chi_{k\ell}(\phi_u,\phi_v)
= e^{\,i(k\phi_u+\ell\phi_v)},\qquad k,\ell\in\mathbb Z .
}
Along the Hopf fiber $(\alpha=\beta)$, $\chi_{k\ell}$ carries a gauge charge
$k+\ell$. The diagonal sector $k+\ell=0$ is fiber–invariant and probes the base
(shape) structure at fixed $(Q_0,Q_2)$, while $k+\ell\neq0$ resolves coherence
along the fiber.

\paragraph*{Averaging on the Hopf torus.}
Let $d\sigma$ denote the induced area element on $T_{Q_0,Q_2}$.
The natural normalized average of a function $f$ over the torus is
\bae{
\langle f\rangle_{T_{Q_0,Q_2}}
:=2\int_0^{2\pi}\!\frac{\dd\phi_u}{2\pi}\int_0^{2\pi}\!\frac{\dd\phi_v}{2\pi}\; f ,
}
so that $\langle1\rangle_{T_{Q_0,Q_2}}=2$. 
Since the level sets of $(Q_0,Q_2)$ are precisely the Hopf tori $T_{Q_0,Q_2}$,
the coarea formula yields
\bae{
\int_{\mathbb R^4}\! f\,\dd\mu(\btheta,\bptheta)
= (2\pi\lambdabar)^2\int^\infty_0\!\mathrm{d}Q_0\int^{+Q_0}_{-Q_0}\!\mathrm{d}Q_2\;
\langle f\rangle_{T_{Q_0,Q_2}} .
}
Then we define the Wigner-torus coefficients labeled by $(k,\ell)$ as
\bae{
\mathcal{W}_{k\ell}(Q_0,Q_2)
= \frac{1}{2}\langle W\chi^*_{k\ell}\rangle_{T_{Q_0,Q_2}}\,.
}
In particular, $\mathcal{W}_{00}(Q_0,Q_2)=\langle W\rangle_{T_{Q_0,Q_2}}$ and this obeys the normalization rule
\bae{
1&=\int_{\mathbb R^4}\! W\,\dd\mu(\btheta,\bptheta),\cr
&=(2\pi\lambdabar)^2\int_0^\infty\!\mathrm{d}Q_0\int_{-Q_0}^{+Q_0}\!\mathrm{d}Q_2\;
\mathcal{W}_{00}(Q_0,Q_2),
}
namely, the Wigner function is the torus zero-mode.
On each Hopf torus, there is a series expansion of the Wigner function as
\bae{
\label{eq:W_to_rhoKM_chi}
W(Q_0,Q_2,\phi_u,\phi_v)
= \sum_{k,\ell\in\mathbb Z}
\mathcal{W}_{k\ell}(Q_0,Q_2)\,\chi_{k\ell}(\phi_u,\phi_v),
}
due to the orthogonality condition  $\langle\chi_{k\ell}\chi_{k'\ell'}^* \rangle = 2\delta_{kk'}\delta_{\ell\ell'}$.
\paragraph*{Gauge invariance of $\mathcal{W}_{k\ell}$}
For fixed $(Q_0,Q_2)$, the phases $(\phi_u,\phi_v)\in[0,2\pi)^2$ parametrize the
Hopf torus and generate a natural $\mathrm{U}(1)\times\mathrm{U}(1)$ action
\bae{
u\mapsto e^{i\alpha}u,\qquad v\mapsto e^{i\beta}v,
\label{eq:U1xU1_action}
}
under which $(Q_0,Q_2)$ are invariant.  Then Eq.~\eqref{eq:W_to_rhoKM_chi} carries the phase factor
$e^{i(k\alpha+\ell\beta)}$ which comes only from $\chi_{k\ell}$, leaving $\rho_{k\ell}$ invariant under arbitrary phase shifts. In the WFS, the cross-Wigner function is not
invariant but transforms covariantly as a charged field on the torus:
\bae{
W\big[\Psi^{\rm LG}_{j,s},\Psi^{\rm LG}_{j',s'}\big] \mapsto\ e^{i(\Delta_u\alpha+\Delta_v\beta)}\,W\big[\Psi^{\rm LG}_{j,s},\Psi^{\rm LG}_{j',s'}\big]\,,
\label{eq:W_gauge_covariant}
}
i.e.\ its ``gauge charges'' are $(\Delta_u,\Delta_v)$.
Recalling the cross-Wigner function of the LG modes, 
we obtain the selection rule and the explicit radial closed form:
\bae{
\mathcal{W}^{\rm LG}_{k\ell;js}
= R^{(1)}_{\scalebox{0.5}{$j+s,j+s-k$}}(\sqrt{Q_0{+}Q_2})\;
R^{(1)}_{\scalebox{0.5}{$j-s,j-s-\ell$}}(\sqrt{Q_0{-}Q_2})\,.
\label{eq:rho_kl_closed_factorised}
}
Equation.~\eqref{eq:rho_kl_closed_factorised} provides an analytic
closed form of the Hopf-torus observables $\mathcal{W}_{k\ell}$ in terms of associated
Laguerre polynomials, with the entire gauge (angular) structure reduced to the
Kronecker selection rule $(k,\ell)=(\Delta_u,\Delta_v)$. 
This is equivalent to the condition that  $(j,s)$ must be within the domain $-j+{\rm max}(0,k)\leq s \leq j - {\rm max}(0,\ell)$ given $(k,\ell)$, which originates from the constraint $-j'\leq s'\leq j'$ and the selection rule. One must be careful that the domain is null at $j < ({\rm max}(0,k)+{\rm max}(0,\ell))/2$. Then Eq.~\eqref{eq:W_to_rhoKM_chi} is represented as
\bae{
\label{eq:W_to_rhoKM_LG_chi}
&W
= \sum_{k,\ell\in\mathbb Z} \sum_{j\geq \frac{{\rm max}(0,k)+{\rm max}(0,\ell)}{2}} \cr
&\left\{\sum^{j-{\rm max}(0,\ell)}_{s=-j+{\rm max}(0,k)}c_{(j,s),(j-\frac{k+\ell}{2},s-\frac{k-\ell}{2})}
\mathcal{W}^{\rm LG}_{k\ell;js}\,\chi_{k\ell}\right\}\,. \cr
}
In Fig.~\ref{fig:analytic_band_matrix}, we plot the first 81 combinations of the WFS in Eq.~\eqref{eq:W_to_rhoKM_LG_chi}.
We find that the simultaneous flip: $n_u \mapsto n_v$ and $n'_u \mapsto n'_v$ provides the flip $Q_2 \mapsto -Q_2$, exhibiting the chiral symmetry of the WFS.

After all,  for a given pair of LG modes, a single torus harmonic carries all the
gauge-covariant phase information, while $\mathcal{W}_{k\ell}$ packages it into a
gauge-invariant Hopf-torus observable.

\paragraph*{Statistical isotropy and selection rule.}
Statistical isotropy of the galaxy field means invariance of the Wigner function
under a common physical rotation
$(\btheta,\bptheta)\mapsto(R_\alpha\btheta,R_\alpha\bptheta)$ with
$R_\alpha\in SO(2)$.
In the Hopf parametrisation this induces
$\phi_u\mapsto\phi_u-\alpha$ and $\phi_v\mapsto\phi_v+\alpha$,
while $(Q_0,Q_2)$ remain invariant.
Hence isotropy implies
$W(Q_0,Q_2,\phi_u-\alpha,\phi_v+\alpha)=W(Q_0,Q_2,\phi_u,\phi_v)$ for all $\alpha$.
The torus harmonics transform as
$\chi_{k\ell}(\phi_u-\alpha,\phi_v+\alpha)
=\chi_{k\ell}(\phi_u,\phi_v)\,e^{i(\ell-k)\alpha}$,
so the Wigner-torus coefficients
$\mathcal{W}_{k\ell}(Q_0,Q_2)=\tfrac12\langle W\chi^*_{k\ell}\rangle_{T_{Q_0,Q_2}}$
can be nonzero only if this phase is trivial for all $\alpha$.
Therefore statistical isotropy enforces the diagonal selection rule
$\mathcal{W}_{k\ell}(Q_0,Q_2)=0$ for $k\neq \ell$,
with only the spin-neutral modes surviving. For an ensemble of
statistically isotropic fields this statement holds in expectation, while
individual realizations may exhibit small off-diagonal components due to
cosmic variance, masking, or noise.

\paragraph*{Parity symmetry.}
Parity invariance corresponds to an orientation-reversing transformation
$\btheta\mapsto\mathcal{P}\btheta$ with $\det\mathcal{P}=-1$, under which the
canonical complex variables transform as
$\tilde{\theta}\mapsto-\tilde{\theta}$ and $\tilde{p}\mapsto-\tilde{p}$.
With the definitions
$u=(\tilde{\theta}^*+i\tilde{p}^*)/\sqrt2$ and $v=(\tilde{\theta}+i\tilde{p})/\sqrt2$,
this induces
$(u,v)\mapsto(-u,-v)$ and, therefore, 
$(\phi_u,\phi_v)\mapsto(\phi_u+\pi,\phi_v+\pi)$,
while the invariants $(Q_0,Q_2)$ remain unchanged.
The torus harmonics transform as
$\chi_{k\ell}(\phi_u+\pi,\phi_v+\pi)
=\chi_{k\ell}(\phi_u,\phi_v)\,(-1)^{k+\ell}$,
so parity invariance of the Wigner function implies
\[
\mathcal{W}_{k\ell}(Q_0,Q_2)=(-1)^{k+\ell}\,\mathcal{W}_{k\ell}(Q_0,Q_2).
\]
Hence all modes with odd $k+\ell$ must vanish, while even $k+\ell$ modes are
unconstrained.
When combined with statistical isotropy (which enforces $k=\ell$), only
even-$k$ diagonal components $\mathcal{W}_{kk}$ survive, and any odd-$k$ or
off-diagonal contribution signals parity violation or residual systematics.
\paragraph*{Chiral symmetry.}
Chiral (handedness-reversing) symmetry corresponds to an orientation-reversing
operation in phase space that preserves the isotropic invariant $Q_0$ while
flipping the pseudoscalar invariant $Q_2$.
This is equivalent to the exchange
$(u,v)\mapsto(v,u)$ and hence
$(\phi_u,\phi_v)\mapsto(\phi_v,\phi_u)$, while
\[
Q_0=\tfrac{1}{2}(|u|^2+|v|^2)\mapsto Q_0,
\qquad
Q_2=\tfrac{1}{2}(|u|^2-|v|^2)\mapsto -Q_2,
\]
reflecting the sign flip of the pseudoscalar
$Q_2$.
Under this transformation the torus harmonics satisfy
$\chi_{k\ell}(\phi_v,\phi_u)=\chi_{\ell k}(\phi_u,\phi_v)$, so invariance of the
Wigner function implies the constraint
\[
\mathcal{W}_{k\ell}(Q_0,Q_2)
=
\mathcal{W}_{\ell k}(Q_0,-Q_2).
\]
This relation provides a sharp selection rule for chiral symmetry: any
antisymmetric component under $(k,\ell)\leftrightarrow(\ell,k)$ accompanied by
$Q_2\rightarrow -Q_2$ directly signals intrinsic chirality or pseudoscalar
contamination in the phase-space structure.

Let us mention the edge as described by the zeros of $(u,v)$.
(i) boundary $|Q_2|=Q_0$.
The Hopf torus collapses to a single Hopf circle.
Although $\sqrt{Q_0^2-Q_2^2}\to0$, the torus area shrinks at the same rate, so the
coarea decomposition remains finite for integrable $W$.
(ii) Apex $Q_0=Q_2=0$.
Here $u=v=0$ and $S^3$ collapses to a point.
This set has measure zero and does not contribute to the bounded $W$. If required,
$\mathcal{W}_{00}$ may be defined by a smooth limiting procedure.
\subsection{Morphology modes and rotation modes}
\label{ssec:morphology_rotation_modes}

Let us recall two independent charge operators using a pair of Schwinger
harmonic oscillators $(\hat a_u,\hat a_v)$,
\bae{
\hat T_0 - \frac{1}{2}\hat I
&\equiv
\frac{1}{2}\!\left(\hat a_u^\dagger\hat a_u+\hat a_v^\dagger\hat a_v\right),
\cr
\hat T_2
&\equiv
\frac{1}{2}\!\left(\hat a_u^\dagger\hat a_u-\hat a_v^\dagger\hat a_v\right),
}
which generate two commuting $\mathrm{U}(1)$ transformations.
Their actions on the ladder operators are
\bae{
e^{-i\alpha\hat T_0}\hat a_u e^{+i\alpha\hat T_0}
&=e^{+i\alpha}\hat a_u,
\qquad
e^{-i\alpha\hat T_0}\hat a_v e^{+i\alpha\hat T_0}
=e^{+i\alpha}\hat a_v,
\cr
e^{-i\beta\hat T_2}\hat a_u e^{+i\beta\hat T_2}
&=e^{+i\beta}\hat a_u,
\qquad
e^{-i\beta\hat T_2}\hat a_v e^{+i\beta\hat T_2}
=e^{-i\beta}\hat a_v.
}
Since $[\hat T_0,\hat T_2]=0$, the corresponding charges can be
simultaneously diagonalized.
The two charge modes generated by $\hat{T}_J = \exp{(iJ(\phi_u+\phi_v)\hat{T}_0)}$ and $\hat{T}_K = \exp{(iK(\phi_u-\phi_v)\hat{T}_2)}$ bring the independent charge states. We dub $J\ (J=0,\pm 1, \pm2,\dots)$ and $K\ (K=0,\pm 1, \pm2,\dots)$ as morphology modes and rotational modes, respectively. It is well known that the Fourier power spectrum, corresponding to the
marginal of the Wigner function over configuration space, retains only
even $J$ modes while all $K$ modes survive.
This reflects the fact that the Fourier power alone does not uniquely
determine the image, and that odd morphology modes require higher-order
configuration-space correlations.
The WFS formalism makes this structure explicit: for each mode,
the morphology and rotation charges are given by
$J=\Delta_u+\Delta_v$ and $K=\Delta_u-\Delta_v$.

The effect of PSF blurring can be understood
most transparently in the Heisenberg picture, taking into account the morphology modes and the rotational modes. Following the formulation introduced in Sec.~\ref{ssec:systematics_noise_wigner}, we describe the decoherence of the morphology and the rotational features of a galaxy image separately for whether the PSF is deterministic or stochastic.

\subsubsection{Deterministic PSF}\label{sssec:deterministic_psf_charges}

For a deterministic PSF acting coherently on the complex field amplitude,
the forward map is a single-operator channel
\bae{
\hat\rho_O
=
\mathcal E_{\rm coh}(\hat\rho_I)
=
\hat K\,\hat\rho_I\,\hat K^\dagger,
}
where $\hat K= K_{\rm PSF}(\hat{\bptheta})$ is the optical transfer function
(OTF) acting multiplicatively in Fourier space, equivalently, a function of
the momentum operator $\hat{\bptheta}$.
Charge preservation is, therefore,  controlled by commutators with $\hat K$.

The morphology generator $\hat T_0$ does not commute with $\hat K$ in
general,
\bae{
[\hat T_0,\hat K]\neq 0,
}
reflecting the fact that coherent PSF blur induces mode coupling among
different morphology charges $J$.
For a Gaussian PSF, $\hat K\propto \exp(-\varepsilon\,\hat p^\dagger\hat{p})$,
this noncommutativity is governed by pair creation and annihilation
operators $\hat a_u\hat a_v$ and $\hat a_u^\dagger\hat a_v^\dagger$, and its
magnitude scales with the PSF width.

By contrast, the rotation generator $\hat T_2$ is preserved if the PSF is
isotropic. Since under rotations
$\hat{p}\to e^{+i\beta}\hat{p}$, rotational covariance requires
$\hat K=f(\hat{p}^\dagger\hat{p})$, equivalently
$\tilde K_{\rm PSF}(\bptheta)=\tilde K_{\rm PSF}(|\bptheta|)$.
Thus,
\bae{
[\hat T_2,\hat K]=0
\quad\Longleftrightarrow\quad
\text{isotropic PSF}.
}
Anisotropic coherent PSFs generically mix different rotation charges $K$.

\subsubsection{Stochastic PSF}
A translationally invariant PSF corresponds to a random displacement
channel,
\bae{
\mathcal K(\hat\rho)
&=
\int \dd^2\Delta\,
K(\Delta)\,
\hat T(\Delta)\hat\rho\hat T^\dagger(\Delta) \cr
\hat T(\Delta)
&=
\exp\!\left[
-\frac{i}{\lambdabar}
\big(\Delta\,\hat\pi^\dagger+\Delta^\ast\hat\pi\big)
\right],
}
with $\hat\pi=\tfrac{i}{\sqrt2}(\hat a_u^\dagger-\hat a_v)$.
In the Heisenberg picture,
\bae{
\mathcal K^\dagger(\hat O)=\int \dd^2\Delta\,K(\Delta)\,
\hat T^\dagger(\Delta)\hat O\hat T(\Delta)\,.
}
Applying this map to the morphology generator $\hat T_0$, one finds
\bae{
\mathcal K^\dagger(\hat T_0)
=
\hat T_0
+\frac{\langle|\Delta|^2\rangle_{\rm PSF}}{2\lambdabar^2},
}
for a centered PSF with $\langle\Delta\rangle_{\rm PSF}=0$. Note that $\langle A\rangle_{\rm PSF} \equiv \int \dd^2\Delta\,K(\Delta)\,
A(\Delta)$.
Thus PSF blurring induces an effective diffusion in morphology space,
leading to the suppression of large-$J$ morphology modes.
In the limit $\lambdabar\to0$ with
$\langle|\Delta|^2\rangle_{\rm PSF}/\lambdabar^2\to0$,
all the morphological information becomes asymptotically recoverable.

By contrast, the rotation generator obeys
\bae{
\mathcal K^\dagger(\hat T_2)=\hat T_2
}
for any centered PSF.
However, preservation of individual rotation modes $K$ requires the
stronger condition of rotational covariance of the kernel,
$K(\Delta)=K(|\Delta|)$.
Anisotropic PSFs generically mix different $K$ sectors, while leaving
the morphology diffusion unaffected.

This separation clarifies the distinct physical roles of the two charge
types: morphology modes quantify the sensitivity of image structure to
translational diffusion, while rotation modes probe anisotropy.
As shown in Sec.~\ref{ssec:weak_lensing_response_shear} and Sec.~\ref{ssec:weak_lensing_response_flexion}, weak gravitational
lensing couples these modes according to well-defined selection rules,
enabling targeted estimators for parity-violating and higher-spin
signals.
\subsection{Correspondence between WFS and BiPoSHs representation}
We connect the WFS to the BiPoSHs representation of the angular two-point correlation function. More specifically, we derive the analytic expression of $A^{LM}_{\ell_1\ell_2}$ in Eq.~\eqref{eq:ALM_l1l2} in terms of the WFS. Here in this section, a character $\ell$ denotes a multipole moment. Let us start with the analytic expansion of the Laguerre-Gaussian mode $\Psi^{\rm LG}_{j,s}(\btheta;\boldsymbol{n}_c)$ in terms of the spherical harmonics $Y_{\ell m}(\boldsymbol{n})$, where the small-angle approximation leads to  $\boldsymbol{n} = \boldsymbol{n}_c + \btheta + {\cal O}(\theta^2)$.
At the small-angle limit around the
the centroid position $\boldsymbol{n}_c$, there is the asymptotic expression of $Y_{\ell m}$ as
\bae{
\label{eq:SH_small_angle_limit}
Y_{\ell m} &\approx {\mathcal{B}_{\ell, |m|}}J_{|m|}\left(\tau_\ell \theta\right)e^{im\varphi}\,,\cr
\mathcal{B}_{\ell,|m|} &= (-1)^{|m|}
\sqrt{\frac{\tau_\ell}{2\pi}}\sqrt{\frac{(\ell+|m|)!}{(\ell-|m|)!}}\tau^{-m}_{\ell}\,,
}
where $J_a(x)$ is the Bessel function of the $a$-th kind and $\tau_\ell = \ell +1/2$. Note that the angle dependence is factorized by the exponential function and thus we have a constraint $m = 2s$ as a consequence of orthogonality. Then we obtain the following expression
\bae{
\label{eq:LG_to_SH_at_small_angle}
\Psi^{\rm LG}_{j,s}(\btheta;\boldsymbol{n}_c) \approx \sum_{\ell} \psi^{(j,s)}_\ell B^{-1}_{\ell,2|s|} Y_{\ell,2s}(\boldsymbol{n})\,.
}
By using the integral identity in the radial direction (see the full derivation in Appendix.~\ref{app:LG_SH_small_angle_limit}), the coefficient is analytically derived as
\bae{
\psi^{(j,s)}_{\ell} = \frac{\tau_\ell(-1)^{j-|s|}}{2}\left|\Psi^{\rm LG}_{j,s}\left(\frac{\tau_\ell \sigma}{2}\right)\right|\,.
}
We find that $\psi^{(j,s)}_\ell$ provide a low-pass filter that localizes the multipole mode by $\ell \lesssim \sigma^{-1}$.
Let us plug Eq.~\eqref{eq:LG_to_SH_at_small_angle} into Eq.~\eqref{eq:wigner-def}, replacing the Wigner function response to that of the spherical harmonics. After straightforward computation, we obtain
\bae{
\label{eq:ALM_l1l2_cjsjpsp}
\mathcal{A}^{LM}_{\ell_1 \ell_2} &=  \sum_{j,s, j',s'}\langle c_{(j,s),(j',s')} \rangle_{\rm cosmo.} \cr &\times \psi^{(j,s)}_{\ell_1} \psi^{(j,s')}_{\ell_2} B^{-1}_{\ell_1,2|s|} B^{-1}_{\ell_2,2|s'|}
\langle \ell_1 2s \ell_2 2s' |LM\rangle\,.\cr
}
One can extend the relation in Eq.~\eqref{eq:ALM_l1l2_cjsjpsp} into a generic pair of fields $(\psi_a(\btheta),\psi_b(\btheta))$, fixing the coefficient $c_{(j,s),(j',s')}$. 
\subsection{Correspondence between the WFS and FPFS}\label{ssec:correspondence_btw_WFS_FPFS}
We derive the correspondence between the WFS and the FPFS \cite{LiEtAl2018}.  
The Fourier power function is expressed by the WFS expansion as
\bae{
\label{eq:FP_LG_bilinear_form}
F(\bk) &= \sum_{(j,s),(j',s')} c_{(j,s),(j',s')} \cr
& \qquad \times \widetilde{\Psi}^{\mathrm{LG}}_{j,s}\Big(|\bk|,\varphi_{\bk};\sigma\Big) \widetilde{\Psi}^{\mathrm{LG}*}_{j',s'}\Big(|\bk|,\varphi_{\bk};\sigma\Big)\,.
}
The FPFS is defined as
\bae{
\label{eq:FPFS_modes}
M_{j,s} = \int{\dd^2\bk F(\bk)\widetilde{\Psi}^{\mathrm{LG}*}_{j,s}\Big(|\bk|,\varphi_{\bk};\sigma\Big)}\,.
}
Inserting Eq.~\eqref{eq:FP_LG_bilinear_form} into Eq.~\eqref{eq:FPFS_modes}, we obtain an analytic expression 
\begin{widetext}
\bae{
\label{eq:FPFS_WFS_correspondence}
M_{j_1,s_1} = & \pi^3\sigma^4\sum_{(j_2,s_2),(j_3,s_3)}\delta_{s_2-s_3-s_1,0}c_{(j_2, s_2),(j_3,s_3)}\mathcal{N}_{j_1,s_1}\mathcal{N}_{j_2,s_2}\mathcal{N}_{j_3,s_3} \left(\frac{2}{3}\right)^{1+j_1+j_2+j_3} \cr
&\times \frac{(-1)^{2j_2 - (|s_1|+|s_2|+|s_3|)}\Gamma\left[j_1+j_2+j_3+1\right]}{(j_1-|s_1|)!(j_2-|s_2|)!(j_3-|s_3|)!}
\times F^{(3)}_B\left[-j_1-|s_1|,-j_2-|s_2|,-j_3-|s_3|;-(j_1+j_2+j_3);\frac{3}{2},\frac{3}{2},\frac{3}{2}\right]\,. \cr
}
\end{widetext}
Note that we use a concise analytic notation of the integral of the triple multiplication of the Laguerre-Gaussian mode following Eq.~(11) in \cite{LG_integrals_2001}. Here $F^{(3)}_B$ is a $B$-type Lauricella hypergeometric function for the three variables. 

\begin{figure*}[p]
  \centering
  \includegraphics[width=\textwidth,height=\textheight,keepaspectratio]{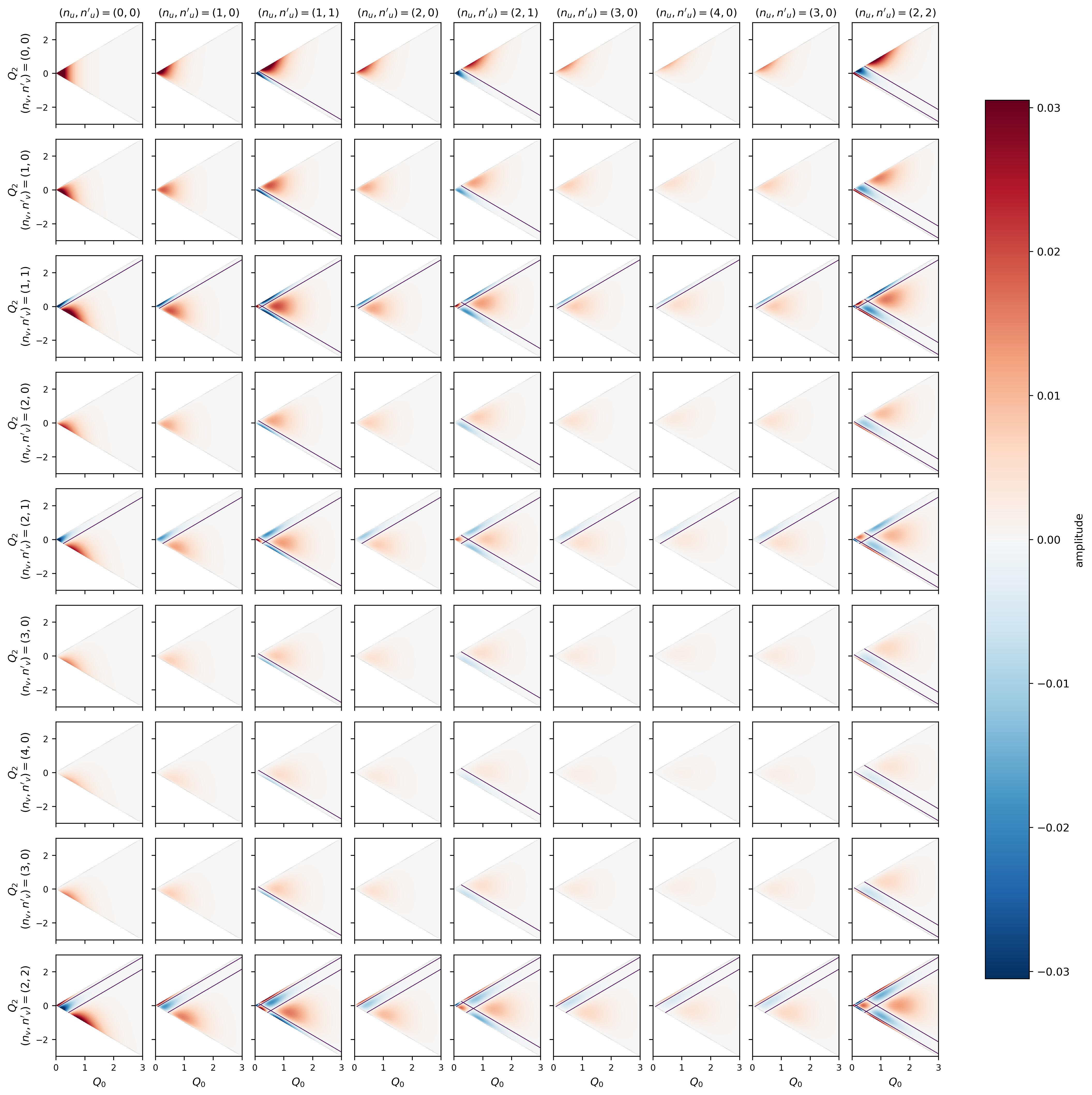}
  \caption{
    Analytic band matrix in the $(Q_0,Q_2)$ plane.
    Columns correspond to $(n_u,n_u')$ and rows to $(n_v,n_v')$.
    Each panel shows the radial product
    $R_{n_u n_u'}(\sqrt{Q_0+Q_2})\,R_{n_v n_v'}(\sqrt{Q_0-Q_2})$,
    masked to the physical wedge $|Q_2|\le Q_0$.
  }
  \label{fig:analytic_band_matrix}
\end{figure*}
\clearpage
%
%
From this expression, we find that the coefficients of the FPFS are a mixture of the fundamental modes of the image state in the WFS. At the lowest modes, the expression in Eq.~\eqref{eq:FPFS_WFS_correspondence} can be simplified and it is still convenient to employ the image analysis such as shape measurements, at which the one of the lowest shear response function in the FPFS is sufficient.  Given such a method of local morphology marginalization, we employ the WFS to estimate the weak-lensing flexion and the galaxy shape statistic correlations. We will aim to show our findings in the following section. The modes in Fig.~\ref{fig:analytic_band_matrix} are distinguished in terms of the morphological charge $J$ and the rotational mode $K$. We emphasize that systematic oscillatory patterns on $(Q_0,Q_2)$ play a critical role in characterizing the phase information of the Wigner function of a galaxy image. Table.~\ref{tab:KM_grouping_abs_plain} summarizes $(J,K)$. Note that the original modes $(j,s)$ are recovered by
\bae{
j &= \frac{n_u + n_v}{2}\,,\ s = \frac{n_u - n_v}{2}\,, \cr
j' &= \frac{n'_u + n'_v}{2} = j - \frac{J}{2}\,,\ s' = \frac{n'_u - n'_v}{2} = s - \frac{K}{2}\,, \cr
}
\hspace{-2.0cm}
\begin{table}[h]
\centering
\footnotesize
\setlength{\tabcolsep}{3pt}
\renewcommand{\arraystretch}{1.05}

\caption{Grouping of the $9\times9$ index combinations into distinct $(J,K)$ sectors, sorted by increasing $(|J|,|K|)$.
We use $k=n_u-n_u'$, $\ell=n_v-n_v'$, $J=k+\ell$, and $K=k-\ell$.
}
\label{tab:KM_grouping_abs_plain}

\begin{tabular}{rrrrrll}
\toprule
$J$ & $K$ & $k$ & $\ell$ & count & $(n_u,n_u')$ choices & $(n_v,n_v')$ choices \\
\midrule

0  & 0  & 0 & 0 & 9 & (0,0), (1,1), (2,2) & (0,0), (1,1), (2,2) \\

\midrule
1  & -1 & 0 & 1 & 6 & (0,0), (1,1), (2,2) & (1,0), (2,1) \\
1  &  1 & 1 & 0 & 6 & (1,0), (2,1)        & (0,0), (1,1), (2,2) \\

\midrule
2  & -2 & 0 & 2 & 3 & (0,0), (1,1), (2,2) & (2,0) \\
2  &  0 & 1 & 1 & 4 & (1,0), (2,1)        & (1,0), (2,1) \\
2  &  2 & 2 & 0 & 3 & (2,0)               & (0,0), (1,1), (2,2) \\

\midrule
3  & -3 & 0 & 3 & 3 & (0,0), (1,1), (2,2) & (3,0) \\
3  & -1 & 1 & 2 & 2 & (1,0), (2,1)        & (2,0) \\
3  &  1 & 2 & 1 & 2 & (2,0)               & (1,0), (2,1) \\
3  &  3 & 3 & 0 & 3 & (3,0)               & (0,0), (1,1), (2,2) \\

\midrule
4  & -4 & 0 & 4 & 3 & (0,0), (1,1), (2,2) & (4,0) \\
4  & -2 & 1 & 3 & 2 & (1,0), (2,1)        & (3,0) \\
4  &  0 & 2 & 2 & 1 & (2,0)               & (2,0) \\
4  &  2 & 3 & 1 & 2 & (3,0)               & (1,0), (2,1) \\
4  &  4 & 4 & 0 & 3 & (4,0)               & (0,0), (1,1), (2,2) \\

\midrule
5  & -3 & 1 & 4 & 2 & (1,0), (2,1)        & (4,0) \\
5  & -1 & 2 & 3 & 1 & (2,0)               & (3,0) \\
5  &  1 & 3 & 2 & 1 & (3,0)               & (2,0) \\
5  &  3 & 4 & 1 & 2 & (4,0)               & (1,0), (2,1) \\

\midrule
6  & -2 & 2 & 4 & 1 & (2,0)               & (4,0) \\
6  &  0 & 3 & 3 & 1 & (3,0)               & (3,0) \\
6  &  2 & 4 & 2 & 1 & (4,0)               & (2,0) \\

\midrule
7  & -1 & 3 & 4 & 1 & (3,0)               & (4,0) \\
7  &  1 & 4 & 3 & 1 & (4,0)               & (3,0) \\

\midrule
8  &  0 & 4 & 4 & 1 & (4,0)               & (4,0) \\

\bottomrule
\end{tabular}
\end{table}

\section{Cosmological information in the Wigner function}\label{sec:cosmo_info_wigner}
We describe how cosmological information emerges in the Wigner function and its angular correlations. We exemplify the linear response of weak lensing i.e. shear \cite{Mandelbaum2018} and flexions \cite{Bacon:2005qr}, cosmic birefringence\cite{2022NatRP...4..452K}, and higher-moment of galaxy shapes \cite{2024PhRvD.109f3541P,2025arXiv250908787K} in the language of the Wigner function and the WFS. Throughout this section, we will argue that the response function is well defined at the local level of the image plane, implying a possibility to handle image systematics and noise at a local level. Note that we represent here the ensemble with $\langle\cdots\rangle$ without mentioning the subscript that denotes the ensemble is taken over cosmological realizations. 

\subsection{Shear response}
\label{ssec:weak_lensing_response_shear}

We derive the linear response to weak gravitational lensing in the Wigner function. We work in the weak-lensing regime and retain terms up to second order of spatial derivatives of the
lens potential $\Phi(\btheta)$, i.e.\ cosmic shear, following the derivation in \cite{Zhang2008}.
In the case of the shear, where the matrix ${\mathrm{A}}_{ij} = \delta_{ij} + \Phi_{ij}\ (i,j = (x,y))$ \footnote{We follow the convention \cite{Zhang2008}, which is the inverse of  the standard convention that treats $A$ here to $A^{-1}$. }and we define the convergence $\kappa = (\Phi_{xx} + \Phi_{yy})/2$, and $\gamma = \gamma_1 + i\gamma_2 = (\Phi_{xx}-\Phi_{yy})/2 +  i \Phi_{xy}$, where $\Phi_{xy} = \Phi_{yx}$.
We obtain $\delta \btheta = (\kappa\I+\Gam\big)\btheta^I $ and $\delta \bptheta = -(\kappa\I+\Gam\big)\bptheta^I$. We shall denote the Lie derivative of the Wigner function as $\mathcal{L}_z W \equiv W^I(z)-W^S(z)$. Hereafter, we omit the superscript $I$ for concise presentation. If we change the canonical variables to $(|u|, |v|, \phi_u, \phi_v)$, $\mathcal{L}_z W$ is rewritten as
\bae{
\mathcal{L}_z W
=
-\delta |u|\,\partial_{|u|}W
-\delta |v|\,\partial_{|v|}W
-\delta\phi_u\,\partial_{\phi_u}W
-\delta\phi_v\,\partial_{\phi_v}W ,
\label{eq:deltaW_hopf}
}
where
\bae{
\label{eq:delta_uv}
\delta u \equiv u' - u = \gamma^* u^*, \qquad
\delta v \equiv v' - v = \gamma\, v^* .
}
and
\bae{
\label{eq:delta_uv_Re_Im}
\delta |u| &= |u|\Re (\gamma e^{2i\phi_u})\,,\ \delta |v| = |v|\Re (\gamma^*e^{2i\phi_v})\,, \cr
\delta\phi_u
&= -\Im\!\left(\gamma\,e^{i2\phi_u}\right)\,,\ \delta\phi_v
= -\Im\!\left(\gamma^*\,e^{i2\phi_v}\right).
}

These expressions provide a complete and closed description of the linear shear
response in the $(u,v)$ variables and in the $(Q_0,Q_2,\phi_u,\phi_v)$
representation, which will be used throughout the remainder of this work.

Plugging Eqs.~\eqref{eq:delta_uv} and ~\eqref{eq:delta_uv_Re_Im} into Eq.~\eqref{eq:deltaW_hopf}, the linear response of the weak-lensing shear is derived for the observable $\mathcal{W}_{k\ell}$ as
\bae{
&\mathcal{W}^I_{k\ell}(Q_0,Q_2) - \mathcal{W}^S_{k\ell}(Q_0,Q_2)
= A^{\gamma}_{k\ell} \gamma + A^{\gamma^\ast}_{k\ell}\gamma^*\,,\cr
A^{\gamma}_{k\ell} &= -\frac{1}{2}\left[|u|\partial_{|u|}-(k-2)\right]\mathcal{W}^S_{k-2,\ell} \cr
&\qquad -\frac{1}{2}\left[|v|\partial_{|v|}+(\ell+2)\right]\mathcal{W}^S_{k,\ell+2}\,, \cr
A^{\gamma^\ast}_{k\ell} &= -\frac{1}{2}\left[|u|\partial_{|u|}+(k+2)\right]\mathcal{W}^S_{k+2,\ell} \cr
&\qquad -\frac{1}{2}\left[|v|\partial_{|v|}-(\ell-2)\right]\mathcal{W}^S_{k,\ell-2}\,. \cr
}
Applying the mode decomposition in terms of the WFS, one may find a shear estimator with the WFS. We will leave the problem of constructing a practical shear estimator in phase space for our future research.
\subsection{Flexion response}\label{ssec:weak_lensing_response_flexion}
We derive the flexion response as follows. 
Following the convention of the flexion in \cite{2006MNRAS.365..414B,2022MNRAS.516..668O}
\bae{
\delta \theta = \frac{1}{4}(2\mathcal{F}|\theta|^2 + \mathcal{F}^*\theta^2 + \mathcal{G}\theta^{*2} ) \,.
}
We can derive the corresponding variation of the complex conjugate momentum as
\bae{
\delta p = -\frac{\mathcal{F}}{2}(\theta p^* + \theta^* p)  -\frac{\mathcal{F^*}}{2}\theta p - \frac{\mathcal{G}}{2}\theta^* p^*\,.
}
Here we denote $\mathcal{F} = [\Phi_{xxx} + \Phi_{xyy} + i(\Phi_{xxy} + \Phi_{yyy})]/2$ and $\mathcal{G} =[\Phi_{xxx} -3 \Phi_{xyy} + i(3\Phi_{xxy} - \Phi_{yyy})]/2$, where $\Phi_{xxy} = \Phi_{xyx} = \Phi_{yxx}$.
With a lengthy but straightforward computation, we obtain
\bae{
&\mathcal{W}^I_{k\ell}(Q_0,Q_2) - \mathcal{W}^S_{k\ell}(Q_0,Q_2) \cr
&= A^{\mathcal{F}}_{k\ell} \mathcal{F} + A^{\mathcal{F}^\ast}_{k\ell}\mathcal{F}^\ast + A^{\mathcal{G}}_{k\ell} \mathcal{G} + A^{\mathcal{G}^\ast}_{k\ell} \mathcal{G}^\ast\,,
}
where the coefficients are explicitly given in Eq.~\eqref{eq:flexion_response_coefficients}, as we omit them for concise presentation in the main text. 

\subsection{Signals of parity violation}\label{ssec:signals_parity_violation}
The possibility of detecting the cosmic birefringence to test the parity symmetry of the Universe has been discussed \cite{MinamiKomatsu2020,2025arXiv250406709N}. In the Wigner function, the birefringence appears as
\bae{
\mathcal{W}^I_{k\ell}(Q_0,Q_2) - \mathcal{W}^S_{k\ell}(Q_0,Q_2) = 
\omega(k-\ell)\mathcal{W}^S_{k\ell}\,.
}
One needs to take the ensemble of the source shape, keeping the modes $k \neq \ell$ to detect the birefringence, selecting a certain set of galaxy samples correlating with the polarized emission \cite{2020PhRvL.125v1301M,2025arXiv250406709N}. 

The parity-violating feature can be inspected in terms of the shape correlation function with its trispectra \cite{2025arXiv250908787K}.
We define the spin-$2$ parity-even and parity-odd combinations of the
Hopf-projected Wigner coefficients 
$\mathcal{W}_{k\ell}$ as
\bae{
\mathcal{W}^{E}_{k\ell}
&\equiv
\frac{1}{2}\left(\mathcal{W}_{k\ell}+\mathcal{W}_{\ell k}\right), \cr
\mathcal{W}^{B}_{k\ell}
&\equiv
\frac{1}{2i}\left(\mathcal{W}_{k\ell}-\mathcal{W}_{\ell k}\right),
\qquad |k-\ell|=2,
}
where the condition $|k-\ell|=2$ selects the spin-$2$ sector associated
with galaxy shapes.
By construction, $\mathcal{W}^{E}_{k\ell}$ is even under parity, while
$\mathcal{W}^{B}_{k\ell}$ is odd.

The basic parity-odd observable is the cross-correlation
\bae{
\langle \mathcal{W}^{E}_{k\ell}\,\mathcal{W}^{B\,*}_{k'\ell'} \rangle
=
\frac{1}{4i}
\Big\langle
\big(\mathcal{W}_{k\ell}+\mathcal{W}_{\ell k}\big)
\big(\mathcal{W}^{*}_{k'\ell'}-\mathcal{W}^{*}_{\ell'k'}\big)
\Big\rangle .
\label{eq:rhoE_rhoB_def}
}

Since the Wigner function is bilinear in the galaxy image field $\psi$,
the projected coefficient $\mathcal{W}_{k\ell}$ is itself quadratic in $\psi$.
Consequently, the correlator in Eq.~\eqref{eq:rhoE_rhoB_def} is a
four-point function of the underlying galaxy field.

For a parity-symmetric sky, the two-point functions are parity even,
implying that the disconnected contribution vanishes identically in the
parity-odd channel,
\bae{
\langle \mathcal{W}^{E}_{k\ell}\,\mathcal{W}^{B\,*}_{k'\ell'} \rangle_{\rm disc} = 0 .
}
As a result, any nonzero ensemble expectation value of
$\langle \mathcal{W}^{E}_{k\ell}\mathcal{W}^{B\,*}_{k'\ell'} \rangle$ must arise solely
from the connected four-point function, i.e.\ from a parity-violating
angular trispectrum.

\subsection{Galaxy shape response to spin modes }\label{ssec:gal_shape_response_spin_modes}
We summarize a generating-functional formulation of galaxy shape moments
and their spin (helicity) decomposition, formulated directly in phase space.
The starting point is the image density matrix $\hat\rho$ and its Wigner
function $W(\btheta,\bptheta)$, which are in one-to-one correspondence with
the Weyl characteristic function.
Derivatives of $\chi$ at the origin generate phase-space moments of the
state. Explicitly,
\bae{
&\left.
\frac{\partial^{n+m}\chi}
{\partial \eta_{a_1}\cdots\partial \eta_{a_n}\,
 \partial \zeta_{b_1}\cdots\partial \zeta_{b_m}}
\right|_{\boldsymbol{\eta}=\boldsymbol{\zeta}=\mathbf{0}}
= \cr
&\Big(\frac{i}{\lambdabar}\Big)^{n+m}
\int \dd \mu
W\,
\theta_{a_1}\cdots\theta_{a_n}\,
p_{b_1}\cdots p_{b_m}\,, \cr
\label{eq:generated_moments_helicity_section}
}
which correspond to Weyl-ordered (fully symmetrized) operator moments.
Noting the phase-space coordinates as complex variables, we obtain
\bae{
\left.
\partial_{\eta^*}^{\,p}\partial_{\eta}^{\,q}
\partial_{\zeta^*}^{\,r}\partial_{\zeta}^{\,s}
\chi
\right|_{0}
=
\Big(\frac{i}{\lambdabar}\Big)^{p+q+r+s}
\int \dd \mu\,
W\,
\theta^p(\theta^*)^q\,p^r(p^*)^s.
\label{eq:helicity_moment_generator}
}
Under a physical rotation by $\alpha$,
$(\theta,p) \mapsto e^{ i\alpha}(\theta,p)$, implying that the
moment in Eq.~\eqref{eq:helicity_moment_generator} carries definite helicity
\bae{
n=(p+r)-(q+s).
\label{eq:helicity_counting}
}
In particular, the pure configuration-space spin-$n$ moments follow from
\bae{
\left.\partial_{\eta^*}^{\,n}\chi\right|_{0}
&=
\Big(\frac{i}{\lambdabar}\Big)^n
\int \dd \mu\;
W\,\theta^{n}, \cr
\left.\partial_{\eta}^{\,n}\chi\right|_{0}
&=
\Big(\frac{i}{\lambdabar}\Big)^n
\int \dd \mu\;
W\,(\theta^*)^n.
\label{eq:pure_spin_n_moments}
}
After integrating over $\bptheta$, these reduce to the standard spin-$n$
moments of the image field.

Writing the Wigner function as $W=W(\uu,\vv,\phi_u,\phi_v)$ with phase-space
measure $\dd\mu$, it is natural to define a Hopf-spinor characteristic
function by coupling sources directly to $(u,v)$,
\bae{
\chi_H(J_u,J_v)
:=
\int \dd\mu\;
W\,
e^{\!\ 
J_u u + J_v v + J_u^\ast u^\ast + J_v^\ast v^\ast
}.
\label{eq:hopf_characteristic_def}
}
Derivatives generate Hopf--spinor moments,
\bae{
\left.
\partial_{J_u}^{m}\partial_{J_v}^{n}
\partial_{J_u^\ast}^{r}\partial_{J_v^\ast}^{s}
\chi_H
\right|_{0}
=
\int \dd\mu\;
W\;
u^{m}v^{n}(u^\ast)^{r}(v^\ast)^{s},
\label{eq:hopf_generated_moments_def}
}
which carry torus charges $(m-r,n-s)$ under $(\phi_u,\phi_v)$ shifts. The
physical spin content follows from the embedding of spatial rotations into
the Hopf phases in the WFS formalism.

\section{Practical usage in image analyses}
\label{sec:practical_usage_in_image_analyses}
In this section, we describe how to apply the WFS in practical image analyses. We argue for a choice of the parameters $\sigma, \lambdabar$. Here, $\sigma$ plays a role in determining the typical size of a galaxy image from the data, and thus we should define $\sigma$ in a data-dependent way. As for the weak gravitational lensing, we take care of the fact that $\sigma$ and the lensing convergence $\kappa$ cannot be disentangled from one another, which forces us to measure a reduced shear $\tilde{\gamma} = \gamma/(1-\kappa)$. $\lambdabar$ is relevant to the resolution of galaxy images, given the diffraction limit of a telescope/an array of interferometers or the seeing from the ground. $\lambdabar$ is thus chosen by the property of the PSF. 
We also discuss how to treat the point spread function in the framework of the WFS in terms of an algebraic notation. As we will show later, we point out that image deconvolution from the PSF can be described with operators in phase space or the reduced space of $(Q_0, Q_2)$.

\subsection{Choice of $\sigma$}\label{ssec:choice_of_sigma}
We introduce the two complementary ways to determine $\sigma$ as follows.
We introduce the second moments of the galaxy image as follows. Let $\btheta_c$ be the centroid of the image intensity $I(\btheta)=|\psi|^2$, and define the second moments as
\bae{
\{\boldsymbol{\mathrm{Q}}\}_{ij}&=\frac{\int \dd^2 \btheta (\theta_i-\theta_{c,i})(\theta_j-\theta_{c,j})I\,}{\int \dd^2 \btheta I\,}\,,
}
The first way to determine $\sigma$ is the method of the second-moment matching to a circular ground state of the LG modes, i.e., the Gaussian function $\exp{[-|\btheta-\btheta_c|^2/\sigma^2]}$. We equalize $\sigma$ to the scale $\sigma_g$ that is computed from the trace of the second moments as
\bae{
\sigma = \sigma_g =\sqrt{\tfrac12\,\mathrm{Tr}({\boldsymbol{\mathrm{Q}}})}\,.
}
This definition becomes exact when the galaxy image is well approximated by a Gaussian profile. Note that this way of determining the image size has been widely used in the measurement of the cosmic shear \cite{2003MNRAS.343..459H, LiEtAl2018}. 

The second way to determine $\sigma$ is to infer $\sigma$ from the ground–mode maximization. Project $\psi$ onto the orthonormal LG modes $\{\Psi^\mathrm{LG}_{j,s}(\btheta-\btheta_c;\sigma)\}$  up to order $J_{\max}$. We introduce the function of $\sigma$ as
\bae{
f_{00}(\sigma)=\frac{|\psi_{00}(\sigma)|^2}{\sum_{j\le J_{\max}}\sum_{s=-j}^j |\psi_{j,s}(\sigma)|^2}.
}
We choose $\sigma_\ast=\arg\max_{\sigma>0} f_{00}(\sigma)$. For a nearly Gaussian image $\sigma_\ast\simeq \sigma_g$, for structured morphology, $\sigma_\ast$ adapts to concentrate power in the lowest modes, improving stability for truncated expansions.
Note that we prescribe the scale cut to the image, so that the whole analysis can compute the WFS in a stable way. To make sure of these requirements, we follow a scale cut with a top-hat selection function as
\bae{
T(\btheta) = 
\begin{cases}
    1 \qquad |\btheta-\btheta_c| < r_{\rm cut} \\
    0 \qquad |\btheta-\btheta_c| \geq r_{\rm cut}
\end{cases}
}
where $r_{\rm cut} = \alpha_{\rm cut} \sigma_g$ (or alternatively $\alpha_{\rm cut} \sigma_*$). In practice, $\alpha_{\rm cut} \sim 3-5$ to one galaxy image, isolating the image field from the diffused surroundings.

\subsection{Choice of \texorpdfstring{$\lambdabar$}{lambda-bar}}\label{ssec:choice_lambdabar}
The parameter $\lambdabar$ fixes the canonical relation between the image-plane angle $\btheta$
and its conjugate momentum $\bptheta$. In practice, $\lambdabar$ sets the phase-space cell area, like $(2\pi\lambdabar^2)$ is regard as one resolvable unit: smaller cells overfit noise, while
larger cells smear morphology (spirals, bars, and bulges). Most importantly, $\lambdabar$ provides the quantization scale of a galaxy image, where the LG eigenmodes take part, as well as the WFS. In what follows, we argue what $\lambdabar$ should be.

Our Fourier convention is
\bae{
\tilde{\psi}(\bk) \;=\; \int \dd^2\btheta\, \psi(\btheta)\,e^{-i\,\bk\cdot\btheta}\,,
\qquad
\bk \;=\; \bptheta/\lambdabar.
}
Let $\sigma_\theta$ denote a characteristic angular resolution (e.g.\ PSF width) and
$\sigma_p$ is a characteristic half-bandwidth in $\bptheta$-space from the OTF, which is defined as the Fourier transform of the PSF as
\bae{
K_{\rm OTF}(\bptheta/\lambdabar) \equiv \int \,\dd^2\btheta K_{\rm PSF}(\btheta)\,e^{-\frac{i\bptheta\cdot\bxi}{\lambdabar}}\,.
}
For a reference Gaussian, $\sigma_\theta \sigma_k = 1$ with $\,\sigma_k$ the width of $\tilde\psi(\bk)$,
hence $\sigma_p=\lambdabar\,\sigma_k$ and
\bae{
\lambdabar \;=\; \sigma_\theta\,\sigma_p\;\,.
}
Thus matching $\lambdabar$ to the instrument/data amounts to matching the resolution $\times$ bandwidth product.
In the absence of additional systematic blurring---such as atmospheric turbulence for ground-based telescopes---an astronomical image reaches the diffraction limit set by the finite aperture, where any astronomical images are inevitably unresolved.  
For a circular pupil of diameter $D$, the incoherent PSF is given by the Airy pattern,
\[
I(\theta)
\propto
\left[\frac{2J_1(x)}{x}\right]^2,
\qquad
x \equiv \frac{\pi D}{\lambda_{\rm phys}}\,\theta ,
\]
where $J_1$ is the Bessel function of the first kind.
The full width at half maximum (FWHM) is determined by the solution
\(
[2J_1(x_{1/2})/x_{1/2}]^2 = 1/2
\),
which yields $x_{1/2}\simeq1.6163$ and hence
\[
{\rm FWHM}
=2\theta_{1/2}
=\frac{2x_{1/2}}{\pi}\frac{\lambda_{\rm phys}}{D}
\simeq1.028\,\frac{\lambda_{\rm phys}}{D}.
\]
Approximating the Airy core by a Gaussian of equal FWHM then gives
\[
\sigma_\theta
=
\frac{{\rm FWHM}}{2\sqrt{2\ln2}}
\simeq0.437\,\frac{\lambda_{\rm phys}}{D}.
\]

The OTF of a clear circular pupil has compact support and vanishes for angular spatial frequencies larger than
\[
\kappa_c \simeq \frac{2\pi D}{\lambda_{\rm phys}} ,
\]
so a characteristic momentum-space bandwidth may be taken as
\(
\sigma_p \simeq \kappa_c/2 = \pi D/\lambda_{\rm phys}.
\)
The corresponding phase-space area is , therefore, 
\bae{
\lambdabar
\;\equiv\;
\sigma_\theta\,\sigma_p
\;\simeq\;
0.437\,\pi
\;\simeq\;
1.37
\quad (\mathcal{O}(1)),
}
demonstrating that a diffraction-limited image naturally occupies a phase-space cell of order unity. In real observations, however, other systematics add to the diffraction limits, making the image smoother. In what follows, we write down several systematics:
\paragraph*{Seeing-limited.}
  The image is further blurred by atmospheric turbulence for ground-based telescopes, reaching the seeing limit. In this case, one should take $\sigma_\theta$ from the measured seeing FWHM and $\sigma_p$ from the empirical modulation transfer function (MTF) half-power
  or effective support --- set $\lambdabar=\sigma_\theta\sigma_p$.
\paragraph*{Radio interferometry.}
Radio interferometry overcomes the atmospheric seeing limit by observing at long wavelengths and by synthesizing an effective aperture from widely separated antennas.
In this case, however, the measured sky modes are no longer continuous: the interferometer samples the Fourier plane only at discrete baseline vectors, leading to the well-known problem of incomplete $uv$ coverage \citep{2017isra.book.....T}.
For an array with maximum projected baseline $b_{\max}$, the synthesized beam sets the characteristic angular resolution
$
\sigma_\theta \sim \lambda_{\rm phys}/{b_{\max}},
$
up to a numerical factor that depends on the detailed baseline distribution and weighting scheme.
In Fourier space, the sampled visibility function has compact support bounded by the longest baseline, implying a cutoff in angular spatial frequency
$
\kappa_c \simeq 2\pi b_{\max}/\lambda_{\rm phys}.
$
A representative momentum-space bandwidth may, therefore, be taken as
\(
\sigma_p \simeq \kappa_c/2 .
\)
The resulting phase-space area then satisfies
\[
\lambdabar
\;\equiv\;
\sigma_\theta\,\sigma_p
\;\sim\;
\mathcal{O}(1),
\]
indicating that, despite incomplete mode sampling, the synthesized image produced by a radio interferometer occupies a phase-space cell comparable to that of a diffraction-limited optical system.

\paragraph*{Additional instrumental blurring mechanisms.}
Astronomical images are further degraded by several
instrumental effects that suppress high-spatial-frequency modes.
In Fourier space, these effects act multiplicatively on the OTF, while in the image plane, they broaden the effective PSF.
To leading order, these blurs are well approximated as Gaussian and may be
combined in quadrature \citep{Goodman2005,2005PASP..117..594K}.
The first one is pointing jitter. Random telescope pointing fluctuations during an exposure are well described
by a Gaussian angular displacement with rms width $\sigma_{\theta,\rm jit.}$
\citep{1998aoad.book.....H}.
The resulting image is the convolution of the diffraction-limited PSF with
the jitter kernel. 

Second, the pixel response matters in image detection. Finite pixel size introduces an additional blur corresponding to the pixel
response function.
For square pixels of angular width $p$, the image is convolved with a
top-hat kernel, yielding an OTF factor
$
K_{\rm OTF, pix.}
=
\operatorname{sinc}\!\left(\frac{\kappa_x p}{2}\right)
\operatorname{sinc}\!\left(\frac{\kappa_y p}{2}\right),
$
where $\operatorname{sinc}(x)=\sin x/x$ \citep{Goodman2005}.
For engineering estimates, the pixel blur is often replaced by a Gaussian
with equivalent variance,
$
\sigma_{\theta,\rm pix.}
\;\approx\;p/\sqrt{12},
$
corresponding to the second moment of a square top-hat response.

Third, detector charge diffusion occurs when CCD/CMOS stores the image. In CCD and CMOS detectors, photoelectrons diffuse laterally before collection,
producing an approximately Gaussian broadening of the image
with rms width $\sigma_{\theta,\rm det.}$ \citep{2001sccd.book.....J}.

Treating all
nondiffraction blurring mechanisms as a Gaussian in the image plane, and to
add their variances in quadrature,
\bae{
\sigma_{\theta,\rm tot.}^2
\;\approx\;
\sigma_{\theta,\rm diff.}^2
+
\sigma_{\theta,\rm jit.}^2
+
\sigma_{\theta,\rm pix.}^2
+
\sigma_{\theta,\rm det.}^2 .
}
This prescription defines the effective resolution element of the system
and naturally identifies the minimum resolvable phase-space cell.
Consequently, a Wigner-space structure below this scale should not be
overinterpreted as physically meaningful.

One may determine $\lambdabar$ in a data-driven way from the image. In this case we propose to make use of
\bae{
\sigma_\theta^{2} \;=\; \frac{\int \,\dd^2\btheta |\btheta-\bar\btheta|^{2}\,I(\btheta)}{\int \,\dd^2\btheta I(\btheta)},
\qquad
\sigma_k^{2} \;=\; \frac{\int |\bk|^{2}\,|\tilde\psi(\bk)|^{2}\,\dd^2\bk}{\int |\tilde\psi(\bk)|^{2}\,\dd^2\bk}.
}
With $\bptheta=\lambdabar\,\bk$ we have $\sigma_p=\lambdabar\,\sigma_k$, hence, we enforce the Gaussian reference
$\sigma_\theta\sigma_k=1$ to get $\lambdabar$. With this choice, $\lambdabar=\mathcal{O}(1)$ at the angular scales where galaxy morphology lives, and the Wigner
phase-space analysis resolves exactly what the instrument can support—no less, no more.

\subsection{Existence of optimal recovery}
\label{ssec:entanglement_fidelity_recovery}

In order to quantify how much morphological information is preserved under
systematics and noise, we evaluate the information-theoretic fidelity of a
channel on a certain ensemble of galaxy images, generalizing the single image state for the moment.  Let $\hat{\rho}_I$ denote the
ensemble (source) state on the Hilbert space
$\mathcal{H}_{\lambdabar}$ truncated by $\lambdabar$, for example obtained from a population distribution
$p(\psi)$ as $\hat{\rho}_I=\mathbb{E}_{\psi\sim p}[\,|\psi\rangle\langle\psi|\,]$.
The observation process (PSF, diffusion, detector noise, etc.) is a CPTP map
$\mathcal{E}$, so that $\hat{\rho}_O=\mathcal{E}(\hat{\rho}_I)$.

Let us introduce an abstract reference system $R$ and a purification
$|\Psi_{\hat{\rho}_I}\rangle\in\mathcal{H}_{\lambdabar}\otimes\mathcal{H}_R$
satisfying
$\Tr_R[|\Psi_{\hat{\rho}_I}\rangle\langle\Psi_{\hat{\rho}_I}|]=\hat{\rho}_I$.
The entanglement fidelity of a channel $\mathcal{N}$ acting on the image system
is defined by
\bae{
F_e(\hat{\rho}_I,\mathcal{N})
\equiv
\langle\Psi_{\hat{\rho}_I}|
(\mathcal{N}\otimes\mathbb{I}_R)
\!\left(|\Psi_{\hat{\rho}_I}\rangle\langle\Psi_{\hat{\rho}_I}|\right)
|\Psi_{\hat{\rho}_I}\rangle .
\label{eq:entanglement_fidelity_def}
}
This quantity satisfies $0\le F_e\le 1$ and is independent of the particular
choice of purification.  Operationally, $F_e$ measures the fraction of Schumacher-compressible information of the source $\hat{\rho}_I$
that is preserved by $\mathcal{N}$, including all correlations encoded in
off-diagonal mode coherences.


If $\mathcal{N}$ admits a Kraus representation
$\mathcal{N}(\hat{\rho})=\sum_a \hat{K}_a\hat{\rho}\hat{K}_a^\dagger$,
then the entanglement fidelity admits the useful expression
\bae{
F_e(\hat{\rho}_I,\mathcal{N})
=
\sum_a \big|\Tr(\hat{\rho}_I\,\hat{K}_a)\big|^2 .
\label{eq:entanglement_fidelity_kraus}
}
For a continuous Kraus index $z$, the sum is replaced by an integral,
\bae{
F_e(\hat{\rho}_I,\mathcal{N})
=
\int\!\dd\mu(z)\;
\big|\Tr(\hat{\rho}_I\,\hat{K}(z))\big|^2 .
\label{eq:entanglement_fidelity_continuous}
}


A reconstruction (denoising/deconvolution) is itself a CPTP map
$\mathcal{R}$ applied to the observed state, producing a recovery of an image as
\bae{
\hat{\rho}_{\rm rec}=\mathcal{R}(\hat{\rho}_O)\,.
}
We assess performance at the ensemble level via the entanglement fidelity of
the effective channel
\bae{
\mathcal{N}\equiv \mathcal{R}\circ\mathcal{E},
\qquad
F_e(\hat{\rho}_I,\mathcal{R}\circ\mathcal{E}) .
\label{eq:effective_channel_recovery}
}
Within this framework, an information-theoretically optimal denoiser is a
recovery map that maximizes $F_e$ (or an equivalent monotone) over an allowed
class of physically admissible reconstructions $\mathcal{R}$,
\bae{
\mathcal{R}_\star
\in
\arg\max_{\mathcal{R}\in\mathfrak{R}}
F_e(\hat{\rho}_I,\mathcal{R}\circ\mathcal{E}) .
\label{eq:optimal_recovery_def}
}
The constraint set $\mathfrak{R}$ may incorporate a practical structure, such as
symmetry equivariance, locality in mode space, or computational tractability.

The total observation channel is the composition
\bae{
\mathcal{E}
=
\mathcal{E}_{\rm noise}\circ\mathcal{E}_{\rm PSF},
\qquad
\hat{\rho}_O
=
\mathcal{E}(\hat{\rho}_I),
}
which remains Gaussian and translationally invariant.
In phase space, the combined kernel is again Gaussian,
\bae{
\Pi_{\rm tot}(\btheta,\bptheta)
\propto
\exp\!\left(
-\btheta^\top \mathbf{A}\,\btheta
-
\bptheta^\top \mathbf{B}\,\bptheta
\right),
}
with covariance matrices $\mathbf{A},\mathbf{B}$ determined by $\beta$ and the
noise strengths $\gamma_i$. We expand $W_I$
and $W_K$ in terms of the torus harmonics as 
\bae{
W_{I}
&=
\sum_{(J,K)\in\Zeven}\;
\tilde{\mathcal{W}}_{JK}(\bQ)\,\tilde{\chi}_{JK}(\bphi),\cr
W_{K}
&=
\sum_{(J,K) \in \Zeven}
\tilde{\mathcal{K}}_{JK}(z;\bQ)\,\tilde{\chi}_{JK}(\bphi)\,.
}
Here we define $\Zeven \;\equiv\;
\bigl\{(J,K)\in\mathbb{Z}^2 \,\big|\, J\equiv K \ (\mathrm{mod}\ 2)\bigr\}$, and $\tilde{\mathcal{W}}_{JK} \equiv \mathcal{W}_{\frac{J+K}{2},\frac{J-K}{2}}$ and $\tilde{\chi}_{JK} \equiv \chi_{\frac{J+K}{2},\frac{J-K}{2}}$
. Note that we concisely denote $\bQ = (Q_0,Q_2)$ and $\bphi = (\phi_u,\phi_v)$, The label $z$ denotes the continuous spectrum of the kernel.
The entanglement fidelity can be written as a quadratic form
\bae{
F_e
=&
\sum_{(J,K) \in \Zeven}\sum_{(J',K') \in \Zeven} \cr
&\times
\int\dd \bQ\,\dd \bQ'\;
\mathcal{W}_{JK}(\bQ)\,
\mathcal{M}_{JK,J'K'}(\bQ,\bQ')\,
\mathcal{W}^\ast_{J'K'}(\bQ'),\cr
}
where we have defined the fidelity kernel as
\bae{
\mathcal{M}_{JK,J'K'}(\bQ,\bQ')
\equiv (2\pi)^4
\int\!\dd \mu(z)\;
\mathcal{K}_{JK}(z;\bQ)\,
\mathcal{K}^\ast_{J'K'}(z;\bQ')\,.
\label{eq:fidelity_kernel_def}
}

Let us interpret the above formulation in more detail.
The kernel $\mathcal{M}_{JK,J'K'}$ is a positive semidefinite operator on
the space of WFS coefficients and encodes the combined effect of the
forward imaging channel and the recovery map.
Its spectral properties determine which combinations of morphology and
rotation charges are information-bearing and which are irretrievably
, as follows:
\begin{itemize}
\item Eigenmodes of $\mathcal{M}$ with large eigenvalues correspond to
charge combinations that can be faithfully recovered.
\item Small or vanishing eigenvalues signal directions in
$(J,K,\bQ)$ space where information is erased by PSF blurring and noise.
\item The maximal achievable entanglement fidelity is obtained by
restricting reconstruction to the support of $\mathcal{M}$.
\end{itemize}
For a stationary PSF and homogeneous noise, translational invariance
ensures diagonalization in Fourier space, while rotational covariance
further constrains the kernel to be block-diagonal in the rotation
charge:
\bae{
\mathcal{M}_{JK,J'K'} \;\propto\; \delta_{KK'}\,\mathcal{M}^{(K)}_{JJ'}.
}
In this case, different $K$ sectors are independent, and information loss
is governed primarily by mixing among the morphology charges $J$ within each
rotation sector.
Isotropic PSFs preserve the $K$ labels, whereas anisotropic PSFs induce
off-diagonal couplings between different $J$ blocks.
Within any symmetry-equivariant class of CPTP recovery maps
$\mathcal{R}$, the entanglement-fidelity-optimal reconstruction
corresponds to choosing $\mathcal{R}$ such that the effective kernel
$\mathcal{K}$ is optimally inverted (with regularization) on its support.
In practice this leads to Wiener- or Petz--type \cite{Petz1986} operators that act
block-by-block in $(J,K)$ space.
The WFS basis, therefore, provides a natural coordinate system in which the
fidelity kernel is sparse, symmetry-adapted, and directly interpretable.
The WFS decomposition does not eliminate information
loss---which is fundamentally bounded by $\mathcal{M}$---but renders it
explicit.
By diagonalizing the symmetry structure of $\mathcal{M}$, it enables
recovery maps that saturate the maximum achievable entanglement fidelity
while preserving all symmetry-protected morphology and rotation modes.

\subsection{Feasibility}\label{ssec:feasibility}
We describe how feasible the methodology of image analysis in terms of the WFS is, aligned with the timeline of the present and the forthcoming surveys. At the level of the data compression into the Wigner function, it should be available as it shares the same procedure as the standard Fourier transform, and the mathematics behind the WFS is rigorous.What is novel with the WFS is to make any deformation of images classified as either dynamical deformation i.e. Hamiltonian deformation, or random deformation that dilutes the information of images that can break the symplectic symmetry. For example, gravitational lensing is a dynamical deformation, whereas the intrinsic alignments or PSF blurring are more complex to characterize as a simple dynamical process of deformation with a finite number of degrees of freedom. Besides, it is worth mentioning that the WFS can trace all types of canonical transforms, while the Fourier transform is just one of the canonical transforms of images by using the plane-wave decomposition. In other words, the WFS can treat some dynamical deformation with which the Fourier decomposition becomes complex i.e. nonlinear interactions.  Hence, the WFS is not only the mathematical description of images but also the appropriate representation of images, covering the entire range of physical processes behind the deformation of images. What is nontrivial up to the point when this paper is presented in the WFS is to reconstruct the image of galaxies within the framework of quantum information theory, for which few have addressed similar algorithms in practical image analysis in astronomy, except a couple of conceptual proposals for future telescopes cf.~\cite{2025arXiv250909465M}. Although this may sound too futuristic, it has been rapidly developing in other areas of research, such as quantum image processing \cite{2023Optic..10.1142K, 2019NatCo..10.1865F}. To this end, our methodology should be developed in astronomy, and we need to further examine how much our methodology can be advantageuous in the conventional image processing in astronomy.

\section{Conclusions}\label{sec:conclusion}
We have extended the conventional shapelet analysis for astronomical images into the {\it Wigner function shapelets} (WFSs). The WFS decomposes the Wigner function of the galaxy image directly in the four-dimensional phase space, making the quantum information theory in image processing available. After sorting a useful extraction of the topological information of images via Hopf fibration of the phase space, we have found a useful observable $\mathcal{W}_{k\ell} (Q_0,Q_2)$ with the two rotational-invariant phase-space variables $(Q_0,Q_2)$. We have computed the linear response to the weak gravitational lensing. In addition, we have found a probe for any parity-violating signature in the ensemble average of a pair product of the Wigner function of the two galaxy images. This paper was written to provide a baseline formalism of the WFS. We will validate our methodology in practical image analysis, looking into image simulations as well as real-data processing. 

In astronomy, analyses of shapes of astronomical objects serve to provide physical information about astronomical systems since the pioneering observations carried out over the centuries \cite{Galilei1989, Huygens1659, Cassini1675, Herschel1912, Hubble1926, Hubble1936, Zwicky1957, deVaucouleurs1959}. In the modern context,  the range of studies has expanded to galaxy morphology \cite{SteinmetzNavarro2002,Conselice2014, Buta2010, Lintott2008, Iyer2024}, structure formation \cite{Bag2021}, as well as testing fundamental physics of gravity \cite{EHT2019L1}, The higher-resolution astronomical images can be obtained from many astronomical observations, cf. the Event Horizon Telescope \cite{EHT2019L1}, with more attention to developing shape analysis, e.g., machine learning.
\cite{BartelmannSchneider2001,Kilbinger2015}, while parity-violating effects can rotate polarization or induce cosmic birefringence \cite{LueWangKamionkowski1999,MinamiKomatsu2020}. These science targets motivate estimators that are robust to intrinsic morphology, anisotropic PSFs, masking, and source blending. To this end, we aim to establish the WFS as a novel method of image analyses ahead of the future research, during which some of the fundamental questions in astrophysics and cosmology may be solved.

\begin{acknowledgments}
S.A. thanks Tsutomu Takeuchi for future encouraging applications of the WFS in galaxy image analysis. S.A. also thanks Eiichiro Komatsu and Fumihiro Naokawa for insightful discussions. 
S.A. is supported by the Japan Society for the Promotion of Science
(JSPS) Grants-in-Aid for Scientific Research (KAKENHI) Grant No. JP24K17045.
\end{acknowledgments}

\section*{data availability}
All the mathematical formulae are written, and no external data are used.

\appendix
\section{Classical limit of Wigner function and Wigner transport}\label{app:classical_lim_wigner}
We describe how the classical Liouville equation is derived from the Wigner function and the Wigner transport at the limit of $\lambdabar \rightarrow 0$, where the resolution of the system is infinitesimal. Then we conclude that the Wigner function is proportional to the radiance (or specific intensity), and the Liouville equation we obtain is nothing but the radiative-transfer equation of an image.

Let us assume a semiclassical family of image-plane fields of the form
\bae{
\psi(\btheta) \;=\; a(\btheta)\,\exp\!\Big[\tfrac{i}{\lambdabar}\,S(\btheta)\Big],
}
with smooth real phase $S(\btheta)$ and slowly varying complex amplitude $a(\btheta)$. In other words, we take a WKB approximation for the image field. The Wigner function is expressed by $a(\btheta)$ and $S(\btheta)$ as
\bae{
W(\btheta,\bptheta)
=\frac{1}{(2\pi\lambdabar)^2}\int \dd^2\bxi\;
a\!\Big(\btheta+\tfrac{\bxi}{2}\Big)\,a^*\!\Big(\btheta-\tfrac{\bxi}{2}\Big)\,
e^{\tfrac{i}{\lambdabar}\,\Phi(\btheta,\bptheta;\bxi)},
}
with phase
\bae{
\Phi(\btheta,\bptheta;\bxi)
= S\!\Big(\btheta+\tfrac{\bxi}{2}\Big) - S\!\Big(\btheta-\tfrac{\bxi}{2}\Big) - \bptheta\!\cdot\!\bxi .
}
We rescale the integration variable as $\bxi=\lambdabar\,\boldsymbol{y}$ to obtain
\bae{
W(\btheta,\bptheta)
=\frac{1}{(2\pi)^2}\int \dd^2\boldsymbol{y}\;
a\!\Big(\btheta+\tfrac{\lambdabar}{2}\boldsymbol{y}\Big)
a^*\!\Big(\btheta-\tfrac{\lambdabar}{2}\boldsymbol{y}\Big)\,
e^{i\,\Psi(\btheta,\bptheta;\boldsymbol{y},\lambdabar)},
}
where
\bae{
\Psi(\btheta,\bptheta;\boldsymbol{y},\lambdabar)
=\frac{S(\btheta+\frac{\lambdabar}{2}\boldsymbol{y})-S(\btheta-\frac{\lambdabar}{2}\boldsymbol{y})}{\lambdabar}
-\boldsymbol{y}\!\cdot\!\bptheta .
}
Executing Taylor expansion of $\lambdabar$ at fixed $\btheta$ and using that the difference is odd in $\boldsymbol{y}$ gives
\bae{
\frac{S(\btheta+\frac{\lambdabar}{2}\boldsymbol{y})-S(\btheta-\frac{\lambdabar}{2}\boldsymbol{y})}{\lambdabar}
= \boldsymbol{y}\!\cdot\!\nabla_{\btheta}S(\btheta) \;+\; \mathcal{O}\!\big(\lambdabar^{2}|\boldsymbol{y}|^{3}\big),
}
and similarly
\bae{
a\!\Big(\btheta\pm\tfrac{\lambdabar}{2}\boldsymbol{y}\Big)
= a(\btheta) \;\pm\; \mathcal{O}\!\big(\lambdabar |\boldsymbol{y}|\big).
}
Keeping the leading terms in phase and amplitude,
\bae{
W(\btheta,\bptheta)
= \frac{1}{(2\pi)^2}\int \dd^2\boldsymbol{y}\; |a(\btheta)|^2\,
e^{i\,\boldsymbol{y}\!\cdot\!\big(\nabla_{\btheta}S(\btheta)-\bptheta\big)}
\;+\;\mathcal{O}(\lambdabar).
}
Using the Fourier identity $\frac{1}{(2\pi)^2}\int \dd^2\boldsymbol{y}\; e^{\,i\,\boldsymbol{y}\cdot\boldsymbol{q}}=\delta(\boldsymbol{q})$, we obtain the distributional limit
\bae{
W(\btheta,\bptheta) \;\xrightarrow[\ \lambdabar\to 0\ ]{}\;
L(\btheta,\bptheta) \;=\; |a(\btheta)|^2\,\delta\!\Big(\bptheta-\nabla_{\btheta}S(\btheta)\Big).
}
Thus the Wigner function concentrates on the Lagrangian manifold
$\{(\btheta,\bptheta):\bptheta=\nabla_{\btheta}S(\btheta)\}$ with weight $|a(\btheta)|^2$.
The limit $L(\btheta,\bptheta)$ is a positive Radon measure on phase space: a non-negative generalized density (possibly singular) integrable against smooth test functions.

The measure $L(\btheta,\bptheta)$ is the classical radiance (or specific intensity) on the image-plane phase space: it assigns power per unit projected area per unit solid angle (per unit frequency/wavelength if spectral). In fact, its $\btheta$-marginal recovers the image intensity:
\bae{
\int L(\btheta,\bptheta)\,\dd^2\bptheta \;=\; |a(\btheta)|^2 \;=\; \lim_{\lambdabar\to 0} |\psi(\btheta)|^2 \;\equiv\; I(\btheta),
}
while its $\bptheta$–marginal gives the classical distribution over propagation directions (spatial frequencies). For a finite superposition $\psi=\sum_n a_n(\btheta)\exp[i S_n(\btheta)/\lambdabar]$ one has
$W=\sum_n W_{nn}+\sum_{n\neq n'}W_{nn'}$.
Each diagonal term $W_{nn}$ converges to
$|a_n(\btheta)|^2\,\delta\!\big(\bptheta-\nabla_{\btheta}S_n(\btheta)\big)$.
The cross-terms $W_{nn'}$ carry the rapidly oscillating phase
$\exp[i(S_n-S_{n'})/\lambdabar]$ and vanish in the weak limit by nonstationary phase
(unless $\nabla_{\btheta}S_n=\nabla_{\btheta}S_m$ on a set of measure zero, e.g.\ caustics). Hence
\bae{
W \;\rightharpoonup\; L(\btheta,\bptheta)
= \sum_n |a_n(\btheta)|^2\,\delta\!\big(\bptheta-\nabla_{\btheta}S_n(\btheta)\big),
}
a positive phase-space measure with no residual Wigner negativity.

If $\psi$ evolves under a generator with Weyl symbol $H_W(\btheta,\bptheta)$, the exact Moyal evolution
$\partial_s W=\{H_W,W\}_{\rm MB}$ as shown in Eq.~\eqref{eq:wigner_transport} reduces as $\lambdabar\to 0$ to the Liouville (radiative-transfer) equation
\bae{
\partial_s L \;+\; \{H,L\}_{\rm PB} \;=\; 0,
}
so radiance is transported along Hamiltonian rays. For quadratic Hamiltonian $H$ this transport law already holds exactly for all $\lambdabar$. In the rest of the appendix, we derive that the quadratic Hamiltonian $H$ is derived from the Wigner symbol $H_W$, completing the whole logic to be consistent.

Define the canonical momentum operator
\bae{
\hat{\bptheta} \;=\; -\,i\,\lambdabar\, \nabla_{\boldsymbol{\theta}}, 
}
so that the commutators are $[\hat\theta_i,\hat p_j] = i\,\lambdabar\,\delta_{ij}$ (with $i,j\in\{x,y\}$).
Then
\bae{
\hat H \;=\; -\,\frac{\lambdabar^{2}}{2}\nabla_{\boldsymbol{\theta}}^{2} \;+\; \hat V
\;=\; \frac{\hat{\bptheta}^{\,2}}{2} \;+\; \hat V,
}

If an operator $\hat A$ has kernel $A(\boldsymbol{\theta}_1,\boldsymbol{\theta}_2)$ in the angle basis, its Weyl symbol is
\bae{
A_W(\btheta,\bptheta)
\;=\; 
\int \mathrm{d}^{2}\boldsymbol{\xi}\;
A\!\left(\boldsymbol{\theta}+\tfrac{\boldsymbol{\xi}}{2},\,\boldsymbol{\theta}-\tfrac{\boldsymbol{\xi}}{2}\right)
\,e^{-\frac{i\bptheta\cdot\bxi}{\lambdabar}}.
}
For $(\hat V f)(\boldsymbol{\theta}) = V(\boldsymbol{\theta})f(\boldsymbol{\theta})$, the kernel is
$V(\boldsymbol{\theta}_1,\boldsymbol{\theta}_2)=V(\boldsymbol{\theta}_1)\,\delta(\boldsymbol{\theta}_1-\boldsymbol{\theta}_2)$, hence
\bae{
(V)_W(\btheta,\bptheta) \;=\; V(\btheta) \quad \text{(exact).}
}
For $\hat T=\hat{\bptheta}^{\,2}/2$, the kernel is
\bae{
T(\boldsymbol{\theta}_1,\boldsymbol{\theta}_2)
\;=\; -\,\frac{\lambdabar^{2}}{2}\,\nabla_{\boldsymbol{\theta}_1}^{2}\,\delta(\boldsymbol{\theta}_1-\boldsymbol{\theta}_2).
}
Write $\boldsymbol{\theta}=\tfrac{1}{2}(\boldsymbol{\theta}_1+\boldsymbol{\theta}_2)$ and $\boldsymbol{\xi}=\boldsymbol{\theta}_1-\boldsymbol{\theta}_2$. Since $\delta$ depends only on $\boldsymbol{\xi}$, we obtain
\bae{
T\!\left(\boldsymbol{\theta}+\tfrac{\boldsymbol{\xi}}{2},\,\boldsymbol{\theta}-\tfrac{\boldsymbol{\xi}}{2}\right)
\;=\; -\,\frac{\lambdabar^{2}}{2}\,\nabla_{\boldsymbol{\xi}}^{2}\,\delta(\boldsymbol{\xi}).
}
Thus,
\bae{
\begin{aligned}
(T)_W(\btheta,\bptheta)
&= \int \mathrm{d}^{2}\boldsymbol{\xi}\;\Big[-\,\frac{\lambdabar^{2}}{2}\,\nabla_{\boldsymbol{\xi}}^{2}\delta(\boldsymbol{\xi})\Big]\, e^{-\frac{i\bptheta\cdot\bxi}{\lambdabar}} \\
&= -\frac{\lambdabar^{2}}{2}\int \mathrm{d}^{2}\boldsymbol{\xi}\;\delta(\boldsymbol{\xi})\,
\nabla_{\boldsymbol{\xi}}^{2}\,e^{-\frac{i\bptheta\cdot\bxi}{\lambdabar}} \\
&= \frac{\lambdabar^{2}}{2}\,\Big[\,\,\frac{{\bptheta}^{2}}{\lambdabar^{2}}\Big] \\
&= \frac{\bptheta^2}{2},
\end{aligned}
}
Hence, the kinetic Weyl symbol equals $\bk^2/2$ exactly. Combining the two pieces,
\bae{
H_W(\btheta,\bptheta) \;=\; \frac{\bptheta^2}{2} \;+\; V(\btheta),
}
which is precisely the classical Hamiltonian of geometrical optics on $(\btheta,\bptheta)$. 
Because $H_W$ is quadratic in $\bptheta$ and local in $\btheta$, the Moyal and Poisson brackets coincide exactly, thus the phase-space flow is a linear symplectic (metaplectic) map with no higher-order corrections.

\section{Wigner function for polarized/colored images}\label{sec:wigner_polarisations_colours}
We extend the formalism for the Wigner function of a galaxy image so that it can analyze an image with polarizations/colors. Let us treat the image as a Jones field $\psi(\btheta)\in\mathbb{C}^2$ and define a $2\times2$ matrix-valued cross-Wigner function,
\bae{
\big[W_{ab}^{\rm pol}(z)\big]_{ij}
= \frac{1}{(2\pi\lambdabar)^2}\!\int\!\dd^2\bxi\;\psi_{a,i}(\btheta{+}\bxi/2)\,\psi_{b,j}^*(\btheta{-}\bxi/2)\,e^{-\frac{i\bptheta\cdot\bxi}{\lambdabar}}.
}
For a polarization basis change $U\in{\rm U}(2)$ acting on components, one has
\bae{
W_{Ua,\,Ub}^{\rm pol}(z)=U\,W_{ab}^{\rm pol}(z)\,U^\dagger,
}
so the total-intensity (scalar) cross-Wigner $\mathrm{tr}\,W_{ab}^{\rm pol}$ is basis-invariant. Thus, polarization changes are genuinely unitary in this framework (rotations, phase retarders, Faraday rotation in a lossless model, etc.).

Two images of the same galaxy at different wavelengths need not be related by any unitary $U$ on a common Hilbert space: the sky emission, PSF, and transfer function are chromatic and typically include nonunitary effects (convolution, absorption, and emission). A useful exception is a pure first-order, lossless, paraxial change that is equivalent to a symplectic squeeze. Writing the symplectic squeeze
\; $S_s=\mathrm{diag}(s\,I_2,\,s^{-1}I_2)$ \; with metaplectic lift $\mu(S_s)$, one has
\bae{
W_{\mu(S_s)a,\,\mu(S_s)b}(z)=W_{ab}(S_s^{-1}z).
}
In diffraction-limited optics, the core scale changes $\propto\lambda$. After nondimensionalizing $z$ so that $\lambdabar$ is fixed, choosing $s\approx\sqrt{\lambda_2/\lambda_1}$ models the achromatic scaling between bands. Beyond this ideal case (e.g. band-dependent PSF, dust, and stellar-population changes), no unitary map exists, and the cross-Wigner should be used as a comparative statistic rather than expected to obey simple covariance.

\section{Derivation of the PSF phase-space kernel by direct variable changes}
\label{app:psf_kernel_derivation}

We derive the phase-space representation of PSF blurring directly from the
configuration-space action by changing integration variables.
Let $h(\btheta)$ be the amplitude spread function, and let the morphology kernel
be $\rho_I(\btheta_1,\btheta_2)$.
The PSF acts bilinearly as
\bae{
\rho_O(\btheta_1,\btheta_2)
=
\int \dd^2\btheta_1' \dd^2\btheta_2'\;
h(\btheta_1-\btheta_1')\,
h^*(\btheta_2-\btheta_2')\,
\rho_I(\btheta_1',\btheta_2')\,.
\label{eq:rho_psf_bilinear}
}
Inserting $\rho_O$ into the Wigner transform in Eq.~\eqref{eq:wigner-def}, we obtain
\bae{
\begin{aligned}
W_O(\btheta,\bptheta)
&=
\int \dd^2\bxi\;
e^{-i\bptheta\cdot\bxi/\lambdabar}\,
\rho_O\!\Big(\btheta+\tfrac{\bxi}{2},\,\btheta-\tfrac{\bxi}{2}\Big)
\\
&=
\int \dd^2\btheta_1 \dd^2\btheta_2 \dd^2\bxi\;e^{-i\bptheta\cdot\bxi/\lambdabar}\cr
& \quad \times h\!\Big(\btheta+\tfrac{\bxi}{2}-\btheta_1\Big)\,
h^*\!\Big(\btheta-\tfrac{\bxi}{2}-\btheta_2\Big)\,
\rho_I(\btheta_1,\btheta_2)\,.
\end{aligned}
\label{eq:WO_start}
}
Then, we change the variables in the integral with
\bae{
\btheta' \equiv \frac{\btheta_1+\btheta_2}{2},
\qquad
\boldsymbol{\eta} \equiv \btheta_1-\btheta_2, \qquad \boldsymbol{s} \equiv \bxi-\boldsymbol{\eta}\,.
\label{eq:mean_diff}
}
One can confirm that the Jacobian stays at unity. As the PSF factors depend only on $(\btheta-\btheta',\boldsymbol{s})$ and the phase
splits as
$e^{-i\bptheta\cdot\bxi/\lambdabar}
=
e^{-i\bptheta\cdot\boldsymbol{s}/\lambdabar}\,
e^{-i\bptheta\cdot\boldsymbol{\eta}/\lambdabar}$, thus, $W_O$ is rewritten as
\bae{
\begin{aligned}
W_O(\btheta,\bptheta)
&=
\int \dd^2\btheta'\dd^2\boldsymbol{\eta}\dd^2\boldsymbol{s}\; \cr
& \qquad \times h\!\Big(\btheta-\btheta'+\tfrac{\boldsymbol{s}}{2}\Big)\,
h^*\!\Big(\btheta-\btheta'-\tfrac{\boldsymbol{s}}{2}\Big)\,
e^{-\frac{i\bptheta\cdot\bxi}{\lambdabar}}\,\cr
&\qquad \times
\rho_I\!\Big(\btheta'+\tfrac{\boldsymbol{\eta}}{2},\,\btheta'-\tfrac{\boldsymbol{\eta}}{2}\Big)\,
e^{-\frac{i\bptheta\cdot\bxi}{\lambdabar}}.
\end{aligned}
\label{eq:WO_factorised}
}
At this stage, it is more explicit that the integrals over $\boldsymbol{s}$ and $\boldsymbol{\eta}$, respectively, create the Wigner functions for each kernel $h(\btheta-\btheta')h^*(\btheta-\btheta')$ and $\rho_I(\btheta')$. Hence, the final expression becomes by representing $\btheta = \btheta^I$,
\bae{
W_O(\btheta,\bptheta)
=
\int \dd\mu^I\;
\Pi_{\rm PSF}\!\left(\btheta-\btheta^I,\,\bptheta-\bptheta^I\right)\,
W_I(\btheta^I,\bptheta^I),
\label{eq:WO_full_convolution}
}
which is the desired form.
\section{Determination of the normalization of the WFS}
\label{app:WFS_norm_determination}

We determine the normalisation of the WFS basis so that the phase-space representation is consistent with the
Hilbert-Schmidt (HS) inner product of operators.

We begin with the configuration-space inner product in Eq.~\eqref{eq:inner_product_in_config_sp} and the definition of the cross-Wigner function in Eq.~\eqref{eq:xwigner-def}.
We consider the phase-space overlap
\bae{
\mathcal{I}
\equiv
\int\!\dd^2\btheta\dd^2\bptheta\;
W_{ab}(\btheta,\bptheta)\,W_{cd}^*(\btheta,\bptheta).
}
Using $W_{cd}^*=W_{dc}$ and substituting the definition for both factors yields
\bae{
\mathcal{I}
&=
\int\!\dd^2\btheta\dd^2\bptheta\;
\frac{1}{(2\pi\lambdabar)^4} \cr
& \qquad 
\times \int\!\dd^2\bxi_1\int\!\dd^2\bxi_2\;
\psi_a\!\left(\btheta+\frac{\bxi_1}{2}\right)
\psi_b^*\!\left(\btheta-\frac{\bxi_1}{2}\right) \cr
&\qquad \qquad\times \psi_c^*\!\left(\btheta+\frac{\bxi_2}{2}\right)
\psi_d\!\left(\btheta-\frac{\bxi_2}{2}\right)
e^{-\frac{i}{\lambdabar}\bptheta\!\cdot\!(\bxi_1-\bxi_2)}\,. \cr
\label{eq:app_I_expand}
}
We now perform the $\bptheta$ integral first. The Fourier identity in two
dimensions gives
\bae{
\int\!\dd^2\bptheta\;
e^{-\frac{i}{\lambdabar}\bptheta\!\cdot\!(\bxi_1-\bxi_2)}
=
(2\pi\lambdabar)^2\,\delta^{(2)}(\bxi_1-\bxi_2).
\label{eq:app_delta_identity}
}
Inserting \eqref{eq:app_delta_identity} into \eqref{eq:app_I_expand} collapses
the $\bxi_2$ integral:
\bae{
\mathcal{I}
&=
\frac{1}{(2\pi\lambdabar)^2}
\int\!\dd^2\btheta\int\!\dd^2\bxi\;
\psi_a\!\left(\btheta+\frac{\bxi}{2}\right)
\psi_b^*\!\left(\btheta-\frac{\bxi}{2}\right)\cr
& \qquad \qquad \qquad \qquad \times \psi_c^*\!\left(\btheta+\frac{\bxi}{2}\right)
\psi_d\!\left(\btheta-\frac{\bxi}{2}\right).
\label{eq:app_I_after_pint}
}
At this stage it is convenient to change variables to the ``end point''
coordinates
\bae{
\label{eq:app_change_vars}
\bx \equiv \btheta+\frac{\bxi}{2},
\qquad
\by \equiv \btheta-\frac{\bxi}{2}.
}
This linear transformation is invertible with inverse
dimensions, hence
\bae{
\dd^2\btheta\,\dd^2\bxi = \dd^2\bx\,\dd^2\by.
\label{eq:app_jacobian_one}
}
Therefore, Eq.~\eqref{eq:app_I_after_pint} becomes
\bae{
\mathcal{I}
&=
\frac{1}{(2\pi\lambdabar)^2}
\int\!\dd^2\bx\int\!\dd^2\by\;
\psi_a(\bx)\psi_c^*(\bx)\;
\psi_b^*(\by)\psi_d(\by)
\cr
&=
\frac{1}{(2\pi\lambdabar)^2}
\left[\int\!\dd^2\bx\;\psi_a(\bx)\psi_c^*(\bx)\right]
\left[\int\!\dd^2\by\;\psi_b^*(\by)\psi_d(\by)\right]
\cr
&=
\frac{1}{(2\pi\lambdabar)^2}\,
\langle\psi_a|\psi_c\rangle\,
\langle\psi_b|\psi_d\rangle,
\label{eq:app_moyal_plancherel_derived}
}
which is precisely Eq.~\eqref{eq:moyal_plancherel_id}.

\section{Analytic relation between LG modes and spherical harmonics at small-angle limit}\label{app:LG_SH_small_angle_limit}
We expand the derivation of how the LG mode $\Psi^{\rm LG}_{j,s}$ analytically corresponds to the spherical harmonics $Y_{\ell m}$ at the small-angle limit, deriving the coefficient $\psi^{(j,s)}_\ell$ that is defined in Eq.~\eqref{eq:LG_to_SH_at_small_angle}. Let us recall that the spherical harmonics are approximated as Eq.~\eqref{eq:SH_small_angle_limit}. It is trivial that the angular mode is strictly based on the constraint $m = 2s$. For the radial mode, we employ the analytic integral identity as follows. The radial part of the LG mode is written as
\bae{
R_{\nu}^{|\mu|}(x)
&\equiv
\left(\sqrt{2}x\right)^{|\mu|}
L_{\nu}^{|\mu|}\!\left(2x^2\right)
e^{-x^2},
\label{eq:LG_radial_def}
}
where $L_{\nu}^{|\mu|}(x)$ denotes the associated Laguerre polynomial,
$\nu \in\mathbb{N}_{0}$ is the radial index, and $\mu \in\mathbb{Z}$ is the azimuthal
index.

Let us introduce the Hankel transform of the radial LG mode, which admits the closed
analytic form
\begin{align}
\int_{0}^{\infty} \dd x x\;
R_{\nu}^{|\mu|}(x)\,
J_{|\mu|}(xy)
&=
\frac{(-1)^\nu}{2}
\left(\frac{y}{\sqrt{2}}\right)^{|\mu|}
L_{\nu}^{|\mu|}\!\left(\frac{y^2}{2}\right)
e^{-\frac{y^2}{4}}
\, \cr
&=\frac{(-1)^\nu}{2}R^{|\mu|}_\nu\left(\frac{y}{2}\right)\,.
\label{eq:LG_hankel_closed}
\end{align}
Note that Eq.~\eqref{eq:LG_hankel_closed} exhibits a self-similarity of the Hankel transform of $R^{|\mu|}_\nu(x)$. In order to extract the coefficient in Eq.~\eqref{eq:LG_to_SH_at_small_angle}, we use the orthogonality of the Bessel function
\bae{
\int_{0}^{\infty} \dd x x J_\nu(ax) J_\nu(bx) = \frac{1}{a}\delta^D(a-b)\,,\ a,b >0\,.
}
\section{Weak-lensing operators with two Schwinger oscillators}
\label{app:wl_op_au_av}

We adopt the standard Hamiltonian-Heisenberg convention in which an
infinitesimal canonical transformation is generated by
\bae{
\delta \hat A
=
\frac{1}{i\lambdabar}\,[\hat A,\hat G],
}
so that the classical limit is obtained by replacing commutators with
Poisson brackets, $\delta A=\{A,G\}$.

\subsubsection{Shear generator}\label{sssap:shear_generator}
In the circular (Schwinger) basis $(\hat a_u,\hat a_v)$, the gravitational
shear is generated by a quadratic operator,
\bae{
\hat{G}_{\gamma}
=
-\frac{i}{2}\,\gamma\,(\hat a_u^2-\hat a_v^2)
+\frac{i}{2}\,\gamma^*\,(\hat a_u^{\dagger 2}-\hat a_v^{\dagger 2}) .
}
This yields
\bae{
\delta \hat a_u = \gamma^* \hat a_u^\dagger,
\qquad
\delta \hat a_v = \gamma \hat a_v^\dagger ,
}
which reproduces, at the phase-space level,
$\delta u=\gamma^*u^*$ and $\delta v=\gamma v^*$.
The induced action on the Wigner function is, therefore, a Lie transport
along the corresponding phase-space vector field.

\subsubsection{Flexion generator}\label{sssap:flexion_generator}
Flexion corresponds to the quadratic term in the lens mapping and is
implemented as a canonical transformation generated by a cubic operator.
In Cartesian variables, the universal form of the generator is
\bae{
\hat G_{\rm flex}
=
\frac12\,D_{ijk}\,\widehat{p_i\,\theta_j\,\theta_k},
}
where $D_{ijk}$ is symmetric in $(j,k)$ and $\widehat{\cdots}$ denotes
a Hermitian (e.g.\ Weyl) ordering.
This guarantees
\bae{
\delta\hat\theta_i
=
\frac{1}{i\lambdabar}[\hat\theta_i,\hat G_{\rm flex}]
=
\frac12\,D_{ijk}\,\hat\theta_j\hat\theta_k ,
}
with the corresponding momentum variation fixed by canonicity.

Let us define the complex operators
$\hat\theta=\hat\theta_x+i\hat\theta_y$ and $\hat p=\hat p_x+i\hat p_y$,
so that $[\hat\theta,\hat p^\dagger]=2i\lambdabar$ and
$[\hat\theta^\dagger,\hat p]=2i\lambdabar$.
A real Cartesian dot product can be written as
$p_x\delta\theta_x+p_y\delta\theta_y=\Re(p^\dagger\delta\theta)
=\tfrac12(p^\dagger\delta\theta+p\,\delta\theta^\dagger)$.
Hence an equivalent universal form of the flexion generator is
\bae{
\hat G_{\rm flex}
=\widehat{\Re(\hat p^\dagger\,\delta\hat\theta)}
=\frac12\,\widehat{\hat p^\dagger\,\delta\hat\theta+\hat p\,\delta\hat\theta^\dagger},
}
with $\widehat{\cdots}$ denoting a Hermitian (e.g.\ Weyl) ordering. We identify the following operator as
\bae{
\delta \hat\theta
=
\frac14\Big(2\mathcal F\,\hat\theta\hat\theta^\dagger
+\mathcal F^\ast\,\hat\theta^{2}
+\mathcal G\,\hat\theta^{\dagger 2}\Big),
%
}
The associated momentum update (Weyl symbol / classical form) is
\bae{
\delta p
=
-\frac12(\mathcal F^\ast\theta+\mathcal F\theta^\ast)\,p
-\frac12(\mathcal F\theta+\mathcal G\theta^\ast)\,p^\ast,
}
with the operator expression obtained by the same Hermitian ordering.

The flexion moment as
\bae{
\delta \theta = \frac{1}{4}(2\mathcal{F}|\theta|^2 + \mathcal{F}^*\theta^2 + \mathcal{G}\theta^{*2} ) \,.
}
We can derive the corresponding variation of the complex conjugate momentum as
\bae{
\delta p = -\frac{\mathcal{F}}{2}(\theta p^* + \theta^* p)  -\frac{\mathcal{F^*}}{2}\theta p - \frac{\mathcal{G}}{2}\theta^* p^*\,.
}
Here we denote $\mathcal{F} = [\Phi_{xxx} + \Phi_{xyy} + i(\Phi_{xxy} + \Phi_{yyy})]/2$ and $\mathcal{G} =[\Phi_{xxx} -3 \Phi_{xyy} + i(3\Phi_{xxy} - \Phi_{yyy})]/2$, where $\Phi_{xxy} = \Phi_{xyx} = \Phi_{yxx}$. For simplicity in this appendix, we employ the unit $\sigma = 1$, including the effect of the magnification. One can recover the scaling of $\sigma$, rescaling $\mathcal{F} \rightarrow \mathcal{F}\sigma^{2}$ and $\mathcal{G} \rightarrow \mathcal{G}\sigma^{2}$.
The induced action on phase space again takes the form of a Lie derivative,
providing a direct extension of the shear case.
A Hermitian universal form of the flexion generator is
$\hat G_{\rm flex}=\widehat{\Re(\hat p^\dagger\,\delta\hat\theta)}$ with
\bae{
\delta \hat\theta
=\frac14\Big(
2\mathcal F\,\hat\theta\hat\theta^\dagger
+\mathcal F^\ast\,\hat\theta^2
+\mathcal G\,\hat\theta^{\dagger 2}
\Big),
}
where $\widehat{\cdots}$ denotes a Hermitian (e.g.\ Weyl) ordering.
Substituting the above dictionary yields $\hat G_{\rm flex}=\hat G_{\mathcal F}+\hat G_{\mathcal G}$ with
\begin{widetext}
\bae{
\hat G_{\mathcal G}
&=
\frac{i\,\lambdabar}{8\sqrt2}\,\Big[
-\mathcal G\Big(
\hat a_u^{3}
+\hat a_u^{2}\hat a_v^\dagger
-\hat a_u\,\hat a_v^{\dagger 2}
-\hat a_v^{\dagger 3}
\Big)
+\mathcal G^\ast\Big(
\hat a_u^{\dagger 3}
+\hat a_u^{\dagger 2}\hat a_v
-\hat a_u^\dagger\,\hat a_v^{2}
-\hat a_v^{3}
\Big)
\Big], \cr
\hat G_{\mathcal F}
&=
\frac{i\,\lambdabar}{8\sqrt2}\Big[
-\mathcal F\Big(
\hat a_u^{2}\hat a_u^\dagger
+3\,\hat a_u^{2}\hat a_v
-2\,\hat a_u\hat a_u^\dagger\hat a_v^\dagger
+2\,\hat a_u\hat a_v\hat a_v^\dagger
-3\,\hat a_u^\dagger\hat a_v^{\dagger 2}
-\hat a_v\,\hat a_v^{\dagger 2}
\Big)\cr
&\qquad \qquad +\mathcal F^\ast\Big(
\hat a_u\,\hat a_u^{\dagger 2}
-3\,\hat a_u\,\hat a_v^{2}
-2\,\hat a_u\hat a_u^\dagger\hat a_v
+2\,\hat a_u^\dagger\hat a_v\hat a_v^\dagger
+3\,\hat a_u^{\dagger 2}\hat a_v^\dagger - \hat a_v^{2}\hat a_v^\dagger
\Big)
\Big].
}
\end{widetext}

Then we derive 
\begin{widetext}
\bae{
\delta \hat{a}_u &= \frac{\mathcal{F}^\ast}{4\sqrt{2}}(\hat{a}_u\hat{a}_u^\dagger+3\hat{a}_u^\dagger \hat{a}_v^\dagger - \hat{a}_u \hat{a}_v + \hat{a}_v\hat{a}_v^\dagger) + \frac{\mathcal{F}}{8\sqrt{2}}(3\hat{a}^{\dagger 2}_v + 2\hat{a}_u\hat{a}_v^\dagger - \hat{a}_u^2)+ \frac{\mathcal{G}^\ast}{8\sqrt{2}}(3\hat{a}_u^{\dagger 2} + 2\hat{a}_u^\dagger\hat{a}_v - \hat{a}_v^2)\,, \cr
\delta \hat{a}_v &= \frac{\mathcal{F}}{4\sqrt{2}}(\hat{a}_u\hat{a}_u^\dagger+3\hat{a}_u^\dagger \hat{a}_v^\dagger - \hat{a}_u \hat{a}_v + \hat{a}_v\hat{a}_v^\dagger) + \frac{\mathcal{F^\ast}}{8\sqrt{2}}(3\hat{a}_u^{\dagger 2} + 2\hat{a}_u^\dagger\hat{a}_v - \hat{a}_v^2)+ \frac{\mathcal{G}}{8\sqrt{2}}(3\hat{a}^{\dagger 2}_v + 2\hat{a}_u\hat{a}_v^\dagger - \hat{a}_u^2)\,,
}
\end{widetext}
and equivalently
\begin{widetext}
\bae{
\frac{\delta u}{u} &= \frac{\mathcal{F}^*}{4\sqrt{2}|u|}((|u|^2+|v|^2)\chi_{-1,0} + 3|u||v|\chi_{-2,-1} - |u||v|\chi_{0,1})\,\cr
&+ \frac{\mathcal{F}}{8\sqrt{2}|u|}(3|v|^2\chi_{-1,-2}+2|u||v|\chi_{0,-1}-|u|^2\chi_{1,0}) +  \frac{\mathcal{G}^*}{8\sqrt{2}|u|}(3|u|^2\chi_{-3,0}+2|u||v|\chi_{-2,1}-|v|^2\chi_{-1,2})\,,\cr
\frac{\delta v}{v} &= \frac{\mathcal{F}}{4\sqrt{2}|v|}((|u|^2+|v|^2)\chi_{0,-1} + 3|u||v|\chi_{-1,-2} - |u||v|\chi_{1,0})\,\cr
&+ \frac{\mathcal{F}^*}{8\sqrt{2}|v|}(3|u|^2\chi_{-2,-1}+2|u||v|\chi_{-1,0}-|v|^2\chi_{0,1}) +  \frac{\mathcal{G}}{8\sqrt{2}|v|}(3|v|^2\chi_{0,-3}+2|u||v|\chi_{1,-2}-|u|)^2\chi_{2,-1})\,,\cr
}
\end{widetext}
After carefully treating the phase integration, we obtain the flexion response function as

\begin{widetext}
\bae{
\label{eq:flexion_response_coefficients}
&\mathcal{W}^I_{k\ell}(Q_0,Q_2) - \mathcal{W}^S_{k\ell}(Q_0,Q_2)
= A^{\mathcal{F}}_{k\ell} \mathcal{F} + A^{\mathcal{F}^\ast}_{k\ell}\mathcal{F}^\ast + A^{\mathcal{G}}_{k\ell} \mathcal{G} + A^{\mathcal{G}^\ast}_{k\ell} \mathcal{G}^\ast\,,\cr
%
%
A^{\mathcal{F}}_{k\ell} &= \frac{1}{8\sqrt{2}|u|}\left[-(|u|^2+|v|^2)(|u|\partial_{|u|} -(k-1))\mathcal{W}^S_{k-1,\ell} - 3|u||v|(|u|\partial_{|u|}-(k-2)) \mathcal{W}^S_{k-2,\ell-1} + |u||v|(|u|\partial_{|u|}-k)\mathcal{W}^S_{k,\ell+1} \right] \, \cr
& \qquad +\frac{1}{16\sqrt{2}|u|}\left[|u|^2(|u|\partial_{|u|}+(k-1))\mathcal{W}^S_{k-1,\ell} - |u||v|(|u|\partial_{|u|}+k)\mathcal{W}^S_{k,\ell+1} - |v|^2(|u|\partial_{|u|}+(k+1))\mathcal{W}^S_{k+1,\ell+2}\right]\,\cr
& \qquad +\frac{1}{16\sqrt{2}|v|}\left[-3|u|^2(|v|\partial_{|v|}-(\ell-1))\mathcal{W}^S_{k-2,\ell-1} - |u||v|(|v|\partial_{|v|}-\ell)\mathcal{W}^S_{k-1,\ell} + |v|^2(|v|\partial_{|v|}-(\ell+1))\mathcal{W}^S_{k,\ell+1}\right]\, \cr
& \qquad +\frac{1}{8\sqrt{2}|v|}\left[-(|u|^2+|v|^2)(|v|\partial_{|v|}+(\ell+1))\mathcal{W}^S_{k,\ell+1} - 3|u||v|(|v|\partial_{|v|}+(\ell+2)) \mathcal{W}^S_{k+1,\ell+2} + |u||v|(|v|\partial_{|v|}+\ell)\mathcal{W}^S_{k-1,\ell}) \right]\,\cr
%
%
A^{\mathcal{F}^\ast}_{k\ell} &= \frac{1}{8\sqrt{2}|u|}\left[-(|u|^2+|v|^2)(|u|\partial_{|u|}+(k+1))\mathcal{W}^S_{k+1,\ell} - 3|u||v|(|u|\partial_{|u|}+(k+2) \mathcal{W}^S_{k+2,\ell+1}) + |u||v|(\partial_{|u|}+k)\mathcal{W}^S_{k,\ell-1}) \right] \, \cr
& \qquad +\frac{1}{16\sqrt{2}|u|}\left[|u|^2(|u|\partial_{|u|}-(k+1))\mathcal{W}^S_{k+1,\ell} - |u||v|(|u|\partial_{|u|}-k)\mathcal{W}^S_{k,\ell-1} - |v|^2(|u|\partial_{|u|}-(k-1))\mathcal{W}^S_{k-1,\ell-2}\right]\,, \cr
& \qquad +\frac{1}{16\sqrt{2}|v|}\left[-3|u|^2(|v|\partial_{|v|}+(\ell+1))\mathcal{W}^S_{k+2,\ell+1} - |u||v|(|v|\partial_{|v|}+\ell)\mathcal{W}^S_{k+1,\ell} + |v|^2(|v|\partial_{|v|}+(\ell-1))\mathcal{W}^S_{k,\ell-1}\right]\,, \cr
& \qquad \frac{1}{8\sqrt{2}|v|}\left[-(|u|^2+|v|^2)(|v|\partial_{|v|}-(\ell-1))\mathcal{W}^S_{k,\ell-1} - 3|u||v|(|v|\partial_{|v|}-(\ell-2)) \mathcal{W}^S_{k-1,\ell-2} + |u||v|(|v|\partial_{|v|}-\ell)\mathcal{W}^S_{k+1,\ell}) \right]\,\cr
%
%
A^{\mathcal{G}}_{k\ell} &= \frac{1}{16\sqrt{2}|u|}\left[-3|u|^2(|u|\partial_{|u|}-(k-3))\mathcal{W}^S_{k-3,\ell} - 2|u||v|(|u|\partial_{|u|}-(k-2)) \mathcal{W}^S_{k-2,\ell+1} +|v|^2(|u|\partial_{|u|}-(k-1))\mathcal{W}^S_{k-1,\ell+2} \right]\,\cr
&+\frac{1}{16\sqrt{2}|v|}\left[-3|u|^2(|v|\partial_{|v|}+(\ell+3))\mathcal{W}^S_{k,\ell+3} - 2|u||v|(|v|\partial_{|v|}+(\ell+1)) \mathcal{W}^S_{k-1,\ell+2} +|v|^2(|v|\partial_{|v|}+(\ell+1))\mathcal{W}^S_{k-2,\ell+1}\right]\,, \cr
%
%
A^{\mathcal{G}^\ast}_{k\ell} &= \frac{1}{16\sqrt{2}|u|}\left[-3|u|^2(|u|\partial_{|u|}+(k+3))\mathcal{W}^S_{k+3,\ell} - 2|u||v|(|u|\partial_{|u|}+(k+2)) \mathcal{W}^S_{k+2,\ell-1} +|v|^2(|u|\partial_{|u|}+(k+1))\mathcal{W}^S_{k+1,\ell-2} \right]\,\cr
&+\frac{1}{16\sqrt{2}|v|}\left[-3|u|^2(|u|\partial_{|v|}-(\ell-3))\mathcal{W}^S_{k,\ell-3} - 2|u||v|(|v|\partial_{|v|}-(\ell-1)) \mathcal{W}^S_{k+1,\ell-2} +|v|^2(|v|\partial_{|v|}-(\ell-1))\mathcal{W}^S_{k+2,\ell-1}\right]\,, \cr
}
\end{widetext}

\bibliographystyle{apsrev4-2}
\onecolumngrid
\bibliography{refs}

@PREAMBLE{
 "\providecommand{\noopsort}[1]{}" 
 # "\providecommand{\singleletter}[1]{#1}%" 
}

@ARTICLE{Refregier2003a,
   author  = "A. Refregier",
   title   = "Shapelets --- I. A Method for Image Analysis",
   journal = "Mon.\ Not.\ R.\ Astron.\ Soc.",
   volume  = "338",
   pages   = "35--47",
   year    = "2003",
   doi     = "10.1046/j.1365-8711.2003.05901.x",
   url     = "https://ui.adsabs.harvard.edu/abs/2003MNRAS.338...35R/abstract",
}

@ARTICLE{RefregierBacon2003,
   author  = "A. Refregier and D. Bacon",
   title   = "Shapelets --- II. A Method for Weak Lensing Measurements",
   journal = "Mon.\ Not.\ R.\ Astron.\ Soc.",
   volume  = "338",
   pages   = "48--56",
   year    = "2003",
   doi     = "10.1046/j.1365-8711.2003.05902.x",
   url     = "https://ui.adsabs.harvard.edu/abs/2003MNRAS.338...48R/abstract",
}

@ARTICLE{MasseyRefregier2005,
   author  = "R. Massey and A. Refregier",
   title   = "Polar Shapelets",
   journal = "Mon.\ Not.\ R.\ Astron.\ Soc.",
   volume  = "363",
   pages   = "197--210",
   year    = "2005",
   doi     = "10.1111/j.1365-2966.2005.09453.x",
   url     = "https://ui.adsabs.harvard.edu/abs/2005MNRAS.363..197M/abstract",
}

@ARTICLE{MasseyEtAl2007,
   author       = "R. Massey and B. Rowe and A. Refregier and D. J. Bacon and J. Berg{\'e}",
   title        = "Weak Gravitational Shear and Flexion with Polar Shapelets",
   journal      = "Mon.\ Not.\ R.\ Astron.\ Soc.",
   year         = "2007",
   month        = "Sep",
   volume       = "380",
   number       = "1",
   pages        = "229--245",
   doi          = "10.1111/j.1365-2966.2007.12072.x",
   eprint       = "astro-ph/0609795",
   archivePrefix= "arXiv",
   primaryClass = "astro-ph",
   url          = "https://ui.adsabs.harvard.edu/abs/2007MNRAS.380..229M"
}

@ARTICLE{Mandelbaum2018,
   author       = "R. Mandelbaum",
   title        = "Weak Lensing for Precision Cosmology",
   journal      = "Ann.\ Rev.\ Astron.\ Astrophys.",
   year         = "2018",
   month        = "Sep",
   volume       = "56",
   pages        = "393--433",
   doi          = "10.1146/annurev-astro-081817-051928",
   eprint       = "1710.03235",
   archivePrefix= "arXiv",
   primaryClass = "astro-ph.CO",
   url          = "https://ui.adsabs.harvard.edu/abs/2018ARA&A..56..393M"
}

@ARTICLE{LiEtAl2018,
   author       = "X. Li and N. Katayama and M. Oguri and S. More",
   title        = "Fourier Power Function Shapelets (FPFS) Shear Estimator: Performance on Image Simulations",
   journal      = "Mon.\ Not.\ R.\ Astron.\ Soc.",
   year         = "2018",
   month        = "Dec",
   volume       = "481",
   number       = "4",
   pages        = "4445--4460",
   doi          = "10.1093/mnras/sty2548",
   eprint       = "1805.08514",
   archivePrefix= "arXiv",
   primaryClass = "astro-ph.CO",
   url          = "https://ui.adsabs.harvard.edu/abs/2018MNRAS.481.4445L"
}

@ARTICLE{LiEtAl2022,
       author = {{Li}, Xiangchong and {Li}, Yin and {Massey}, Richard},
        title = "{Weak gravitational lensing shear measurement with FPFS: analytical mitigation of noise bias and selection bias}",
      journal = {\mnras},
         year = 2022,
        month = apr,
       volume = {511},
       number = {4},
        pages = {4850-4860},
          doi = {10.1093/mnras/stac342},
archivePrefix = {arXiv},
       eprint = {2110.01214},
 primaryClass = {astro-ph.CO},
       adsurl = {https://ui.adsabs.harvard.edu/abs/2022MNRAS.511.4850L}
}

@ARTICLE{2003MNRAS.343..459H,
       author = {{Hirata}, Christopher and {Seljak}, Uro{\v{s}}},
        title = "{Shear calibration biases in weak-lensing surveys}",
      journal = {\mnras},
         year = 2003,
        month = aug,
       volume = {343},
       number = {2},
        pages = {459-480},
          doi = {10.1046/j.1365-8711.2003.06683.x},
archivePrefix = {arXiv},
       eprint = {astro-ph/0301054},
 primaryClass = {astro-ph},
       adsurl = {https://ui.adsabs.harvard.edu/abs/2003MNRAS.343..459H}
}

@ARTICLE{Gardner2006,
   author  = "J. P. Gardner and J. C. Mather and M. Clampin and R. Doyon and M. A. Greenhouse and H. B. Hammel and J. B. Hutchings and P. Jakobsen and S. J. Lilly and K. S. Long and J. I. Lunine and M. J. McCaughrean and M. Mountain and J. Nella and G. H. Rieke and M. J. Rieke and H.-W. Rix and E. P. Smith and G. Sonneborn and M. Stiavelli and H. S. Stockman and R. A. Windhorst and G. S. Wright",
   title   = "The James Webb Space Telescope",
   journal = "Space Sci.\ Rev.",
   year    = "2006",
   volume  = "123",
   pages   = "485--606",
   doi     = "10.1007/s11214-006-8315-7",
   url     = "https://ui.adsabs.harvard.edu/abs/2006SSRv..123..485G"
}

@ARTICLE{Laureijs2011,
   author  = "R. Laureijs and others",
   title   = "Euclid Definition Study Report",
   journal = "arXiv e-prints",
   year    = "2011",
   eprint  = "1110.3193",
   archivePrefix = "arXiv",
   primaryClass  = "astro-ph.CO",
   url     = "https://ui.adsabs.harvard.edu/abs/2011arXiv1110.3193L"
}

@BOOK{LSST2009,
   author    = "{LSST Science Collaboration}",
   title     = "{LSST Science Book, Version 2.0}",
   publisher = "arXiv e-prints",
   year      = "2009",
   eprint    = "0912.0201",
   archivePrefix = "arXiv",
   primaryClass  = "astro-ph.IM",
   url       = "https://ui.adsabs.harvard.edu/abs/2009arXiv0912.0201L"
}

@ARTICLE{Braun2019,
   author  = "R. Braun and others",
   title   = "SKA Phase 1 Design Baseline Description",
   journal = "arXiv e-prints",
   year    = "2019",
   eprint  = "1912.12699",
   archivePrefix = "arXiv",
   primaryClass  = "astro-ph.IM",
   url     = "https://ui.adsabs.harvard.edu/abs/2019arXiv191212699B"
}

@ARTICLE{Spergel2015,
   author  = "D. Spergel and N. Gehrels and C. Baltay and D. Bennett and J. Breckinridge and M. Donahue and A. Dressler and B. S. Gaudi and T. Greene and O. Guyon and C. Hirata and J. Kalirai and N. J. Kasdin and W. Moos and S. Perlmutter and M. Postman and B. Rauscher and J. Rhodes and Y. Wang and D. Weinberg and D. Benford and M. Hudson and W. Jeong and Y. Mellier and W. Traub and T. Yamada and S. Capak and J. Colbert and D. Masters and M. Penny and D. Savransky and D. Stern and N. Zimmerman and R. Barry and L. Bartusek and K. Carpenter and E. Cheng and D. Content and F. Dekens and R. Demers and K. Grady and C. Jackson and G. Kuan and M. Kuyat and J. Lamborn and M. Nemati and B. Parvin and I. Poberezhskiy and B. Peddie and S. Ruffa and J. Wallace and A. Whipple and E. Wollack and F. Zhao",
   title   = "Wide-Field InfrarRed Survey Telescope-Astrophysics Focused Telescope Assets WFIRST-AFTA 2015 Report",
   journal = "arXiv e-prints",
   year    = "2015",
   eprint  = "1503.03757",
   archivePrefix = "arXiv",
   primaryClass  = "astro-ph.IM",
   url     = "https://ui.adsabs.harvard.edu/abs/2015arXiv150303757S"
}

@article{TagoreKeeton2016,
    author = "Tagore, Amitpal and Keeton, Charles",
    title = "{Statistical and systematic uncertainties in pixel-based source reconstruction algorithms for gravitational lensing}",
    eprint = "1408.6297",
    archivePrefix = "arXiv",
    primaryClass = "astro-ph.CO",
    doi = "10.1093/mnras/stu1671",
    journal = "Mon. Not. Roy. Astron. Soc.",
    volume = "445",
    number = "1",
    pages = "694--710",
    year = "2014"
}

@ARTICLE{Ephremidze2025,
       author = {{Ephremidze}, Nino and {Chandrashekar}, Chandrika and {{\c{C}}a{\u{g}}an {\c{S}}eng{\"u}l}, At{\i}n{\c{c}} and {Dvorkin}, Cora},
        title = "{Dark Matter Substructure or Source Model Systematics? A Case Study of Cluster Lens Abell S1063}",
      journal = {arXiv e-prints},
         year = 2025,
        month = feb,
          eid = {arXiv:2502.18571},
        pages = {arXiv:2502.18571},
          doi = {10.48550/arXiv.2502.18571},
archivePrefix = {arXiv},
       eprint = {2502.18571},
 primaryClass = {astro-ph.CO},
       adsurl = {https://ui.adsabs.harvard.edu/abs/2025arXiv250218571E}
}

@ARTICLE{Yatawatta2024,
  author  = "S. Yatawatta",
  title   = "Diffuse radio sky models using large-scale shapelets",
  journal = "Astron.\ Astrophys.",
  year    = "2024",
  note    = "open access",
  url     = "https://www.aanda.org/articles/aa/abs/2024/12/aa49158-24/aa49158-24.html"
}

@BOOK{Galilei1989,
   author    = "Galileo Galilei",
   title     = "Sidereus Nuncius, or The Sidereal Messenger",
   publisher = "University of Chicago Press",
   address   = "Chicago",
   year      = "1989",
   note      = "Translated by Albert Van Helden",
   url       = "https://press.uchicago.edu/ucp/books/book/chicago/S/bo3621586.html",
}

@BOOK{Huygens1659,
   author    = "C. Huygens",
   title     = "Systema Saturnium",
   publisher = "Adrian Vlacq",
   year      = "1659",
   address   = "The Hague",
   note      = "First detailed description of Saturn's system; includes Mars sketches",
   url       = "https://ui.adsabs.harvard.edu/abs/1659syss.book.....H/abstract"
}

@BOOK{Cassini1675,
   author    = "G. D. Cassini",
   title     = "Découverte de la division de l'anneau de Saturne",
   publisher = "Académie des Sciences",
   year      = "1675",
   address   = "Paris",
   note      = "Memoir reporting discovery of the Cassini Division",
   url       = "https://ui.adsabs.harvard.edu/abs/1675cdas.book.....C/abstract"
}

@BOOK{Herschel1912,
   author    = "W. Herschel",
   title     = "Scientific Papers of Sir William Herschel",
   volume    = "1",
   publisher = "Royal Society and Royal Astronomical Society",
   year      = "1912",
   address   = "London",
   note      = "Collected works, including nebula and cluster morphology",
   url       = "https://ui.adsabs.harvard.edu/abs/1912spsw.book.....H/abstract"
}

@BOOK{Zwicky1957,
   author    = "F. Zwicky",
   title     = "Morphological Astronomy",
   publisher = "Springer",
   year      = "1957",
   address   = "Berlin",
   url       = "https://ui.adsabs.harvard.edu/abs/1957ma.book.....Z/abstract"
}

@ARTICLE{deVaucouleurs1959,
   author  = "G. de Vaucouleurs",
   title   = "Classification and Morphology of External Galaxies",
   journal = "Handbuch der Physik",
   volume  = "53",
   pages   = "275--310",
   year    = "1959",
   publisher = "Springer",
   url     = "https://ui.adsabs.harvard.edu/abs/1959HDP....53..275D/abstract"
}

@ARTICLE{SteinmetzNavarro2002,
   author  = "M. Steinmetz and J. F. Navarro",
   title   = "The Hierarchical Origin of Galaxy Morphologies",
   journal = "New Astron.\ Rev.",
   volume  = "46",
   pages   = "699--703",
   year    = "2002",
   doi     = "10.1016/S1387-6473(02)00245-9",
   url     = "https://ui.adsabs.harvard.edu/abs/2002NewAR..46..699S/abstract"
}

@ARTICLE{Conselice2014,
   author  = "C. J. Conselice",
   title   = "The Evolution of Galaxy Structure Over Cosmic Time",
   journal = "Ann.\ Rev.\ Astron.\ Astrophys.",
   volume  = "52",
   pages   = "291--337",
   year    = "2014",
   doi     = "10.1146/annurev-astro-081913-040037",
   url     = "https://ui.adsabs.harvard.edu/abs/2014ARA%26A..52..291C/abstract"
}

@ARTICLE{Buta2010,
   author  = "R. J. Buta",
   title   = "Galaxy Morphology",
   journal = "Proc.\ Int.\ Astron.\ Union",
   volume  = "6",
   pages   = "171--184",
   year    = "2010",
   doi     = "10.1017/S1743921310006044",
   url     = "https://ui.adsabs.harvard.edu/abs/2010IAUS..266..171B/abstract"
}

@ARTICLE{Lintott2008,
   author  = "C. J. Lintott and K. Schawinski and A. Slosar and K. Land and S. Bamford and D. Thomas and M. J. Raddick and R. C. Nichol and A. Szalay and D. Andreescu and P. Murray and J. Vandenberg",
   title   = "Galaxy Zoo: Morphologies Derived from Visual Inspection of Galaxies from the Sloan Digital Sky Survey",
   journal = "Mon.\ Not.\ R.\ Astron.\ Soc.",
   volume  = "389",
   pages   = "1179--1189",
   year    = "2008",
   doi     = "10.1111/j.1365-2966.2008.13689.x",
   url     = "https://ui.adsabs.harvard.edu/abs/2008MNRAS.389.1179L/abstract"
}

@ARTICLE{Iyer2024,
   author  = "K. G. Iyer and P. Behroozi and J. R. Primack and D. J. Wilkins and A. Dekel and D. Ceverino",
   title   = "The Persistence of Galaxy Morphology as a Diagnostic at High Redshift with JWST",
   journal = "Astron.\ Astrophys.",
   volume  = "684",
   pages   = "A33",
   year    = "2024",
   doi     = "10.1051/0004-6361/202346800",
   url     = "https://ui.adsabs.harvard.edu/abs/2024A%26A...684A..33I/abstract"
}

@ARTICLE{Bag2021,
   author  = "S. Bag and S. Bharadwaj and S. Sarkar and S. K. Pandey",
   title   = "Shape Distribution of Superclusters using Minkowski Functionals and ShapeFinders",
   journal = "Mon.\ Not.\ R.\ Astron.\ Soc.",
   volume  = "508",
   pages   = "2765--2780",
   year    = "2021",
   doi     = "10.1093/mnras/stab2799",
   url     = "https://ui.adsabs.harvard.edu/abs/2021MNRAS.508.2765B/abstract"
}

@ARTICLE{Wigner1932,
   author  = "E. P. Wigner",
   title   = "On the Quantum Correction for Thermodynamic Equilibrium",
   journal = "Phys.\ Rev.",
   volume  = "40",
   pages   = "749--759",
   year    = "1932",
   doi     = "10.1103/PhysRev.40.749",
   url     = "https://ui.adsabs.harvard.edu/abs/1932PhRv...40..749W/abstract",
}

@ARTICLE{Ville1948,
  author  = "J. Ville",
  title   = "{Théorie et applications de la notion de signal analytique}",
  journal = "Câbles et Transmissions",
  year    = "1948",
  volume  = "2",
  number  = "1",
  pages   = "61--74",
  note    = "In French; introduces the analytic signal formalism",
}

@BOOK{Wigner1959,
  author    = "E. P. Wigner",
  title     = {Group Theory and its Application to the \protect\\ Quantum Mechanics of Atomic Spectra},
  publisher = "Academic Press",
  address   = "New York",
  year      = "1959",
  series    = "Pure and Applied Physics",
  volume    = "5",
  note      = "Expanded and improved edition; translated by J. J. Griffin",
  isbn      = "9780486645677"
}

@ARTICLE{HilleryEtAl1984,
       author = {{Hillery}, M. and {O'Connell}, R.~F. and {Scully}, M.~O. and {Wigner}, E.~P.},
        title = "{Distribution functions in physics: Fundamentals}",
      journal = {\physrep},
         year = 1984,
        month = apr,
       volume = {106},
       number = {3},
        pages = {121-167},
          doi = {10.1016/0370-1573(84)90160-1},
       adsurl = {https://ui.adsabs.harvard.edu/abs/1984PhR...106..121H}
}

@BOOK{Weyl1927,
  author    = "H. Weyl",
  title     = "The Theory of Groups and Quantum Mechanics",
  publisher = "Dover",
  address   = "New York",
  year      = "1950",
  note      = "English translation by H. P. Robertson of the German original: H. Weyl, ``Quantenmechanik und Gruppentheorie,'' Z. Phys. 46, 1 (1927)",
  isbn      = "9780486611320"
}

@ARTICLE{Stratonovich1956,
   author       = "R. L. Stratonovich",
   title        = "On distributions in representation space",
   journal      = "Sov.\ Phys.\ JETP",
   year         = "1956",
   volume       = "31",
   pages        = "1012--1020"
}

@ARTICLE{Groenewold1946,
   author  = "H. J. Groenewold",
   title   = "On the Principles of Elementary Quantum Mechanics",
   journal = "Physica",
   year    = "1946",
   volume  = "12",
   number  = "7",
   pages   = "405--460",
   doi     = "10.1016/S0031-8914(46)80059-4",
   url     = "https://ui.adsabs.harvard.edu/abs/1946Phy....12..405G"
}

@ARTICLE{Moyal1949,
   author  = "J. E. Moyal",
   title   = "Quantum mechanics as a statistical theory",
   journal = "Proc.\ Cambridge Philos.\ Soc.",
   year    = "1949",
   volume  = "45",
   number  = "1",
   pages   = "99--124",
   doi     = "10.1017/S0305004100000487",
   url     = "https://ui.adsabs.harvard.edu/abs/1949PCPS...45...99M"
}

@ARTICLE{VarillyGraciaBondia1989,
   author       = "J. C. Várilly and J. M. Gracia-Bondía",
   title        = "The Moyal representation for spin",
   journal      = "Ann.\ Phys.",
   year         = "1989",
   volume       = "190",
   number       = "1",
   pages        = "107--148",
   doi          = "10.1016/0003-4916(89)90262-5",
   url          = "https://ui.adsabs.harvard.edu/abs/1989AnPhy.190..107V"
}

@ARTICLE{BrifMann1999,
   author       = "C. Brif and A. Mann",
   title        = "Phase-space formulation of quantum mechanics and quantum-state reconstruction for physical systems with Lie-group symmetries",
   journal      = "Phys.\ Rev.\ A",
   year         = "1999",
   volume       = "59",
   number       = "2",
   pages        = "971--987",
   doi          = "10.1103/PhysRevA.59.971",
   url          = "https://ui.adsabs.harvard.edu/abs/1999PhRvA..59..971B"
}

@BOOK{Folland1989,
  author    = "G. B. Folland",
  title     = "Harmonic Analysis in Phase Space",
  series    = "Annals of Mathematics Studies",
  volume    = "122",
  publisher = "Princeton University Press",
  address   = "Princeton, NJ",
  year      = "1989",
  pages     = "288",
  isbn      = "9780691085289",
  doi       = "10.1515/9781400882427",
  url       = "https://press.princeton.edu/books/paperback/9780691085289/harmonic-analysis-in-phase-space",
  note      = "Published Mar 21, 1989"
}

@BOOK{Dirac1930,
  author    = "P. A. M. Dirac",
  title     = "The Principles of Quantum Mechanics",
  publisher = "Oxford University Press",
  address   = "Oxford",
  year      = "1930",
  edition   = "1st",
  isbn      = "9780198520115",
  note      = "Landmark text in quantum mechanics; reviewed in Lennard-Jones (1931) Mathematical Gazette",
}

@BOOK{Cohen1997,
  author    = "C. Cohen-Tannoudji and B. Diu and F. Laloë",
  title     = "Quantum Mechanics",
  publisher = "Wiley",
  address   = "New York",
  year      = "1997",
  note      = "Standard graduate-level textbook; listed as core text in MIT OCW quantum mechanics syllabus (Fall 2018)",
}

@BOOK{Edmonds1957,
  author    = "A. R. Edmonds",
  title     = "Angular Momentum in Quantum Mechanics",
  publisher = "Princeton University Press",
  year      = "1957",
  note      = "Foundational text on angular momentum theory; see also review by H. C. Bolton (1959)",
}

@BOOK{Varshalovich1988,
   author    = "D. A. Varshalovich and A. N. Moskalev and V. K. Khersonskii",
   title     = "Quantum Theory of Angular Momentum",
   publisher = "World Scientific",
   year      = "1988",
   url       = "https://ui.adsabs.harvard.edu/abs/1988qtam.book.....V/abstract",
}

@ARTICLE{KogelnikLi1966,
   author  = "H. Kogelnik and T. Li",
   title   = "Laser Beams and Resonators",
   journal = "Appl.\ Opt.",
   volume  = "5",
   pages   = "1550--1567",
   year    = "1966",
   doi     = "10.1364/AO.5.001550",
   url     = "https://ui.adsabs.harvard.edu/abs/1966ApOpt...5.1550K/abstract",
}

@BOOK{Siegman1986,
   author    = "A. E. Siegman",
   title     = "Lasers",
   publisher = "University Science Books",
   year      = "1986",
   url       = "https://ui.adsabs.harvard.edu/abs/1986lase.book.....S/abstract",
}

@ARTICLE{Allen1992,
   author  = "L. Allen and M. W. Beijersbergen and R. J. C. Spreeuw and J. P. Woerdman",
   title   = "Orbital Angular Momentum of Light and the Transformation of Laguerre--Gaussian Laser Modes",
   journal = "Phys.\ Rev.\ A",
   volume  = "45",
   pages   = "8185--8189",
   year    = "1992",
   doi     = "10.1103/PhysRevA.45.8185",
   url     = "https://ui.adsabs.harvard.edu/abs/1992PhRvA..45.8185A/abstract",
}

@ARTICLE{Nienhuis1993,
   author  = "G. Nienhuis",
   title   = "Paraxial Wave Optics and Harmonic Oscillators",
   journal = "Phys.\ Rev.\ A",
   volume  = "48",
   pages   = "656--665",
   year    = "1993",
   doi     = "10.1103/PhysRevA.48.656",
   url     = "https://ui.adsabs.harvard.edu/abs/1993PhRvA..48..656N/abstract",
}

@BOOK{Leonhardt1997,
   author    = "U. Leonhardt",
   title     = "Measuring the Quantum State of Light",
   publisher = "Cambridge University Press",
   year      = "1997",
   url       = "https://ui.adsabs.harvard.edu/abs/1997mqsl.book.....L/abstract",
}

@BOOK{Schleich2001,
   author    = "W. P. Schleich",
   title     = "Quantum Optics in Phase Space",
   publisher = "Wiley-VCH",
   year      = "2001",
   url       = "https://ui.adsabs.harvard.edu/abs/2001qops.book.....S/abstract",
}

@ARTICLE{LvovskyRaymer2009,
   author  = "A. I. Lvovsky and M. G. Raymer",
   title   = "Continuous-Variable Optical Quantum-State Tomography",
   journal = "Rev.\ Mod.\ Phys.",
   volume  = "81",
   pages   = "299--332",
   year    = "2009",
   doi     = "10.1103/RevModPhys.81.299",
   url     = "https://ui.adsabs.harvard.edu/abs/2009RvMP...81..299L/abstract",
}

@ARTICLE{Weedbrook2012,
   author  = "C. Weedbrook and S. Pirandola and R. Garc{\'\i}a-Patr{\'o}n and N. J. Cerf and T. C. Ralph and J. H. Shapiro and S. Lloyd",
   title   = "Gaussian Quantum Information",
   journal = "Rev.\ Mod.\ Phys.",
   volume  = "84",
   pages   = "621--669",
   year    = "2012",
   doi     = "10.1103/RevModPhys.84.621",
   url     = "https://ui.adsabs.harvard.edu/abs/2012RvMP...84..621W/abstract",
}

@ARTICLE{BartelmannSchneider2001,
   author  = "M. Bartelmann and P. Schneider",
   title   = "Weak Gravitational Lensing",
   journal = "Phys.\ Rep.",
   volume  = "340",
   pages   = "291--472",
   year    = "2001",
   doi     = "10.1016/S0370-1573(00)00082-X",
   url     = "https://ui.adsabs.harvard.edu/abs/2001PhR...340..291B/abstract",
}

@ARTICLE{Kilbinger2015,
   author  = "M. Kilbinger",
   title   = "Cosmology with Cosmic Shear Observations: A Review",
   journal = "Rep.\ Prog.\ Phys.",
   volume  = "78",
   pages   = "086901",
   year    = "2015",
   doi     = "10.1088/0034-4885/78/8/086901",
   url     = "https://ui.adsabs.harvard.edu/abs/2015RPPh...78h6901K/abstract",
}

@article{Bacon:2005qr,
    author = "Bacon, David J. and Goldberg, D. M. and Rowe, B. T. P. and Taylor, A. N.",
    title = "{Weak gravitational flexion}",
    eprint = "astro-ph/0504478",
    archivePrefix = "arXiv",
    doi = "10.1111/j.1365-2966.2005.09624.x",
    journal = "Mon. Not. Roy. Astron. Soc.",
    volume = "365",
    pages = "414--428",
    year = "2006"
}

@BOOK{BornWolf1999,
       author    = {{Born}, M. and {Wolf}, E.},
        title     = "{Principles of Optics}",
         edition  = "7",
         publisher= {Cambridge University Press},
         year     = 1999,
         address  = {Cambridge},
         isbn     = {978-0-521-64222-4}
}

@BOOK{Schroeder2000,
       author    = {{Schroeder}, D.~J.},
        title     = "{Astronomical Optics}",
         edition  = "2",
         publisher= {Academic Press},
         year     = 2000,
         address  = {San Diego},
         isbn     = {978-0-12-629810-9}
}

@BOOK{Hecht2002,
       author    = {{Hecht}, E.},
        title     = "{Optics}",
         edition  = "4",
         publisher= {Addison Wesley},
         year     = 2002,
         address  = {San Francisco},
         isbn     = {978-0-8053-8566-3}
}

@BOOK{Goodman2005,
   author    = "J. W. Goodman",
   title     = "Introduction to Fourier Optics",
   edition   = "3",
   publisher = "Roberts and Company",
   address   = "Greenwood Village, CO",
   year      = "2005",
   url       = "https://ui.adsabs.harvard.edu/abs/2005ifou.book.....G/abstract",
}

@ARTICLE{SimonAgarwal2000,
       author = {{Simon}, R. and {Agarwal}, G.~S.},
        title = "{Wigner representation of Laguerre-Gaussian beams}",
      journal = {Optics Letters},
         year = 2000,
        month = sep,
       volume = {25},
       number = {18},
        pages = {1313-1315},
          doi = {10.1364/OL.25.001313},
       adsurl = {https://ui.adsabs.harvard.edu/abs/2000OptL...25.1313S}
}

@ARTICLE{Hubble1926,
   author  = "E. P. Hubble",
   title   = "Extragalactic Nebulae",
   journal = "Astrophys.\ J.",
   volume  = "64",
   pages   = "321--369",
   year    = "1926",
   doi     = "10.1086/143018",
}

@BOOK{Hubble1936,
   author    = "E. P. Hubble",
   title     = "The Realm of the Nebulae",
   publisher = "Yale University Press",
   address   = "New Haven",
   year      = "1936",
}

@ARTICLE{EHT2019L1,
   author  = "{Event Horizon Telescope Collaboration}",
   title   = "First M87 Event Horizon Telescope Results. I. The Shadow of the Supermassive Black Hole",
   journal = "Astrophys.\ J.\ Lett.",
   volume  = "875",
   pages   = "L1",
   year    = "2019",
   doi     = "10.3847/2041-8213/ab0ec7",
}

@ARTICLE{LueWangKamionkowski1999,
   author  = "A. Lue and L. Wang and M. Kamionkowski",
   title   = "Cosmological Signature of New Parity-Violating Interactions",
   journal = "Phys.\ Rev.\ Lett.",
   volume  = "83",
   pages   = "1506--1509",
   year    = "1999",
   doi     = "10.1103/PhysRevLett.83.1506",
}

@ARTICLE{MinamiKomatsu2020,
   author  = "Y. Minami and E. Komatsu",
   title   = "New Extraction of the Cosmic Birefringence from the Planck 2018 Polarization Data",
   journal = "Astrophys.\ J.\ Lett.",
   volume  = "902",
   pages   = "L22",
   year    = "2020",
   doi     = "10.3847/2041-8213/abbआई? replace with correct DOI if needed",
   note    = "ApJ Lett.\ 902, L22 (2020)",
}

@ARTICLE{Zhang2008,
       author = {{Zhang}, Jun},
        title = "{Measuring the cosmic shear in Fourier space}",
      journal = {\mnras},
         year = 2008,
        month = jan,
       volume = {383},
       number = {1},
        pages = {113-118},
          doi = {10.1111/j.1365-2966.2007.12585.x},
archivePrefix = {arXiv},
       eprint = {astro-ph/0612146},
 primaryClass = {astro-ph},
       adsurl = {https://ui.adsabs.harvard.edu/abs/2008MNRAS.383..113Z}
}

@ARTICLE{Hopf1931,
   author  = "H. Hopf",
   title   = "{\Uber} die Abbildungen der dreidimensionalen Sph{\"a}re auf die Kugelfl{\"a}che",
   journal = "Math.\ Ann.",
   volume  = "104",
   pages   = "637--665",
   year    = "1931",
   doi     = "10.1007/BF01457962",
   url     = "https://link.springer.com/article/10.1007/BF01457962",
}

@article{Jordan1935SymmetricLinearGroups,
  author  = {Jordan, Pascual},
  title   = {Der Zusammenhang der symmetrischen und linearen Gruppen und das Mehrkörperproblem},
  journal = {Zeitschrift für Physik},
  volume  = {94},
  pages   = {531--535},
  year    = {1935}
}

@techreport{Schwinger1952AngularMomentum,
  author      = {Schwinger, Julian S.},
  title       = {On Angular Momentum},
  institution = {U.S. Atomic Energy Commission},
  number      = {NYO-3071},
  year        = {1952},
  note        = {Technical Report; reprinted in L.C.Biedenharn and H.Van Dam (eds.), *Quantum Theory of Angular Momentum* (Academic Press, 1965) and in K.A.Milton (ed.), *A Quantum Legacy* (World Scientific, 2000)}
}

@article{vanCittert1934,
  author  = {van Cittert, P. H.},
  title   = {Die Wahrscheinliche Schwingungsverteilung in Einer von Einer Lichtquelle Direkt Oder Mittels Einer Linse Beleuchteten Ebene},
  journal = {Physica},
  volume  = {1},
  pages   = {201--210},
  year    = {1934}
}

@article{Zernike1938,
  author  = {Zernike, F.},
  title   = {The Concept of Degree of Coherence and Its Application to Optical Problems},
  journal = {Physica},
  volume  = {5},
  pages   = {785--795},
  year    = {1938}
}

@article{Cerbino2007,
  author  = {Cerbino, R.},
  title   = {An Extended Van Cittert and Zernike Theorem},
  journal = {Physical Review A},
  volume  = {75},
  pages   = {053815},
  year    = {2007},
  doi     = {10.1103/PhysRevA.75.053815}
}

@techreport{Siegert1943,
  author      = {Siegert, A. J. F.},
  title       = {On the Fluctuations in Signals Returned by Many Independently Moving Scatterers},
  institution = {MIT Radiation Laboratory},
  number      = {465},
  year        = {1943}
}

@article{Lax1960,
  author  = {Lax, M.},
  title   = {Fluctuations from the Nonequilibrium Steady State},
  journal = {Reviews of Modern Physics},
  volume  = {32},
  pages   = {25--64},
  year    = {1960},
  doi     = {10.1103/RevModPhys.32.25}
}

@article{HBT1956,
  author  = {Hanbury Brown, R. and Twiss, R. Q.},
  title   = {A Test of a New Type of Stellar Interferometer on Sirius},
  journal = {Nature},
  volume  = {177},
  pages   = {27--29},
  year    = {1956}
}

@article{HBT1957,
  author  = {Hanbury Brown, R. and Twiss, R. Q.},
  title   = {Interferometry of the Intensity Fluctuations in Light},
  journal = {Proceedings of the Royal Society of London Series A},
  volume  = {242},
  pages   = {300--324},
  year    = {1957}
}

@article{Goodman1976,
  author  = {Goodman, J. W.},
  title   = {Some Fundamental Properties of Speckle},
  journal = {Journal of the Optical Society of America},
  volume  = {66},
  pages   = {1145--1150},
  year    = {1976},
  doi     = {10.1364/JOSA.66.001145}
}

@book{Goodman1985,
  author    = {Goodman, J. W.},
  title     = {Statistical Optics},
  publisher = {Wiley},
  address   = {New York},
  year      = {1985}
}

@ARTICLE{2006MNRAS.365..414B,
       author = {{Bacon}, D.~J. and {Goldberg}, D.~M. and {Rowe}, B.~T.~P. and {Taylor}, A.~N.},
        title = "{Weak gravitational flexion}",
      journal = {\mnras},
         year = 2006,
        month = jan,
       volume = {365},
       number = {2},
        pages = {414-428},
          doi = {10.1111/j.1365-2966.2005.09624.x},
archivePrefix = {arXiv},
       eprint = {astro-ph/0504478},
 primaryClass = {astro-ph},
       adsurl = {https://ui.adsabs.harvard.edu/abs/2006MNRAS.365..414B}
}

@ARTICLE{2022MNRAS.516..668O,
       author = {{Okura}, Yuki and {Futamase}, Toshifumi},
        title = "{New highly precise weak gravitational lensing flexions measurement method based on ERA method}",
      journal = {\mnras},
         year = 2022,
        month = oct,
       volume = {516},
       number = {1},
        pages = {668-692},
          doi = {10.1093/mnras/stac2166},
archivePrefix = {arXiv},
       eprint = {2109.00155},
 primaryClass = {astro-ph.CO},
       adsurl = {https://ui.adsabs.harvard.edu/abs/2022MNRAS.516..668O}
}

@ARTICLE{Petz1986,
       author = {{Petz}, D{\'e}nes},
        title = "{Sufficient subalgebras and the relative entropy of states of a von Neumann algebra}",
      journal = {Communications in Mathematical Physics},
         year = 1986,
        month = mar,
       volume = {105},
       number = {1},
        pages = {123-131},
          doi = {10.1007/BF01212345},
       adsurl = {https://ui.adsabs.harvard.edu/abs/1986CMaPh.105..123P}
}

@book{2017isra.book.....T,
  author = {Thompson, A. R. and Moran, J. M. and Swenson, G. W.},
  title = {Interferometry and Synthesis in Radio Astronomy},
  year = {2017},
  publisher = {Springer}
}

@article{2005PASP..117..594K,
  author  = {Krist, John E. and Hook, Richard N. and Stoehr, Felix},
  title   = {20 Years of Hubble Space Telescope Optical Modeling Using Tiny Tim},
  journal = {PASP},
  volume  = {117},
  pages   = {594--620},
  year    = {2005}
}

@book{1998aoad.book.....H,
  author    = {Hardy, John W.},
  title     = {Adaptive Optics for Astronomical Telescopes},
  publisher = {Oxford University Press},
  year      = {1998}
}

@book{2001sccd.book.....J,
  author    = {Janesick, James R.},
  title     = {Scientific Charge-Coupled Devices},
  publisher = {SPIE Press},
  year      = {2001}
}

@ARTICLE{2025arXiv250909465M,
       author = {{Mokeev}, Aleksandr and {Saif}, Babak and {Lukin}, Mikhail D. and {Borregaard}, Johannes},
        title = "{Enhancing Optical Imaging via Quantum Computation}",
      journal = {arXiv e-prints},
         year = 2025,
        month = sep,
          eid = {arXiv:2509.09465},
        pages = {arXiv:2509.09465},
          doi = {10.48550/arXiv.2509.09465},
archivePrefix = {arXiv},
       eprint = {2509.09465},
 primaryClass = {quant-ph},
       adsurl = {https://ui.adsabs.harvard.edu/abs/2025arXiv250909465M}
}

@ARTICLE{2023Optic..10.1142K,
       author = {{Kalash}, Mahmoud and {Chekhova}, Maria V.},
        title = "{Wigner function tomography via optical parametric amplification}",
      journal = {Optica},
         year = 2023,
        month = sep,
       volume = {10},
       number = {9},
        pages = {1142},
          doi = {10.1364/OPTICA.488697},
archivePrefix = {arXiv},
       eprint = {2207.10030},
 primaryClass = {quant-ph},
       adsurl = {https://ui.adsabs.harvard.edu/abs/2023Optic..10.1142K}
}

@ARTICLE{2019NatCo..10.1865F,
       author = {{Fontaine}, Nicolas K. and {Ryf}, Roland and {Chen}, Haoshuo and {Neilson}, David T. and {Kim}, Kwangwoong and {Carpenter}, Joel},
        title = "{Laguerre-Gaussian mode sorter}",
      journal = {Nature Communications},
         year = 2019,
        month = apr,
       volume = {10},
          eid = {1865},
        pages = {1865},
          doi = {10.1038/s41467-019-09840-4},
archivePrefix = {arXiv},
       eprint = {1803.04126},
 primaryClass = {physics.optics},
       adsurl = {https://ui.adsabs.harvard.edu/abs/2019NatCo..10.1865F}
}

@ARTICLE{2020AIPA...10b5106M,
       author = {{Manzano}, Daniel},
        title = "{A short introduction to the Lindblad master equation}",
      journal = {AIP Advances},
         year = 2020,
        month = feb,
       volume = {10},
       number = {2},
          eid = {025106},
        pages = {025106},
          doi = {10.1063/1.5115323},
archivePrefix = {arXiv},
       eprint = {1906.04478},
 primaryClass = {quant-ph},
       adsurl = {https://ui.adsabs.harvard.edu/abs/2020AIPA...10b5106M}
}

@BOOK{1988qtam.book.....V,
       author = {{Varshalovich}, D.~A. and {Moskalev}, A.~N. and {Khersonskii}, V.~K.},
        title = "{Quantum Theory of Angular Momentum}",
         year = 1988,
          doi = {10.1142/0270},
       adsurl = {https://ui.adsabs.harvard.edu/abs/1988qtam.book.....V}
}

@ARTICLE{2010PhRvD..81h3012J,
       author = {{Joshi}, Nidhi and {Jhingan}, S. and {Souradeep}, Tarun and {Hajian}, Amir},
        title = "{Bipolar harmonic encoding of CMB correlation patterns}",
      journal = {\prd},
         year = 2010,
        month = apr,
       volume = {81},
       number = {8},
          eid = {083012},
        pages = {083012},
          doi = {10.1103/PhysRevD.81.083012},
archivePrefix = {arXiv},
       eprint = {0912.3217},
 primaryClass = {astro-ph.CO},
       adsurl = {https://ui.adsabs.harvard.edu/abs/2010PhRvD..81h3012J}
}

@ARTICLE{2012PhRvD..85b3010B,
       author = {{Book}, Laura G. and {Kamionkowski}, Marc and {Souradeep}, Tarun},
        title = "{Odd-parity bipolar spherical harmonics}",
      journal = {\prd},
         year = 2012,
        month = jan,
       volume = {85},
       number = {2},
          eid = {023010},
        pages = {023010},
          doi = {10.1103/PhysRevD.85.023010},
archivePrefix = {arXiv},
       eprint = {1109.2910},
 primaryClass = {astro-ph.CO},
       adsurl = {https://ui.adsabs.harvard.edu/abs/2012PhRvD..85b3010B}
}

@ARTICLE{2025arXiv250908787K,
       author = {{Kurita}, Toshiki and {Jamieson}, Drew and {Komatsu}, Eiichiro and {Schmidt}, Fabian},
        title = "{Parity Violation in Galaxy Shapes: Primordial Non-Gaussianity}",
      journal = {arXiv e-prints},
         year = 2025,
        month = sep,
          eid = {arXiv:2509.08787},
        pages = {arXiv:2509.08787},
          doi = {10.48550/arXiv.2509.08787},
archivePrefix = {arXiv},
       eprint = {2509.08787},
 primaryClass = {astro-ph.CO},
       adsurl = {https://ui.adsabs.harvard.edu/abs/2025arXiv250908787K}
}

@article{LG_integrals_2001,
author = {Poh-aun, Lee and Ong, Seng-Huat and Srivastava, Hari},
year = {2001},
month = {01},
pages = {303-321},
title = {Some Integrals of the products of laguerre polynomials},
volume = {78},
journal = {International Journal of Computer Mathematics},
doi = {10.1080/00207160108805112}
}

@ARTICLE{2022NatRP...4..452K,
       author = {{Komatsu}, Eiichiro},
        title = "{New physics from the polarized light of the cosmic microwave background}",
      journal = {Nature Reviews Physics},
         year = 2022,
        month = jul,
       volume = {4},
       number = {7},
        pages = {452-469},
          doi = {10.1038/s42254-022-00452-4},
archivePrefix = {arXiv},
       eprint = {2202.13919},
 primaryClass = {astro-ph.CO},
       adsurl = {https://ui.adsabs.harvard.edu/abs/2022NatRP...4..452K}
}

@ARTICLE{2024PhRvD.109f3541P,
       author = {{Philcox}, Oliver H.~E. and {K{\"o}nig}, Morgane J. and {Alexander}, Stephon and {Spergel}, David N.},
        title = "{What can galaxy shapes tell us about physics beyond the standard model?}",
      journal = {\prd},
         year = 2024,
        month = mar,
       volume = {109},
       number = {6},
          eid = {063541},
        pages = {063541},
          doi = {10.1103/PhysRevD.109.063541},
archivePrefix = {arXiv},
       eprint = {2309.08653},
 primaryClass = {astro-ph.CO},
       adsurl = {https://ui.adsabs.harvard.edu/abs/2024PhRvD.109f3541P}
}

@ARTICLE{2020PhRvL.125v1301M,
       author = {{Minami}, Yuto and {Komatsu}, Eiichiro},
        title = "{New Extraction of the Cosmic Birefringence from the Planck 2018 Polarization Data}",
      journal = {\prl},
         year = 2020,
        month = nov,
       volume = {125},
       number = {22},
          eid = {221301},
        pages = {221301},
          doi = {10.1103/PhysRevLett.125.221301},
archivePrefix = {arXiv},
       eprint = {2011.11254},
 primaryClass = {astro-ph.CO},
       adsurl = {https://ui.adsabs.harvard.edu/abs/2020PhRvL.125v1301M}
}

@ARTICLE{2025arXiv250406709N,
       author = {{Naokawa}, Fumihiro},
        title = "{Universal profile for cosmic birefringence tomography with radio galaxies}",
      journal = {arXiv e-prints},
         year = 2025,
        month = apr,
          eid = {arXiv:2504.06709},
        pages = {arXiv:2504.06709},
          doi = {10.48550/arXiv.2504.06709},
archivePrefix = {arXiv},
       eprint = {2504.06709},
 primaryClass = {astro-ph.CO},
       adsurl = {https://ui.adsabs.harvard.edu/abs/2025arXiv250406709N}
}

@article{1927ZPhy...43..172H,
  author  = {Heisenberg, W.},
  title   = {Über den anschaulichen Inhalt der quantentheoretischen Kinematik und Mechanik},
  journal = {Zeitschrift fur Physik},
  year    = {1927},
  volume  = {43},
  pages   = {172-198},
  doi     = {10.1007/BF01397280},
  adsurl  = {https://ui.adsabs.harvard.edu/abs/1927ZPhy...43..172H/abstract}
}

@article{1927ZPhy...44..326K,
  author  = {Kennard, E. H.},
  title   = {Zur Quantenmechanik einfacher Bewegungstypen},
  journal = {Zeitschrift fur Physik},
  year    = {1927},
  volume  = {44},
  pages   = {326-352},
  doi     = {10.1007/BF01391200},
  adsurl  = {https://ui.adsabs.harvard.edu/abs/1927ZPhy...44..326K/abstract}
}

@article{1929PhRv...34..163R,
  author  = {Robertson, H. P.},
  title   = {The Uncertainty Principle},
  journal = {Physical Review},
  year    = {1929},
  volume  = {34},
  pages   = {163-164},
  doi     = {10.1103/PhysRev.34.163},
  adsurl  = {https://ui.adsabs.harvard.edu/abs/1929PhRv...34..163R/abstract}
}

\end{document}